\documentclass[10pt,aps,twocolumn,superscriptaddress,longbibliography]{revtex4}
\usepackage[resetlabels]{multibib}
\usepackage{hyperref}
\usepackage{graphicx, amsmath, verbatim, dsfont, amsfonts, color}
\usepackage{bbm, bm, latexsym, amssymb}
\usepackage[us]{datetime}
\usepackage{color}
\usepackage{tikz, xcolor}

\newcommand{\dagga}{{\phantom{\dagger}}}
\newcommand{\g}[1]{{\bf #1}}

\definecolor{lime}{HTML}{A6CE39}
\DeclareRobustCommand{\orcidicon}{%
	\begin{tikzpicture}
	\draw[lime, fill=lime] (0,0)
	circle [radius=0.16]
	node[white] {{\fontfamily{qag}\selectfont \tiny ID}};
	node[white] {{\fontfamily{cmr}\selectfont \tiny ID}}
	\draw[white, fill=white] (-0.0625,0.095)
	circle [radius=0.007];
	\end{tikzpicture}
	\hspace{-2mm}
}

\foreach \x in {A, ..., Z}{%
	\expandafter\xdef\csname orcid\x\endcsname{\noexpand\href{https://orcid.org/\csname orcidauthor\x\endcsname}{\noexpand\orcidicon}}
}

\newcites{S}{Supplement}

\newcommand{\thetitle}{Signatures of dephasing by mirror-symmetry breaking in weak-antilocalization magnetoresistance across the topological transition in Pb$_{1-x}$Sn$_{x}$Se}

\begin{document}

\title{\thetitle}

\author{Alexander~Kazakov\orcidA}
\email{kazakov@MagTop.ifpan.edu.pl}
\affiliation{International Research Centre MagTop, Institute of Physics, Polish Academy of Sciences, Aleja Lotnikow 32/46, PL-02668 Warsaw, Poland}
\author{Wojciech~Brzezicki\orcidB}
\affiliation{International Research Centre MagTop, Institute of Physics, Polish Academy of Sciences, Aleja Lotnikow 32/46, PL-02668 Warsaw, Poland}
\affiliation{Institute of Theoretical Physics, Jagiellonian University, ulica S. \L{}ojasiewicza 11, PL-30348 Krak\'ow, Poland}
\author{Timo~Hyart\orcidC}
\email{timo.hyart@MagTop.ifpan.edu.pl}
\affiliation{International Research Centre MagTop, Institute of Physics, Polish Academy of Sciences, Aleja Lotnikow 32/46, PL-02668 Warsaw, Poland}
\affiliation{Department of Applied Physics, Aalto University, FI-00076 Aalto, Espoo, Finland}
\author{Bart\l{}omiej~Turowski}
\affiliation{International Research Centre MagTop, Institute of Physics, Polish Academy of Sciences, Aleja Lotnikow 32/46, PL-02668 Warsaw, Poland}
\author{Jakub~Polaczy{\'{n}}ski}
\affiliation{International Research Centre MagTop, Institute of Physics, Polish Academy of Sciences, Aleja Lotnikow 32/46, PL-02668 Warsaw, Poland}
\author{Zbigniew~Adamus}
\affiliation{Institute of Physics, Polish Academy of Sciences, Aleja Lotnikow 32/46, PL-02668 Warsaw, Poland}
\author{Marta~Aleszkiewicz}
\affiliation{Institute of Physics, Polish Academy of Sciences, Aleja Lotnikow 32/46, PL-02668 Warsaw, Poland}
\author{Tomasz~Wojciechowski\orcidG}
\affiliation{International Research Centre MagTop, Institute of Physics, Polish Academy of Sciences, Aleja Lotnikow 32/46, PL-02668 Warsaw, Poland}
\author{Jaroslaw~Z.~Domagala}
\affiliation{Institute of Physics, Polish Academy of Sciences, Aleja Lotnikow 32/46, PL-02668 Warsaw, Poland}
\author{Ond{\v r}ej~Caha}
\affiliation{Department of Condensed Matter Physics, Faculty of Science, Masaryk University, Kotl{\'a}{\v r}sk{\'a} 2, Cz-611 37 Brno, Czech Republic}
\author{Andrei~Varykhalov}
\affiliation{Helmholtz-Zentrum Berlin f{\"u}r Materialien und Energie, Albert-Einstein Strasse 15, D-12489 Berlin, Germany}
\author{Gunther~Springholz}
\affiliation{Institut f\"ur Halbleiter- und Festk\"orperphysik, Johannes Kepler University, Altenbergerstrasse 69, A-4040 Linz, Austria}
\author{Tomasz~Wojtowicz\orcidD}
\affiliation{International Research Centre MagTop, Institute of Physics, Polish Academy of Sciences, Aleja Lotnikow 32/46, PL-02668 Warsaw, Poland}
\author{Valentine~V.~Volobuev\orcidE}
\email{volobuiev@MagTop.ifpan.edu.pl}
\affiliation{International Research Centre MagTop, Institute of Physics, Polish Academy of Sciences, Aleja Lotnikow 32/46, PL-02668 Warsaw, Poland}
\affiliation{National Technical University "KhPI"{}, Kyrpychova Str. 2, 61002 Kharkiv, Ukraine}
\author{Tomasz~Dietl\orcidF}
\email{dietl@MagTop.ifpan.edu.pl}
\affiliation{International Research Centre MagTop, Institute of Physics, Polish Academy of Sciences, Aleja Lotnikow 32/46, PL-02668 Warsaw, Poland}
\affiliation{WPI Advanced Institute for Materials Research, Tohoku University, 2-1-1 Katahira, Aoba-ku, Sendai 980-8577, Japan}

\begin{abstract}
Many conductors, including recently studied Dirac materials, show saturation of coherence length on decreasing temperature. This surprising phenomenon is assigned to external noise, residual magnetic impurities or two-level systems specific to non-crystalline solids. Here,  by considering the SnTe-class of compounds as an example, we show theoretically that breaking of mirror symmetry deteriorates Berry's phase quantization, leading to additional dephasing in weak-antilocalization magnetoresistance (WAL-MR). Our experimental studies of WAL-MR corroborate these theoretical expectations in (111) Pb$_{1-x}$Sn$_x$Se thin film with Sn contents $x$ corresponding to both topological crystalline insulator and topologically trivial phases.  In particular, we find the shortening of the phase coherence length in samples with intentionally broken mirror symmetry. Our results indicate that the classification of quantum transport phenomena into universality classes should encompass, in addition to time-reversal and spin-rotation invariances, spatial symmetries in specific systems.
\end{abstract}

\maketitle

\section{Introduction}

One of the most powerful characterizations of quantum systems is in terms of ten universality classes that correspond to different ways fermionic Hamiltonians transform under time-reversal ${\cal{T}}$, particle-hole and chiral symmetry operations \cite{Beenakker1997, Altland:1997_PRB, Ryu:2010_NJP}.  This generic approach, immune to space symmetry details, allows describing specificities of transport and topological  phenomena in a broad range of normal and superconducting materials \cite{Beenakker1997, Altland:1997_PRB, Ryu:2010_NJP, Evers:2008_RMP}. In the sector of normal conductors, this classification leads to three major experimentally realized cases depending on the presence (+) or the absence (--) of ${\cal{T}}$ and  the spin-rotation invariance ${\cal{S}}$. The instances in the presence of time-reversal symmetry are referred to as orthogonal (++) and symplectic (+--) class, whereas the universality class in the absence of time-reversal symmetry is known as the unitary class \cite{Hikami1980,Beenakker1997}. The unitary class is sometimes divided into subclasses depending on the existence of the spin-rotation symmetry and spin-polarization in systems where the effects of carrier interactions are relevant \cite{Wojtowicz:1986_PRL,Finkelstein:1990_SSR}.

It becomes, however, increasingly clear that this picture is not complete. For instance, weak-antilocalization (WAL) magnetoresistance (MR) described by the Hikami-Larkin-Nagaoka (HLN) formula \cite{Hikami1980} is expected for the symplectic class in the limit of strong spin-orbit scattering, for example, due to: (i) spin-momentum locking of carriers encircling 2D gapless Dirac cones at surfaces of 3D topological materials \cite{Lu:2011_PRB,Garate:2012_PRB,Adroguer:2015_PRB,Wang:2020_PRL}; (ii) a large precession frequency in the interfacial Rashba field compared to the inverse momentum relaxation time \cite{Golub:2016_PRB}; and (iii) a strong Elliott-Yafet mechanism due to sizable mixing of spin states in the carrier wave functions \cite{Chatterjee:2019_PRB}. Surprisingly, however, robust WAL MR is also observed  for graphene \cite{Wu:2007_PRL}.  It has actually been found that because of isospin-momentum locking, the carriers encircling 2D gapless Dirac cones acquire the Berry phase $\varphi = \pi$,  which eliminates backscattering and, thus, results in MR that mimics the symplectic case, even though the spin-orbit interaction is negligible. However, in graphene also the space group symmetries are important because both the intervalley scattering and trigonal warping of the cones, which makes momentum ${\bm p}$ nonequivalent to $-{\bm p}$, suppress the WAL effect \cite{McCann:2006_PRL,Wu:2007_PRL}. Similarly, the theoretical discovery of topological indices associated with crystal point group symmetries \cite{Fu:2011_PRL} showed that the tenfold classification  has to be much extended to incorporate a rather abundant family of topological crystalline insulators (TCIs) and superconductors \cite{Chiu:2016_RMP}. In particular, it was predicted theoretically \cite{Hsieh2012,Safaei:2013_PRB} and confirmed experimentally \cite{Dziawa2012,Tanaka:2012_NP,Xu:2012_NC,Tanaka2013,Polley2014} that the mirror symmetry ${\cal{M}}$ can protect the presence of gapless topological Dirac cones on (001) and (111) surfaces of cubic SnTe-type semiconductors with the inverted band structure.

Here, by taking thin films of cubic SnTe-type semiconductors as an example, we show---combining analytic and numerical approaches developed recently \cite{Brzezicki2019}---that the Berry phase for carriers encircling the Fermi loops is quantized to $\pi$ in these systems once inversion symmetry is broken by, for instance, due to the differences in top and bottom surfaces. This quantization is independent of the Fermi level position in respect to the bulk bandgap, and occurs for the band  arrangement corresponding to topologically nontrivial and trivial phases accessible in Pb$_{1-x}$Sn$_{x}$Se materials with $x>x_c$ and $x <x_c$, respectively, where $x_c=0.16$ \cite{assaf2017magnetooptical,Krizman2018}. Our detailed analysis supported by a direct conductance determination, indicates that WAL MR in these layers might be affected not only by doping with magnetic impurities, as found previously in a range of materials \cite{Liu2012d,Tkac:2019_PRL,Sawicki:1986_PRL,Adhikari:2019_PRB}, but also by breaking ${\cal{M}}$.

In order to test these theoretical predictions, we have grown a series of epitaxial (111) Pb$_{1-x}$Sn$_{x}$Se thin films with the Sn content $0\le x \le 0.3$ capping {\em in-situ} part of the epilayers by amorphous Se. Angle-resolved photoemission spectroscopy (ARPES) confirms that our samples cover both sides of the topological phase transition. Nevertheless, but in agreement with our theoretical expectations, we observe for any $x$ robust WAL MR, well described by the HLN formula in the limit of strong spin-orbit scattering and with the prefactor $\alpha = -1/2$ \cite{Hikami1980}.  This insight elucidates why robust WAL MR has previously been observed not only for topological SnTe epilayers \cite{Adhikari:2019_PRB,Assaf2014,Akiyama2015,Yan:2020_JMST} or Pb$_{0.7}$Sn$_{0.3}$Se quantum wells \cite{Wang:2020_PRB}, but also for non-topological PbTe/(Pb,Eu)Te quantum wells \cite{Peres:2014_JAP}. Importantly, our data reveal striking differences between WAL MR in uncapped films compared to samples in which the amorphous Se surface layer intentionally breaks the mirror symmetry. This breaking of ${\cal{M}}$, relevant as electrons penetrate the cap,  has two consequences revealed here experimentally and explained theoretically: (i) $\tau_{\phi}(T)$ determined from WAL MR saturates below 4\,K in the Se covered samples but not in samples without a Se cap; (ii) the conductance decay with the in-plane magnetic field faster in films covered by Se. Our results, together with those for graphene \cite{McCann:2006_PRL,Wu:2007_PRL}, lead to a rather striking conclusion that specific space symmetries can account for an apparent low-temperature saturation of $\tau_{\phi}(T)$ determined from WAL MR studies. This finding suggests that hidden spatial-symmetry properties, together with decoherence specific to amorphous solids \cite{Afonin:2002_PRB}, might have often been responsible for a hitherto mysterious low-temperature saturation of $\tau_{\phi}(T)$ observed  in many systems \cite{Lin:2002_JPC}, including recently studied Dirac materials \cite{Jing:2016_NS,Islam2019,Rosen:2019_PRB,Nakamura:2020_NC,Wang:2020_PRB}.

\begin{figure}[b]
    \centering
    \includegraphics[width=0.95\columnwidth]{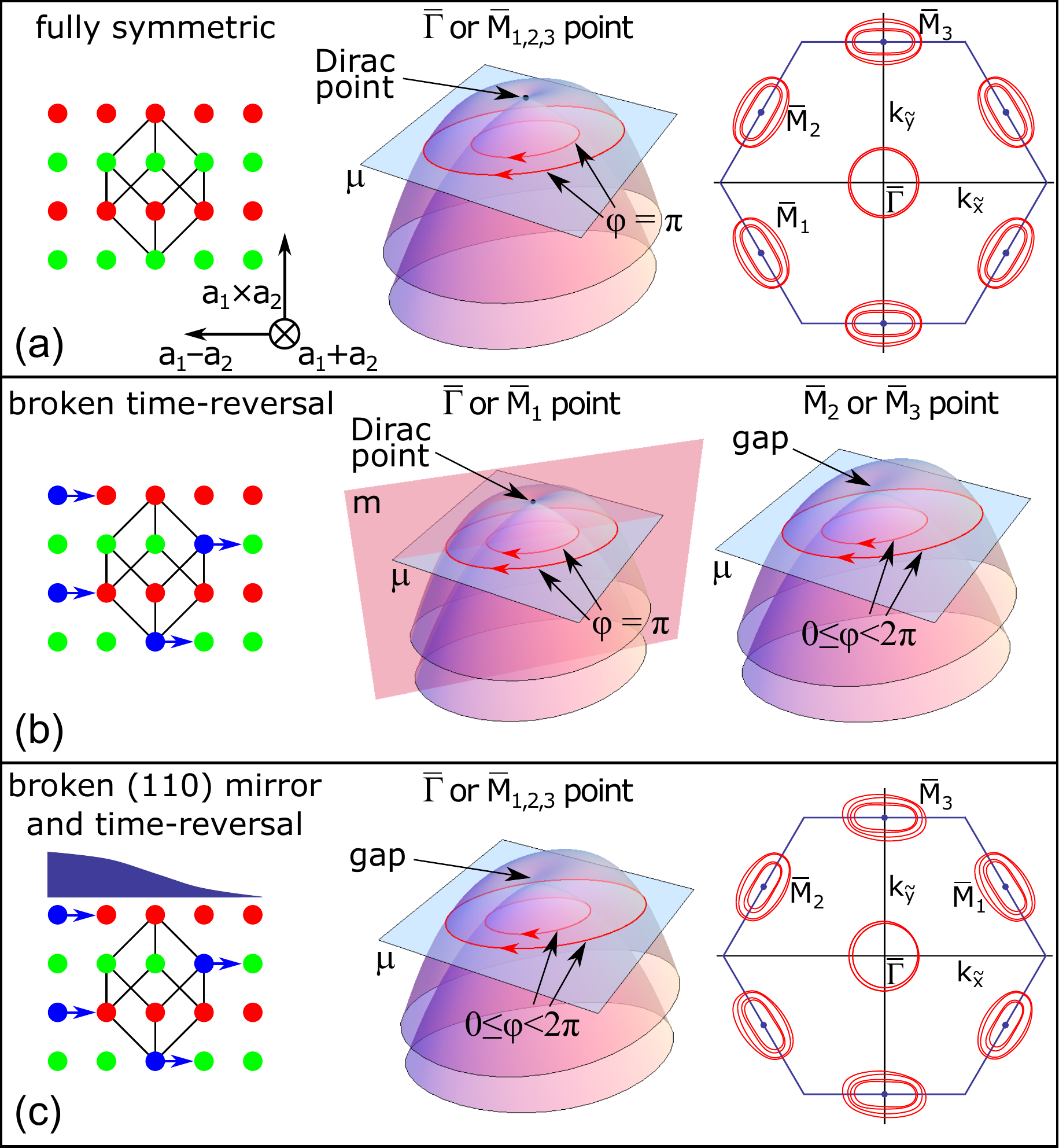}
    \caption{Dependence of Fermi loops and Berry phases $\varphi$ on the presence of time-reversal $\cal T$ and (110) mirror $\cal M$ symmetries for doped Pb$_{1-x}$Sn$_x$Se thin films. (a) If $\cal T$ and $\cal M$ are obeyed the Berry phases of the Fermi loops in subbands of a two-dimensional slab are quantized to $\varphi=\pi$, protecting subband crossings at high-symmetry points $\overline{\Gamma}$ and $\overline{M}_i$ (black dot). This also holds when only $\cal M$ is broken. (b) If $\cal T$ is broken (e.g.~by a Zeeman field shown by arrows) but $\cal M$ is obeyed the Berry phases are quantized to $\pi$  only for Fermi loops going around  a subbands' crossing at  the $\overline{\Gamma}$ and $\overline{M}_1$ points lying in the mirror plane. (c) The Berry phases are arbitrary when both $\cal T$ and $\cal M$ are simultaneously broken (arrows and asymmetric surface layer).}
    \label{fig:loops}
\end{figure}

\section{Theory}
\subsection{Mirror and time-reversal symmetry protected quantization of the Berry phases}

\begin{figure}[bt]
    \includegraphics[width=0.7\columnwidth]{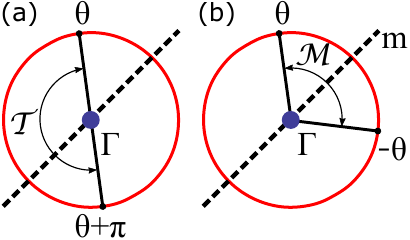}
    \caption{Action of time-reversal $\cal T$ and mirror $\cal M$ symmetries on states belonging to the Fermi loop around $\overline{\Gamma}$ point. Points along the loop are parametrized by the angle $\theta$ in respect to the mirror line (dashed). (a) Time-reversal maps state $|\psi_{\theta}^{(n)}\rangle$ onto $|\psi_{\theta+\pi}^{(n)}\rangle$.  (b) Mirror symmetry maps state $|\psi_{\theta}^{(n)}\rangle$ onto $|\psi_{-\theta}^{(n)}\rangle$.}
    \label{s_fig:T_M}
\end{figure}

We consider the symmetry-enriched berryology of the (111) multilayer system in the SnTe material class \cite{Hsieh2012} (see Supplemental Material, Ref.\, \onlinecite{SM} Sec.\,S1, for more details). By doping the system we obtain Fermi loops around the high-symmetry points $\overline{\Gamma}$ and $\overline{M}_i$ ($i=1, 2, 3$) (see Fig.~\ref{fig:loops}).  The type of doping (electron or hole doping) is not important for our theoretical considerations because both types of doping yield similar Fermi loops and our results  follow from generic symmetry arguments.  In our calculations, all 2D subbands are non-degenerate  along the Fermi loops  because of the spin-orbit interaction and the inversion asymmetry due to the presence of inequivalent surfaces. We show that both time-reversal $\cal T$ and (110) mirror $\cal M$ symmetries  lead to quantization of the Berry phases (see Fig.~\ref{s_fig:T_M}).  Due to three-fold rotational symmetry, there exists also two other mirror symmetries which would lead to equivalent considerations.

In the presence of time-reversal symmetry obeying ${\cal T}^{2}=-1$, we obtain that all the Fermi loops have quantized Berry phase $\varphi=\pi$ [see Fig.\,\ref{fig:loops}(a)]. To prove this we consider the eigenstates $|\psi_{\theta}^{(n)}\rangle$, $\theta\in [-\pi,\pi)$, belonging to the $n$th energy band forming a Fermi loop around $\overline{\Gamma}$ (or $\overline{M}_1$) point (see Fig.~\ref{s_fig:T_M}). Since $\theta$ is periodic variable $|\psi_{\theta}^{(n)}\rangle$ can acquire a Berry phase by a parallel shift along the loop. The gauge-invariant form of the Berry phase is given by,
\begin{widetext}
\begin{equation}
\varphi_{n}=\arg\left[\left\langle \psi_{-\pi}^{(n)}\right|\!\left.\psi_{-\pi+\delta\theta}^{(n)}\right\rangle \!\left\langle \psi_{-\pi+\delta\theta}^{(n)}\right|\!\left.\psi_{-\pi+2\delta\theta}^{(n)}\right\rangle \!\dots\!\left\langle \psi_{\pi-2\delta\theta}^{(n)}\right|\!\left.\psi_{\pi-\delta\theta}^{(n)}\right\rangle \negmedspace\left\langle \psi_{\pi-\delta\theta}^{(n)}\right|\!\left.\psi_{-\pi}^{(n)}\right\rangle \right].
\label{eq:berry}
\end{equation}
\end{widetext}

where $\delta\theta$ is an infinitesimal step in angle $\theta$. In the presence of a time-reversal symmetry  we obtain the states $|\psi_{\theta}^{(n)}\rangle$  for $-\pi \leq \theta < 0$ by diagonalizing the Hamiltonian and we define $0 \leq \theta<\pi$ states as $|\psi_{\theta+\pi}^{(n)}\rangle ={\cal T}|\psi_{\theta}^{(n)}\rangle$. By this construction  we find that most of the phases in Eq.~(\ref{eq:berry}) cancel and we are left with (see Ref.\,\onlinecite{SM}, Sec.\,S2)

\begin{equation}
\varphi_{n} = \arg\left[-\left|\left\langle \psi_{-\delta\theta}^{(n)}\right|\!\left.\psi_{0}^{(n)}\right\rangle \right|^{2}\right]=\pi.
\end{equation}

Hence, in the presence of time-reversal symmetry satisfying ${\cal T}^{2}=-1$ the Berry phases are quantized to $\pi$. Equivalently we can say that the Berry phases are equal to $\pi$ because each Fermi loop encircles a crossing (i.e.~Dirac point) of the subbands  protected by Kramers degeneracy at high-symmetry points of the Brillouin zone (BZ). We point out that for time-reversal symmetry obeying ${\cal T}^2=1$ (no spin-orbit interaction) the Berry phases would be quantized to $0$. Thus, our symmetry analysis of the Berry phases reproduces the well-known result that materials with strong (weak) spin-orbit coupling support WAL (WL) due to belonging to the symplectic (orthogonal) universality classes \cite{Hikami1980,Beenakker1997}. This result holds both in the topologically trivial and non-trivial regimes, and therefore we expect WAL independently of the Sn content in Pb$_{1-x}$Sn$_{x}$Se thin films.

It turns out that the presence of crystalline mirror symmetry $\cal M$ can lead to the quantization of Berry phases even if time-reversal symmetry $\cal T$ is broken, e.g., by a non-zero Zeeman field, as shown in Fig.~\ref{fig:loops}(b). Namely, in the presence of mirror symmetry we obtain the states $|\psi_{\theta}^{(n)}\rangle$ for $0\leq \theta \leq \pi$ by diagonalizing the Hamiltonian and we define $-\pi \leq \theta < 0$ states as $|\psi_{-\theta}^{(n)}\rangle ={\cal M}|\psi_{\theta}^{(n)}\rangle$ (see Fig.~\ref{s_fig:T_M}). From Eq.~(\ref{eq:berry}) we find that most of the phases  cancel due to unitarity of $\cal M$ and we obtain (see Ref.\,\onlinecite{SM}, Sec.\,S2),

\begin{equation}
\varphi_{n}\!=\!\arg\left[\left\langle \psi_{\pi}^{(n)}\right|\!{\cal M}\!\left|\psi_{\pi}^{(n)}\right\rangle \!\left\langle \psi_{0}^{(n)}\right|\!{\cal M}\!\left|\psi_{0}^{(n)}\right\rangle \right],
\end{equation}

where $|\psi_{0}^{(n)}\rangle $ and $|\psi_{\pi}^{(n)}\rangle $ are eigenstates of ${\cal M}$ with eigenvalues $\pm1$.  Thus the product under arg function is either $+1$ or $-1$, and hence the Berry phases for all mirror-symmetric Fermi loops are quantized to $0$ or $\pi$ [Fig.~\ref{fig:loops}(b)]. Again, we find that $\varphi=\pi$ if the Fermi loop encloses a Dirac point i.e.~a crossing of subbands. For weakly broken $\cal T$ the crossings stay inside Fermi loops within the mirror plane. However, the Berry phase changes to $0$ when they move outside at a topological phase transition for strongly broken $\cal T$ (see Ref.\,\onlinecite{SM}, Sec.\,S3).

Finally, we find that the Berry phases are arbitrary when both $\cal T$ and $\cal M$ are simultaneously broken [Fig.~\ref{fig:loops}(c)].  We emphasize that this symmetry analysis is completely generic, but we have also confirmed these findings by explicitly calculating the Berry phases in the presence of specific perturbations  breaking of $\cal T$ and $\cal M$ independently of each other (see Ref.\,\onlinecite{SM}, Sec.\,S3)).

This theoretical analysis leads to two important predictions that can be directly tested experimentally. First, we obtain a similar behavior of the Berry phases for both topologically non-trivial and trivial materials. We expect that similar WAL-like behavior is observed in Pb$_{1-x}$Sn$_{x}$Se alloys independently of the Sn content. Secondly, in realistic condensed matter systems, both $\cal T$ and $\cal M$ symmetries are always weakly broken. Therefore, the deviations of the Berry phases from the quantized values can be increased by intentionally breaking the $\cal T$ and $\cal M$ symmetries more strongly. The crystalline mirror symmetry can be broken in a controllable way by covering the surface of the sample with a suitable material. Amorphous solids have short range order in the sense that the distances between neighboring atoms are similar to those in the crystal, but the translational symmetry is absent, so that there is no long-range order and all point group symmetries are violated in crystallographic sense (the symmetry operation will not result in the same structure). Therefore, the importance of the crystalline mirror symmetry on the WAL effect can be tested, for instance, by proximitizing the sample with an amorphous semiconductor.

We have confirmed both of these predictions experimentally. However, before discussing our experimental findings, we will next calculate  the quantum correction to the conductivity coming from the Cooperon propagator.

\subsection{Quantum correction to the conductivity}
In the vicinity of the band crossings appearing at the high-symmetry points we derive a low-energy 2D Hamiltonian for a single pair of subbands in a form of
\begin{equation}
H_{\vec{k},\vec{\sigma}}=\frac{\hbar^{2}}{2m_{e}}\vec{k}^{2}+\alpha_{\text{so}}\left(\sigma_{x}k_{y}-\sigma_{y}k_{x}\right)+g\sigma_{z}.
\label{eq:ham}
\end{equation}
Here $m_{e}$ is the effective mass of the electron; $\alpha_{\text{so}}$ is an effective spin-orbit-like coupling that arises from breaking of the inversion symmetry, which is always present due to the surface in these samples but may be too small to be seen by ARPES; $g$ is the mass term induced by the breaking of the mirror symmetry and weak breaking of the time-reversal symmetry, and $\sigma$ is an effective pseudospin variable which describes entangled spin and orbital degrees of freedom.

The quantum correction to the conductivity can be written as \cite{Wenk2010}
\begin{equation}
\Delta\sigma_{xx}=-\frac{e^{2}}{\pi\hbar}\frac{D_{e}}{L^{2}}\sum_{\vec{Q}}\sum_{\alpha,\beta=\pm}C_{\alpha\beta\beta\alpha}(\vec{Q})\label{maineq:ds}
\end{equation}
where $D_{e}=v_{F}l/2$ is the 2D diffusion constant, $l=v_{F}\tau$ is the elastic mean-free path, $v_{F}$ is the Fermi velocity, $\tau$ is the elastic scattering time, $L^{2}$ is the area of the sample, and $\alpha,\beta=\pm$ are the pseudospin indices of the Cooperon propagator $C$. For weak disorder ($\tau E_F/\hbar\gg1$ where $ E_F$ is the Fermi energy) the Cooperon propagator can be approximated as
\begin{equation}
C(\vec{Q})=\tau\left(1-\int\frac{d\Omega}{2\pi}\frac{1}{1-i\tau\Sigma/\hbar}\right)^{-1},\label{maineq:coop}
\end{equation}
where
\begin{equation}
\Sigma(\vec{Q})=H_{\vec{Q}-\vec{k},\vec{\sigma}'}-H_{\vec{k},\vec{\sigma}}
\end{equation}
is a $4\times4$ matrix describing two interfering electrons with pseudospins $\vec{\sigma}'$ and $\vec{\sigma}$. The integral in Eq.~(\ref{maineq:coop}) is over all angles of velocity $\vec{v}=\hbar\vec{k}/m_{e}$ on the Fermi surface. By taking lowest order terms in $\vec{Q}$ and $\alpha_{\text{so}}$, we can write the quantum correction to the conductivity in the absence of the magnetic field as
\begin{equation}
\Delta\sigma_{xx}=\frac{e^{2}}{\pi\hbar}\frac{1}{L^{2}}\sum_{\vec{Q}}{\rm Tr}\left[\Gamma\left(\frac{1}{D_{e}\tau_{\phi}}+H_{c}\right)^{-1}\right],
\end{equation}
where $H_c$ is a non-Hermitian Cooperon Hamiltonian
\begin{equation}
H_{c}=\vec{Q}^{2}+2Q_{\text{so}}\vec{Q}\cdot\hat{a}\vec{S}+Q_{\text{so}}^{2}\left(S_{x}^{2}+S_{y}^{2}\right)-i\frac{g}{\hbar D_{e}}\left(\sigma_{z}'-\sigma_{z}\right),\label{eq:Hc}
\end{equation}
$Q_{\text{so}}=2m_{e}\alpha_{\text{so}}/\hbar^{2}$ and $\tau_{\phi}$ is the dephasing time. Singlet and triplet interference is encoded in matrix $\Gamma$ having three $-1$ eigenvalues in the $\sigma \sigma'$ triplet sector and $+1$ in the singlet sector.  The  perpendicular magnetic field can be introduced in the Cooperon Hamiltonian by minimal substitution $\vec{Q}\to\vec{Q}+2e\vec{A}/\hbar$, and the summation in this case should be taken over the Landau levels (see Ref.\,\onlinecite{SM}, Sec.\,S4 for details).

\begin{figure*}
\begin{centering}
\includegraphics[width=1\textwidth]{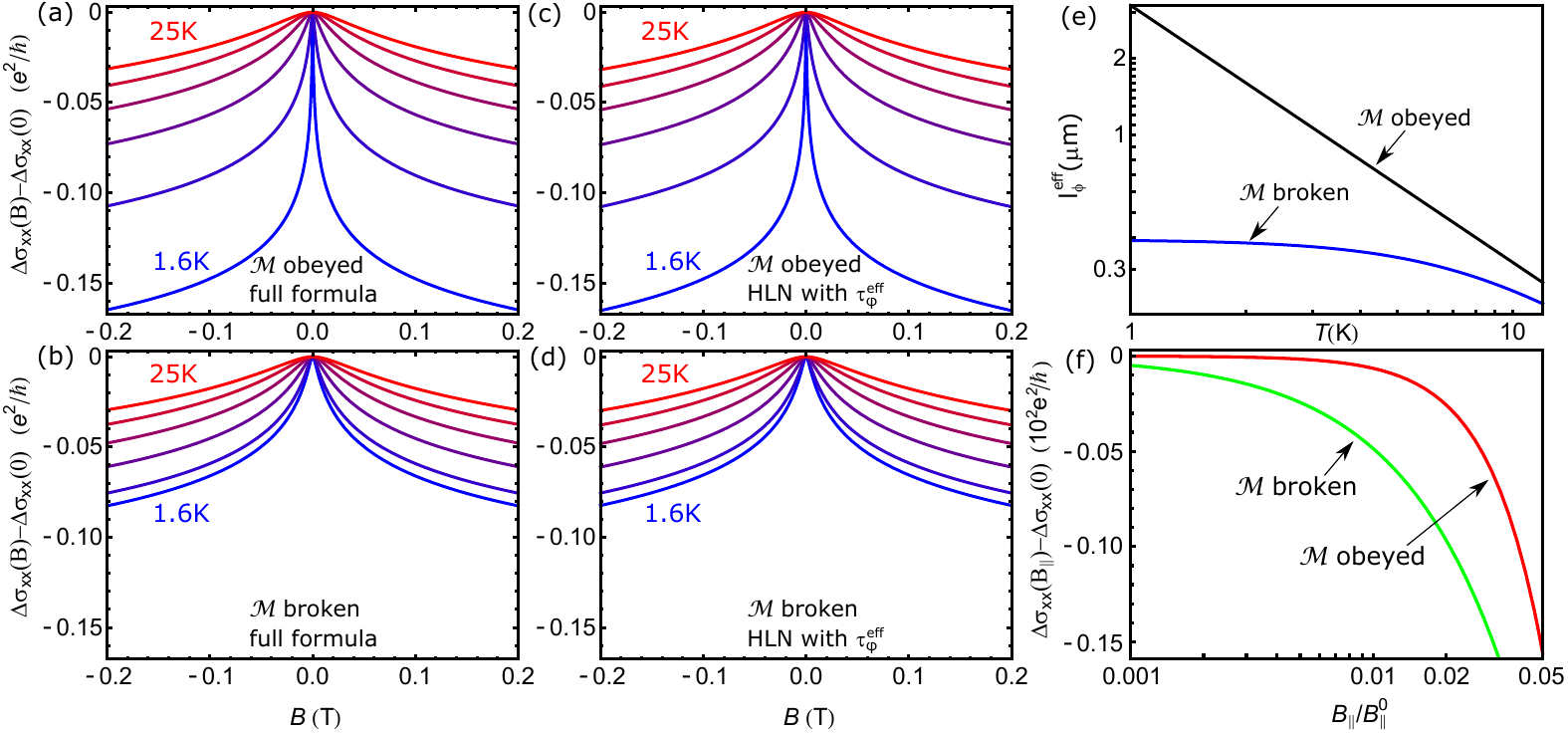}
\par\end{centering}
\caption{(a--d) $\Delta\sigma_{xx}$ as a function $B$ in the fully symmetric and symmetry-broken cases for temperatures $T=1.6, 5, 10, 15, 20, 25$\,K. (a,b)  Numerically calculated quantum correction to the conductivity. (c,d) $\Delta\sigma_{xx}$ calculated from the modified HLN formula (\ref{maineq:hln_eff}) employing the effective dephasing time $\tau_{\phi}^{{\rm eff}}$ [Eq.~(\ref{maineq:teff})]. (e) Effective dephasing length $l_{\phi}^{{\rm eff}}=(D_{e}\tau_{\phi}^{{\rm eff}})^{1/2}$ as function of temperature for fully symmetric and symmetry-broken cases. (f) $\Delta\sigma_{xx}$ as a function of the in-plane field $B_{\parallel}$ in the fully symmetric and symmetry-broken cases.
\label{fig:ds}}
\end{figure*}

The parameter $g=0$ unless both time-reversal and mirror symmetry are simultaneously broken (see Ref.\,\onlinecite{SM}, Sec.\,S4 for details). We can assume that the time-reversal symmetry is always weakly broken by the same amount due to the intrinsic mechanisms, but the breaking of the mirror symmetry is tunable and depends on how the sample is covered. Thus, in uncovered samples we assume that $g \approx 0$ but in the presence of the cover which breaks the mirror symmetry $g$ becomes significantly larger and we assume that it is given by $g=1.75$ meV. Note that the gap opened by $g$ is still too small to be observed by ARPES, but the WAL measurement is a  very sensitive probe of the symmetry-breaking field, so that already such small value of $g$ can dramatically show up in the transport experiments: The resulting dependencies of $\Delta\sigma_{xx}$ on $B$, in the fully symmetric and symmetry-broken cases, are shown in Figs.~\ref{fig:ds}(a) and \ref{fig:ds}(b) (lines labeled $T=1.6$\,K). In this calculation we have used the parameters: $v_{F}=7.3\times10^{5}$ m/s, $l=v_{F}\tau=20$ nm and $l_{\phi}=(D_{e}\tau_{\phi})^{1/2} =2\ \mu$m at 1.6\,K. (In general the phase-coherence length depends on temperature, as discussed below, and therefore we fix here the phase-coherence length at temperature $T=1.6$\,K.)  Moreover, we have estimated that  $Q_{\text{so}}=1.022\times10^{8}$ m$^{-1}$, so that the pseudospin precession length is $l_{\text{so}}=2\pi/Q_{\text{so}}\approx60$\,nm (see Ref.\,\onlinecite{SM}, Sec.\,S4 for details).

The breaking of mirror and tie-reversal symmetries by the symmetry-breaking field $g$ leads to an opening of a small energy gap at the band crossings and to the destruction of the quantization of the Berry phase (see Ref.\,\onlinecite{SM}, Sec.\,S4 for details). Due to the latter reason, the symmetry-breaking field randomizes the phases of the backscattering paths destroying their systematic destructive interference that was caused by the quantization of the Berry phase to $\varphi=\pi$. Therefore,  in the presence of the symmetry-breaking field $g$ there is a new length scale which limits the increase of the WAL effect with lowering temperature (see Fig.~\ref{fig:ds}). Although, the symmetry-breaking field is a quantum-coherent effect the qualitative picture discussed above suggests that it has a similar effect as  phase breaking phenomena described by $l_\phi$. We have confirmed this expectation by demonstrating (see Fig.~\ref{fig:ds}) that the conductivity calculated from the full expression  can be reproduced by a modified HLN formula given by

\begin{eqnarray}
\Delta\sigma_{xx}(B)-\Delta\sigma_{xx}(0)&=&\quad\quad\quad\quad\quad\quad\quad\quad\quad\quad\quad\quad\quad  \nonumber \\  &&\hspace{-3.8cm}
\frac{-e^{2}}{4\pi^{2}\hbar}\bigg[\psi\left(\frac{1}{2}+\frac{1}{4\tau_{\phi}^{{\rm eff}}B}\frac{\hbar}{eD_{e}}\right)-\log\left(\frac{1}{4\tau_{\phi}^{{\rm eff}}B}\frac{\hbar}{eD_{e}}\right)\bigg],
\label{maineq:hln_eff}
\end{eqnarray}

where $\tau_{\phi}^{{\rm eff}}$ is an effective dephasing time

\begin{equation}
\tau_{\phi}^{{\rm eff}}(T)=\frac{1}{D_{e}\left(\frac{1}{l_{\phi}^{2}(T)}+\frac{1}{l_{g}^{2}}\right)}\label{maineq:teff}
\end{equation}

and $l_{g}$ is the new length scale related with the symmetry breaking field $g$

\begin{equation}
l_{g}=2\pi\sqrt{\frac{D_{e}\hbar}{\sqrt{2}g}}.
\label{eq:lg}
\end{equation}

Thus, if the conductivity is measured as a function of perpendicular magnetic field, the symmetry-breaking field shows up as an effective dephasing length as $l_{\phi}^{{\rm eff}}=(D_{e}\tau_{\phi}^{{\rm eff}})^{1/2}$, which saturates at low-temperatures to $l_g$. Additionally, to describe the full temperature dependence observed experimentally we assume that the dephasing length depends on temperature as $l_{\phi}(T)=l_{\phi}(T_{0})T_{0}/T$. The full temperature-dependence of the conductivity from $T=1.6$\,K to $T=25$\,K obtained this way is shown  in Figs.~\ref{fig:ds}(a) and \ref{fig:ds}(b) both in the absence and presence of the symmetry-breaking field. The corresponding temperature dependencies of the effective dephasing lengths $l_{\phi}^{{\rm eff}}$ are shown in Fig.~\ref{fig:ds}(e). In the absence of the symmetry-breaking field  $l_{\phi}^{{\rm eff}}$ continues to increase at low temperatures, whereas in the presence of the symmetry-breaking field  $l_{\phi}^{{\rm eff}}$ saturates to $l_g$ at low temperatures. Although the symmetry-breaking field $g$ is small, it results in a decrease of $l_{\phi}^{{\rm eff}}$ by almost an order of magnitude at $T=1.6$\,K demonstrating that the effect mirror symmetry-breaking is expected to have dramatic experimental consequences.

Within the low-energy theory the effect of  the in-plane field $B_{\parallel}$  (breaking both mirror and time-reversal symmetries) is that it increases the symmetry-breaking field $g$ so that we can write the low-energy Hamiltonian (\ref{eq:ham}) with $g\to\left(g+ g^*\mu_{\rm B} \left|B_{\parallel}\right| \right)\sigma_{z}$. The important scale of the in-plane magnetic field is therefore $B_{\parallel}^{0}=g/g^*\mu_{\rm B}$, where  the symmetry-breaking field due to the in-plane field becomes equal to the symmetry breaking field due to the covering of the sample. Then, in the limit of zero perpendicular magnetic field and $T=T_{0}$ we obtain the result shown in Fig.~\ref{fig:ds}(f) for $\Delta\sigma_{xx}$. For $|B_{\parallel}|<B_{\parallel}^{0}$ the conductivity decreases faster in the symmetry-broken case as a function of the in-plane field.

We point out that in this analysis the effect of the in-plane field can also include orbital effects, because also the orbital effects of the in-plane magnetic field break mirror and time-reversal symmetries, and therefore lead effectively to an increase of $g$ in the low-energy theory. Although the explicit dependence of the energy gap on $|B_{\parallel}|$ can be more complicated the result that the conductivity decreases faster in the symmetry-broken case seems to be relatively robust if the applied in-plane field is reasonably small. (For this result the assumption that the gap increases linearly with $|B_{\parallel}|$ is not necessary.) This result is also consistent with analysis based on Berry phases since the increase of the gap leads to larger deviation of the Berry phases from the quantized value.

\section{Samples growth, characterization, and processing}
\label{sec:samples}
We test the theory on (111) Pb$_{1-x}$Sn$_x$Se 50-nm thick films deposited by molecular beam epitaxy (MBE) on freshly cleaved (111) BaF$_2$ substrates under the base pressure below 10$^{-9}$ mbar. We employ the PREVAC 190 growth chamber equipped with elemental Pb, Sn, and Se sources, whose flux ratio is controlled by a beam flux monitor placed in the substrate position. Typical selenium to metal flux ratio is of the order of 3:2. The structural quality of the film surface is monitored \textit{in-situ} by the reflection high-energy electron diffraction (RHEED). The film growth rate determined by pronounced RHEED oscillations is in the range 0.1-0.3~nm/s. \textit{Ex-situ} x-ray diffraction (XRD) measurements have been performed by PANalytical X'Pert Pro MRD diffractometer with a 1.6\,kW x-ray tube (vertical line focus) with CuK$\alpha_{1}$ radiation ($\lambda=1.5406$\AA), a symmetric 4 $\times$ Ge (220) monochromator and for high resolution measurements a channel-cut Ge(220) analyzer. Atomic force microscopy (AFM) images have been obtained in tapping mode using Veeco Nanoscope IIIa microscope. Additional morphological and composition characterization is accomplished by field emission scanning electron microscopy (FE-SEM) with Neon 40-Auriga Carl Zeiss microscope equipped with energy dispersive x-ray spectroscopy (EDX) system QUANTAX 400 Bruker.  Growth conditions, namely the ratio of the beam fluxes and growth temperature have been thoroughly optimized to obtain high quality thin films with Sn content varying from 0 to 0.40.

Two series of films have been prepared to study the influence of both topological transition and break of the mirror symmetry on the WAL phenomena: the first series consists of bare epilayers A-E, while the samples in the second series (epilayers F-J) are covered by a 100\,nm-thick amorphous and insulating Se cap. In each series, Sn content is varied to drive part of the films through the topological transition at low temperatures. The amorphous Se cap layer has been confirmed to be insulating in a separately checked Se/BaF$_2$ structure. As shown in Ref.\,\onlinecite{SM}, Sec.\,S5, XRD structural analysis of Se covered samples revealed no additional strains compared to the bare epilayers. The Se cap also serves as a protection against contamination for ARPES measurements.

To ensure that the topological transition indeed takes place,  the band structure at the surface of G and D epilayers with trivial and non-trivial compositions, respectively, have been characterized by ARPES at UE112 PGM-2a-1$^{2}$ beamline of BESSY II (Berlin) in the photon energy range 15-90\,eV and horizontal light polarization using a six-axes automated cryomanipulator and a Scienta R8000 electron spectrometer. Typical energy and angular resolutions are better than 20 meV and 0.5$^{\circ}$, respectively.

\begin{figure}[tb]
    \centering
    \includegraphics[width=0.9\columnwidth]{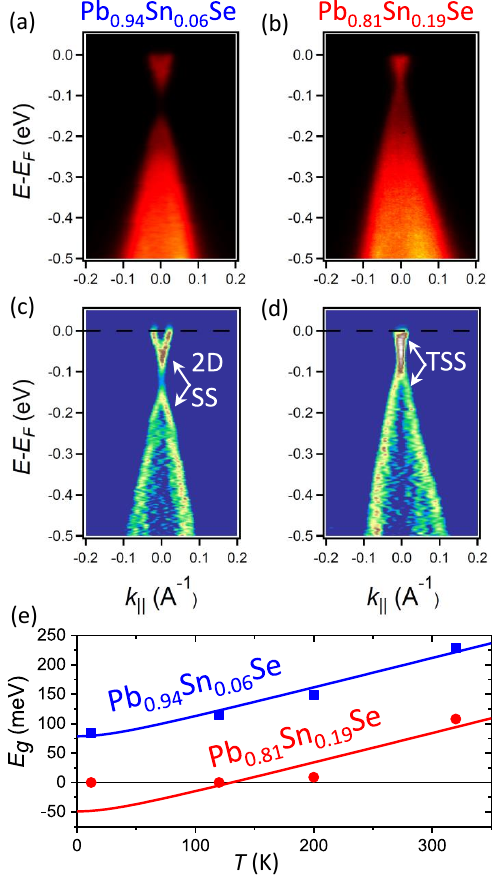}
    \caption{ ARPES results. (a--d) Dispersion and corresponding 2nd derivative of ARPES data taken at 12~K with photon energy of 18~eV in the vicinity of the $\overline{\Gamma}$ point. (a, c) Results trivial Pb$_{0.94}$Sn$_{0.06}$Se (band gap of 84\,meV) and (b, d), topological Pb$_{0.81}$Sn$_{0.19}$Se epilayers (gapless states with Dirac dispersion, as observed previously \cite{Polley2014,Mandal2017,Assaf2017}). Sample surfaces are $n$-type. (c) 2nd derivative plots confirm the presence of gaped precursor surface states in the trivial phase \cite{Wojek2013,Xu:2015_NC}. {\bf e}, Surface band gaps measured at several temperatures (points) are well described, above the topological phase transition, by the semi-empirical Grisar formula \cite{Preier1979} for the bulk band gap (solid lines) thus proving a negligible effect of strains on the band structure of the studied epilayers.}
    \label{fig:arpes}
\end{figure}

Figures \ref{fig:arpes}(a-d) present ARPES spectra and 2nd derivative of ARPES data taken at 12 K with photon energy of 18 eV in the vicinity of the $\bar{\Gamma}$  point. As seen, the results reveal the presence of precursor surface states  \cite{Wojek2013,Xu:2015_NC} in the film with Sn content corresponding to the trivial case [Pb$_{0.94}$Sn$_{0.06}$Se, Figs.~\ref{fig:arpes}(a,c)] and gapless surface states in the topologically nontrivial case [Pb$_{0.81}$Sn$_{0.19}$Se, Figs.~\ref{fig:arpes}(b,d)].  As shown in Fig.~\ref{fig:arpes}(e), the surface band gaps determined here by ARPES at several temperatures are in good agreement with the ARPES data for bulk samples \cite{Dziawa2012} and the semi-empirical Grisar formula \cite{Preier1979}. This agreement demonstrates that residual strains detected by XRD have a minor effect on the band structure. At the same time, the Rashba splitting of bands, appreciable in IV-VI semiconductor epilayers under certain growth conditions \cite{Volobuev2017}, is not significant.

\begin{figure*}[tb]
    \includegraphics[width=1.8\columnwidth]{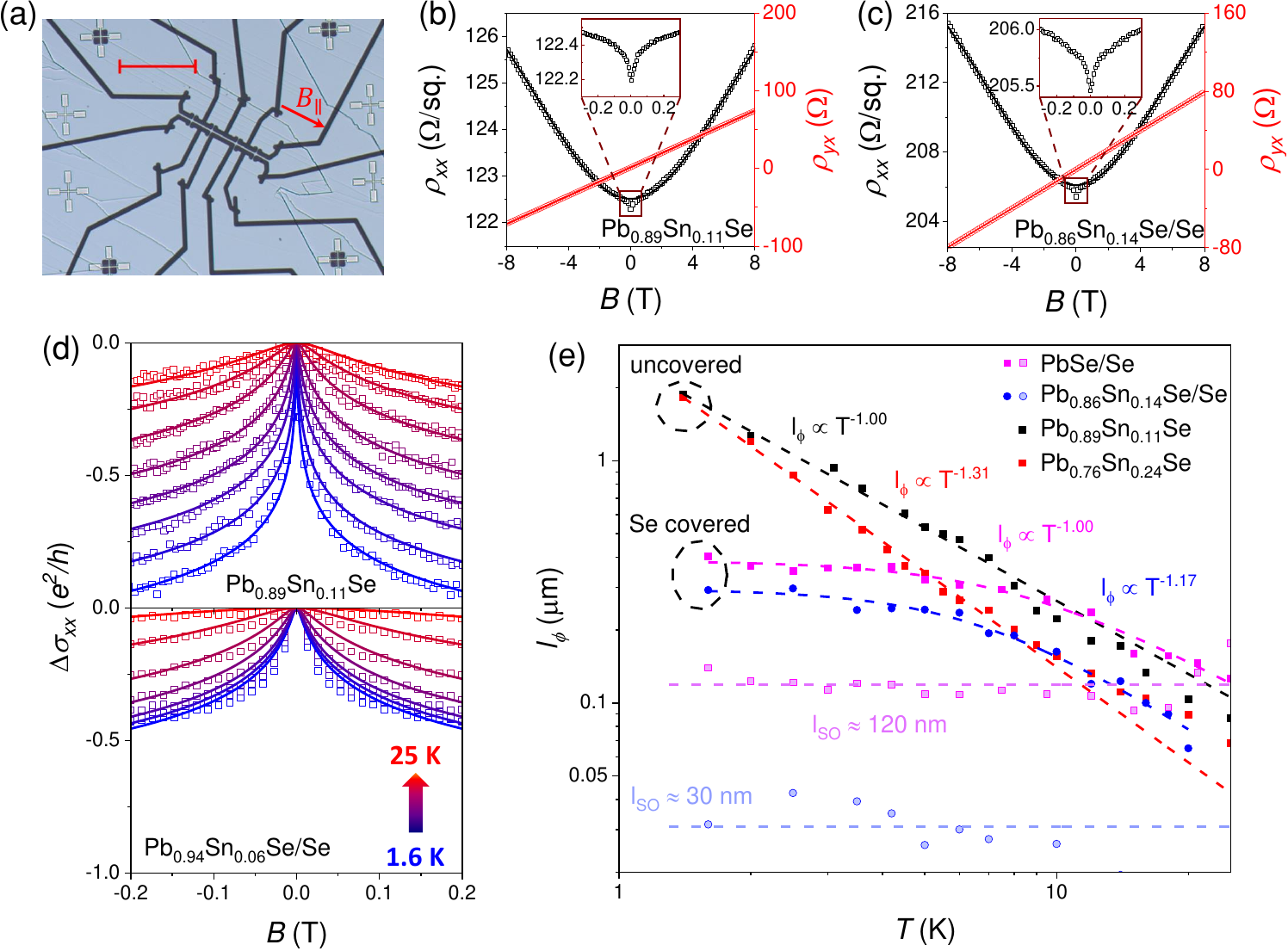}
    \caption{Determination of the phase coherence length $l_{\phi}(T)$ from MR data in the magnetic field perpendicular to the epilayer plane. (a) Photo of a processed Hall-bar (epilayer B, $x_{\text{Sn}} = 0.11$); conducting channel resides on a flat area, avoiding cleavage steps of the BaF$_2$ substrate (scale bar is 200~$\mu$m). (b, c) Longitudinal $\rho_{xx}$ (black symbols) and Hall (red symbols) $\rho_{yx}$ resistivities in high (main figures) and weak fields (insets) in epilayers that are uncovered ({\bf b}) and covered by Se (c). (d) Evolution of the WAL-like low-field MR with increasing temperature in uncovered (upper panel) and Se covered (lower panel) epilayers. Experimental points (empty squares) are fitted to the one-channel HLN expression in the strong spin-orbit approximation (solid lines) treating $l_{\phi}(T)$ as an adjustable parameter. (e) Determined values of $l_{\phi}(T)$ in epilayers uncovered be Se increase down to 1.5~K (black and red), while in Se covered epilayers $l_{\phi}(T)$ saturates at lower temperatures (blue and magenta). Typical values of $l_{\text{so}}(T)$ obtained from the full HLN expression, Ref.\,\onlinecite{SM}, Sec.\,S7.}
    \label{fig:MR}
\end{figure*}

Grown and characterized epilayers have been further processed for magnetoresistance (MR) measurements to the form of Hall bars by e-beam lithography and Br wet etching with the long arm along a $\langle110\rangle$ direction revealed by cleavage of BaF$_2$, as depicted in Fig.~\ref{fig:MR}(a). The in-plane magnetic field was oriented along the current in the tilted field experiments. Resistivity measurements were performed in an 8~T/1.5~K cryostat, using a standard lock-in technique at 20 -- 30~Hz with the excitation current from 1~$\mu$A down to 10~nA at the lowest temperature. We checked that lowering of current down to 1\,nA did not affect positive low-field MR, meaning that the saturation of the phase breaking length at low temperatures we have found in samples covered by Se cannot be explained by Joule heating. Results of resistivity and Hall effect measurements for all studied samples are summarized in Ref.\,\onlinecite{SM}, Sec.\,S6. The carrier mobilities in Se-covered films are typically higher than in the absence of the Se cap, indicating that the deposition of Se does not lead to any kind of reduction of the sample quality.

Despite that ARPES data presented in Fig.~\ref{fig:arpes}(a)-(d) confirm the expected $n$-type character of Pb$_{1-x}$Sn$_x$Se,  a positive sign of the Hall coefficient is observed, pointing out to a relatively large contribution from holes at the interface to BaF$_2$, as found earlier for PbTe/BaF$_2$ epilayers \cite{Kolwas:2013_pss}. High-field parabolic positive MR for the field perpendicular to the film plane, shown in Fig.~\ref{fig:MR}(b,c), is consistent with a multichannel character of charge transport (valleys, 2D subbands, $n$-type and $p$-type layers), whereas a linear component in the highest field suggests and an admixture of the Hall resistance caused by lateral inhomogeneities \cite{Ramakrishnan2017}.

\section{Magnetoresistance in weak magnetic fields: experiment {\em vs.} theory}

Interestingly and crucially for this work, we find the existence of low-field temperature-dependent positive MR in all epilayers regardless of their composition. Except for PbSe, this MR dominates only in the diffusive regime, $l \ll l_B$, where $l$ is the mean free path and $l_B$ is the magnetic length. According to the theory developed here, we assign this MR to the Berry phase quantization brought about by symmetries rather than by a non-trivial character of the topological phase. In particular, the mirror symmetry leads to WAL even if time-reversal symmetry is slightly broken. Within this scenario, and by noting that we expect the phase coherence length $l_{\phi}$ to be greater than the film thickness $d = 50$\,nm, the MR is described by the HLN theory in the limit $l_{\phi} \gg l_{\text{so}}$, where $l_{\text{so}}$ is the spin diffusion length limited be spin-orbit interactions, corresponding to the HLN prefactor $\alpha = -1/2$ \cite{Hikami1980}, as given in Eq.~\ref{maineq:hln_eff} for $l_g \rightarrow \infty$. In our case, as shown in Fig.~\ref{fig:MR}(d) and in supplemental Fig.~S8 \cite{SM} for the full set of the films, magnetoconductance for all epilayers can be fitted by the one-channel formula, treating $l_{\phi}(T)$, as the only fitting parameter. This means that $l_{\phi}$ is longer than length scales characterizing scattering between subbands and valleys (including surface ones in the topological case) \cite{Garate:2012_PRB,Fukuyama:1980_PTPS} as well as between $n$ and $p$-type layers. Alternatively, and more probably, because of short length scales characterizing the $p$-type region, the corresponding WAL or WL MR is shifted to a high field region, so the low field features are solely due to electrons residing closer to the outer surface. It is important recalling that if only one of the parallel layers shows MR, the one channel formula remains valid \cite{Garate:2012_PRB}. As shown in Ref.\,\onlinecite{SM}, Sec.\,S7, by fitting the magnetoconductance data to the full HLN formula we find $l_{\text{so}} \approx l$, which substantiates our conjecture that in our case WAL MR stems from the Berry phase quantization, and not from a sequence of spin rotations in a varying spin-orbit field resulting in $l_{\text{so}} \gg l$. For comparison, we collected in Ref.\,\onlinecite{SM}, Sec.\,S8 the values of $l_{\phi}(T)$ determined previously for various topological insulators and topological crystalline insulators, the latter data are similar to our results on the both sides of the topological phase transition.

While there is no much difference in WAL MR for samples with Sn content $x$ corresponding to different topological phases, our data reveal a striking dissimilarity in the temperature dependence of magnetoconductance in samples covered and uncovered by Se layers, as shown in Fig.~\ref{fig:MR}(d) and S8 \cite{SM}. It is expected that covering of the films by a Se cap may alter the Fermi level position and, thus, change a relative occupancy of electron valleys and subbands in the surface region.  However, previous gating experiments on WAL in 2D systems have not indicated any effects of the Fermi level shift on the {\em temperature} dependence of the coherence length. In Sec.~\ref{sec:samples} we have discussed precautions undertaken in order to eliminate Joule heating of carriers. Furthermore, unintentional magnetic doping cannot be responsible for the difference between capped and uncapped samples because both types of epilayers have been grown in the same MBE chamber. Relatively large Hall bar dimensions (10-100~$\mu$m) exclude finite size effects. Furthermore, as shown in Ref.\,\onlinecite{SM}, Sec.\,S7, fitting of  to the full HLN expression, i.e., containing $l_{\text{so}}$ and the mean free path $l$ explicitly, confirms that $l_{\text{so}} \approx l \ll l_{\phi}$ [Fig.~\ref{fig:MR}(e)], which rules out a cross-over from WAL to WL. Therefore, we assume that the dependence of the WAL magnitude on temperature and the perpendicular magnetic field shown in Fig.~\ref{fig:MR}(d) is solely determined by processes deviating the Berry phase from $\pi$, and controlling the magnitude of $l_{\phi}$. In the samples without Se layers, $l_{\phi}(T)$ follows the power law $T^{-p/2}$, with $p$ ranging from 1.4 to 2.6, which corresponds to electron-phonon dephasing mechanism, without any tendency to saturation down to 1.5~K [Fig.~\ref{fig:MR}(e)]. By contrast, in the Se covered epilayers $l_{\phi}(T)$ tends to saturate at temperatures below $\approx 5$~K. We have fitted $l_{\phi}(T)$ in Se covered epilayers with $(A_0+A_1T^{p})^{-1/2}$, which results in the similar values for $p$, ranging from 1.8 to 2.8 with $l_{\phi}(1.5~K)\approx1-2$~$\mu$m and $150-400$~nm in uncovered and Se covered samples, respectively.

Actually, these striking findings provide a strong support to the theory proposed here [c.f. Figs.~\ref{fig:ds}(a)-(e) and Figs.~\ref{fig:MR}(d)-(e)]. At high temperature, WAL is still governed by the thermally suppressed $l_{\phi}$ even if time-reversal symmetry is weakly broken.   In bare epilayers, WAL is protected by the quantized Berry phase $\varphi=\pi$ due to the $\cal M$, thus $l_{\phi}$ continues to increase with cooling down. In Se capped epilayers there is a different situation: long interference paths, which are relevant for large values of $l_{\phi}$, do not contribute to WAL, since scattering between states at the Fermi level, allowed by mirror symmetry breaking, randomizes the wave function phase $\phi$ and average it to zero. At the same time, we assume that the mirror symmetry breaking by an amorphous cap is stronger than by impurities and defects in high quality crystalline films. Thus, there is a new length scale $l_g$ given in Eq.~\ref{eq:lg} in the Se-covered films, which limits an increase of WAL MR with lowering the temperature, similarly to the effect of spin-disorder scattering and Zeeman splitting considered previously \cite{Hikami1980,Maekawa:1981_JPSJ,Lu:2011_PRL}.

\begin{figure}[tb]
    \includegraphics[width=1.05\columnwidth]{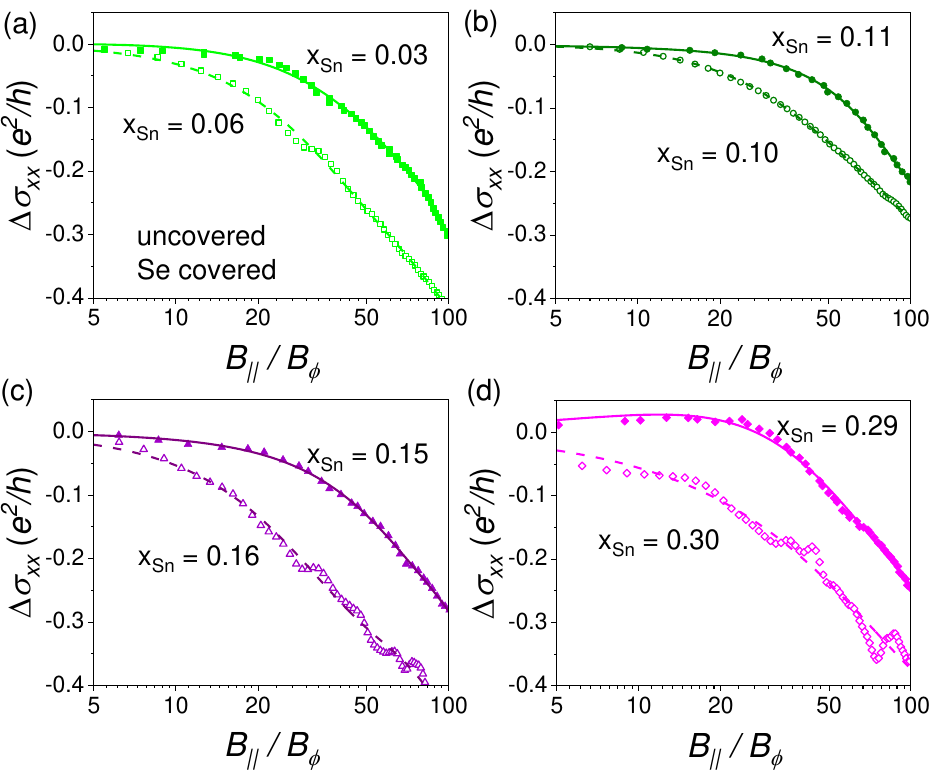}
    \caption{Effect of mirror-symmetry breaking on MR for magnetic field applied parallel to the film plane. Comparison of MR measured at 4.2\,K in the field parallel to the film plane (for which the Zeeman field gives a substantial contribution) for uncovered (full symbols, solid lines) and Se-covered (open symbols, dashed lines) samples with similar $x_{\text{Sn}}$ (lines are guide for the eye); see Ref.\,\onlinecite{SM}, Sec.\,S9 for fitting results obtained employing various formulae for MR in parallel fields; $B_{\phi}=\hbar/4el^2_{\phi}$. }
    \label{fig:parallel field}
\end{figure}

The theory developed here shows explicitly that the application of the in-plane Zeeman field $\vec{h} = g^*\mu_{\text{B}}\vec{B}/2$ leads to a stronger deviation of the Berry phase from the quantized value $\varphi=\pi$ and larger change of conductivity, if reflection symmetry is violated, as shown in Fig.~\ref{fig:ds}(f). In order to test this prediction, we have carried out MR measurements for the magnetic field parallel to the film plane, as in this configuration the role of time-symmetry breaking by the vector potential is reduced \cite{Altshuler1981,Dugaev1984,Beenakker1988}, making the Zeeman effect more important, particularly considering a relatively large magnitude of the electron Land\'e factor, $g^* = 35 \pm 5$ for electrons in PbSe \cite{Bauer1992}. Since also in the parallel configuration \cite{Altshuler1981,Dugaev1984,Beenakker1988} and in the presence of the Zeeman effect \cite{Maekawa:1981_JPSJ,Malshukov:1997_PRB,Glazov:2009_SST}, the WAL magnitude is controlled $\tau_{\phi}$ in weak magnetic fields, we present in Fig.~\ref{fig:parallel field} MR as a function of $B/B_{\phi}$, where $B_{\phi}=\hbar/4el^2_{\phi}$.  As seen, MR is systematically stronger in samples covered by the Se cap, in agreement with the theoretical expectations.

\section{Summary and outlook}

Our theoretical results have demonstrated, taking thin films of SnTe-class of materials as an example, that the inversion asymmetry and mirror symmetry, rather than the topological phase, are essential for the Berry phase quantization to $\pi$ and, hence, to the appearance of the robust WAL MR. This implies  the existence of the hitherto overlooked length scale in quantum coherence phenomena. This new length is associated with the crystal symmetry breaking rather than with a topological phase transition or with the violation of time reversal or spin rotation symmetries considered so far. This offers new prospects in controlling carrier quantum transport by system architectures.

These theoretical expectations have been verified by MR studies on a series of Pb$_{1-x}$Sn$_x$Se epilayers, which have revealed the existence of WAL MR on the both sides of topological phase transitions as well pointed out to striking differences in the dependencies of WAL MR on temperature and the magnetic field for samples with the mirror symmetry maintained compared to the films in which mirror symmetry is intentionally broken, as the electron penetration length into the thick amorphous Se layer is longer than into a native surface oxide in uncapped films.

It is certainly appropriate to analyze carefully other mechanisms that could elucidate a strong influence of amorphous Se overlayers on WAL MR. In principle, a  reduction of $l_{\phi}(T)$ in Se covered samples might be explained by additional decoherence due to the presence of two-level systems in amorphous solids. However, the corresponding theory \cite{Afonin:2002_PRB} suggests that this would not lead to full saturation of $l_{\phi}(T)$ at low temperatures. Furthermore, according to formulae describing WAL MR in the parallel configuration \cite{Altshuler1981,Dugaev1984,Beenakker1988}, a stronger WAL MR for the in-plane magnetic field might result from effectively greater thickness of samples covered by Se.

A fascinating question then arises to what extent surprising saturations of $\tau_{\phi}(T)$ at low temperature observed since decades in many systems \cite{Lin:2002_JPC}, and most recently in Dirac materials \cite{Jing:2016_NS,Islam2019,Rosen:2019_PRB,Nakamura:2020_NC,Wang:2020_PRB}, have been caused by the hidden spatial symmetry breaking, the mechanisms brought into light by our work.

\section*{Acknowledgments}

The International Center for Interfacing Magnetism and Superconductivity with Topological Matter MagTop is supported by the Foundation for Polish Science through the IRA Programme co-financed by EU within SG OP (Grant No. MAB/2017/1). We acknowledge the Helmholtz-Zentrum Berlin for provision of synchrotron radiation beamtime at UE112 PGM-2a-1$^{2}$ of BESSY II under the EU CALIPSO Grant number 312284. W.B. also acknowledges support by Narodowe Centrum Nauki (NCN, National Science Centre, Poland) Project No. 2019/34/E/ST3/00404. G.S. also acknowledges support by Austrian Science Fund, Projects No. P30960-N27 and I 4493-N.

\section*{Author contributions}
A.K. and W.B. contributed equally to this work. W.B and T.H. developed the theory with input from T.D. The samples were grown and characterized by x-ray by V.V.V. with the assistance of B.T. and J.J.D, respectively. A.K. carried out processing and magnetotransport measurements with the assistance of J.P. and Z.A., respectively. V.V.V., O.C. and G.S., with the help of A.V., performed ARPES measurements. AFM data were collected by M.A. and EDX by T.W. The manuscript was written by A.K., W.B., T.H., V.V.V. and T.D. All authors discussed the results and commented on the manuscript. T.D. and T. Wojtowicz supervised the project.

\bibliographystyle{naturemag}

\begin{widetext}
\clearpage
\newpage

\onecolumngrid
\renewcommand{\thefigure}{S\arabic{figure}}
\renewcommand{\thetable}{S\arabic{table}}
\renewcommand{\theequation}{S\arabic{equation}}
\renewcommand{\thepage}{S\arabic{page}}
\renewcommand{\thesection}{S\arabic{section}}

\setcounter{page}{1}
\setcounter{section}{0}
\setcounter{equation}{0}
\setcounter{figure}{0}
\setcounter{table}{0}

\begin{center}
\textbf{\Large Supplementary Information} \\
\vspace{0.2in} \textmd{\Large \thetitle}\\
\end{center}

\section{The multilayer Hamiltonian and the symmetries}\label{Ham}

Our starting point is the tight-binding Hamiltonian for SnTe-material class \citeS{SM_Hsieh2012},

\begin{equation}
  H=m\sum_j (-1)^j\sum_{\g r, \alpha} \hat c^\dagger_{j\alpha}(\g r)\cdot \hat c^\dagga_{j\alpha}(\g r)+\sum_{j,j'}t_{jj'}\sum_{\langle\g r,\g{r'}\rangle,\alpha} \hat c^\dagger_{j\alpha}(\g r)\cdot \hat d_{\g r\g{r'}}\ \hat d_{\g r\g{r'}}\cdot \hat c^\dagga_{j'\alpha}(\g{r'})
  -\sum_j i\lambda \sum_{\g r,\alpha,\beta}\hat c^\dagger_{j\alpha}(\g r)\times \hat c^\dagga_{j\beta}(\g r)\cdot \hat \sigma_{\alpha,\beta},
\end{equation}

where $\hat c_{j \alpha}(\mathbf{r})$ are vectors of fermionic operators corresponding to $p_x$-, $p_y$- and $p_z$-orbitals and the indices denote the sublattice $j\in\{1,2\}$ [(Sn,Pb)/(Te,Se) atoms], spin $\alpha$ and lattice site $\mathbf{r}$. Here $\hat \sigma_{\alpha, \beta}$ is a vector of Pauli matrices, $\hat d_{\g r\g{r'}}$ are unit vectors pointing from $\g r$ to $\g{r'}$ and the next-nearest-neighbour hoppings satisfy $t_{11}=-t_{22}$.

Defining the unit cell as two atoms at positions $(0,0,0)$ and $(0,0,1)$ (taking one interatomic distance as a length unit) and the lattice translation vectors as $\vec{a}_1=(1,0,1)$, $\vec{a}_2=(0,1,1)$ and $\vec{a}_3=(0,0,2)$ we find that the three-dimensional bulk Hamiltonian can be represented in momentum space as \citeS{SM_Brzezicki2019},

\begin{eqnarray}
{\cal H}(\vec{k})&=&m\mathbbm{1}_2\!\otimes\!\mathbbm{1}_3\!\otimes\! \tau_z+t_{12}\!\!\!\sum_{\alpha=x,y,z}\!\!\mathbbm{1}_2
\!\otimes\!\left(\mathbbm{1}_3\!-\!L_{\alpha}^{2}\right)\!\otimes\! h_{\alpha}^{(1)} (\vec{k})+t_{11}\sum_{\alpha\not=\beta}\mathbbm{1}_2\!\otimes\!\left(\mathbbm{1}_3\!-\!\tfrac{1}{2}\left(L_{\alpha}\!+\!\varepsilon_{\alpha\beta}L_{\beta}\right)^{2}\right)
\!\otimes\! h_{\alpha\beta}^{(2)} (\vec{k})\nonumber\\
&&\hspace{-0.2cm} +\sum_{\alpha=x,y,z} \lambda \sigma_{\alpha}\!\otimes L_{\alpha}\otimes\mathbbm{1}_2,
\end{eqnarray}

where $\vec{k}=(k_1,k_2,k_3)$, $\varepsilon_{\alpha\beta}$ is a Levi-Civita symbol, $L_{\alpha}=-i\varepsilon_{\alpha\beta\gamma}$ are the $3\times 3$ angular momentum $L=1$ matrices and spin-orbit coupling is given by $\lambda$. The mass difference between the two sites of the unit cell is encoded in a pseudospin $\tau_z$ Pauli matrix and matrices $h_{\alpha}^{(1)}(\vec{k})$ and $h_{\alpha\beta}^{(2)}(\vec{k})$ describe hopping between nearest neighbors,

\begin{eqnarray}
h_x^{(1)}&=&[\cos k_1+\cos (k_1-k_3)]\tau_x+[-\sin k_1+\sin (k_1-k_3)]\tau_y, \nonumber\\
h_y^{(1)}&=&[\cos k_2+\cos (k_2-k_3)]\tau_x+[-\sin k_2+\sin (k_2-k_3)]\tau_y, \nonumber\\
h_z^{(1)}&=&[1+\cos k_3]\tau_x-\sin k_3\tau_y, \nonumber
\end{eqnarray}
and next-nearest neighbors
\begin{eqnarray}
h_{xy}^{(2)}=2\cos (k_1+k_2-k_3)\tau_z,&\quad&
h_{yx}^{(2)}=2\cos (k_1-k_2)\tau_z,\nonumber\\
h_{xz}^{(2)}=2\cos k_1\tau_z,&\quad&
h_{zx}^{(2)}=2\cos (k_1-k_3)\tau_z,\nonumber\\
h_{yz}^{(2)}=2\cos k_2\tau_z,&\quad&
h_{zy}^{(2)}=2\cos (k_2-k_3)\tau_z.\nonumber
\end{eqnarray}
The multilayer system composed of $N_{\text{L}}$ $(111)$ layers can be obtained from ${\cal H}(\vec{k})$ by replacing quasimomenta $k_3$ by a real-space hopping matrix structure,
\begin{equation}
{\cal H}_{(1,1,1)}(k_{1},k_{2})=
\begin{pmatrix}
{\cal H}_{in} & {\cal H}_{out} & 0 & 0 & 0\\
{\cal H}_{out}^\dagger & {\cal H}_{in} & {\cal H}_{out} & 0 & 0\\
0 & {\cal H}_{out}^\dagger & {\cal H}_{in} & \ddots & 0\\
0 & 0 & \ddots & \ddots & {\cal H}_{out}\\
0 & 0 & 0 & {\cal H}_{out}^\dagger & {\cal H}_{in}
\end{pmatrix},
\label{Hmat}
\end{equation}
where diagonal blocks are given by
\begin{eqnarray}
{\cal H}_{in}(k_1,k_2)&=&m\mathbbm{1}_2\!\otimes\!\mathbbm{1}_3\!\otimes\! \tau_z+t_{12}\!\!\!\sum_{\alpha=x,y,z}\!\!\mathbbm{1}_2
\!\otimes\!\left(\mathbbm{1}_3\!-\!L_{\alpha}^{2}\right)\!\otimes\! h_{\alpha,in}^{(1)} (k_1,k_2)\nonumber\\
&+&t_{11}\sum_{\alpha\not=\beta}\mathbbm{1}_2\!\otimes\!\left[\mathbbm{1}_3\!-\!\tfrac{1}{2}\left(L_{\alpha}\!+\!\varepsilon_{\alpha\beta}L_{\beta}\right)^{2}\right]
\!\otimes\! h_{\alpha\beta,in}^{(2)} (k_1,k_2) +\sum_{\alpha=x,y,z} \lambda \sigma_{\alpha}\!\otimes L_{\alpha}\otimes\mathbbm{1}_2,
\end{eqnarray}
and off-diagonal ones by
\begin{eqnarray}
{\cal H}_{out}(k_1,k_2)&=&t_{12}\!\!\!\sum_{\alpha=x,y,z}\!\!\mathbbm{1}_2
\!\otimes\!\left(\mathbbm{1}_3\!-\!L_{\alpha}^{2}\right)\!\otimes\! h_{\alpha,out}^{(1)} (k_1,k_2)+t_{11}\sum_{\alpha\not=\beta}\mathbbm{1}_2\!\otimes\!\left[\mathbbm{1}_3\!-\!\tfrac{1}{2}\left(L_{\alpha}\!+\!\varepsilon_{\alpha\beta}L_{\beta}\right)^{2}\right]
\!\otimes\! h_{\alpha\beta,out}^{(2)} (k_1,k_2).\quad
\end{eqnarray}
The matrices describing hopping are now given by
\begin{eqnarray}
h_{x,in}^{(1)}=\cos k_1\tau_x-\sin k_1\tau_y,&\quad&
h_{x,out}^{(1)}=\tfrac{1}{2}e^{-ik_1}\tau_x+\tfrac{i}{2}e^{-ik_1}\tau_y, \nonumber\\
h_{y,in}^{(1)}=\cos k_2\tau_x-\sin k_2\tau_y,&\quad&
h_{y,out}^{(1)}=\tfrac{1}{2}e^{-ik_2}\tau_x+\tfrac{i}{2}e^{-ik_2}\tau_y, \nonumber\\
h_{z,in}^{(1)}=\tau_x,&\quad&
h_{z,out}^{(1)}=\tfrac{1}{2}\tau_x+\tfrac{i}{2}\tau_y, \nonumber
\end{eqnarray}

for the nearest neighbors and for the next-nearest neighbors the only non-vanishing matrices are

\begin{eqnarray}
h_{yx,in}^{(2)}=2\cos(k_1-k_2)\tau_z,&\quad&
h_{xy,out}^{(2)}=e^{-i(k_1-k_2)}\tau_z, \nonumber\\
h_{xz,in}^{(2)}=2\cos k_1\tau_z,&\quad&
h_{zx,out}^{(2)}=e^{-i k_1}\tau_z, \nonumber\\
h_{yz,in}^{(2)}=2\cos k_2\tau_z,&\quad&
h_{zy,out}^{(2)}=e^{-i k_2}\tau_z. \nonumber\\
\end{eqnarray}

Additionally, we may add a surface potential term to the Hamiltonian ${\cal H}_{(1,1,1)}(k_{1},k_{2})$ in a form of

\begin{equation}
{\cal V}_{\text{surf}}=V_s \rm diag_N(0,\dots,0,\tfrac{1}{16},\tfrac{1}{8},\tfrac{1}{4},\tfrac{1}{2},1)\!\otimes\!\mathbbm{1}_2\!\otimes\!\mathbbm{1}_3\!\otimes\!\mathbbm{1}_2,
\end{equation}

where  diag$_N(d_1,\dots,d_N)$ means a diagonal matrix with entries given by $d_1,\dots,d_N$ and $V_s$ is the height of the potential.

The important symmetries of the model are mirror reflection symmetry with respect to the $(110)$ plane,
\begin{eqnarray}
{\cal M}{\cal H}_{(1,1,1)}(k_{1},k_{2}){\cal M}^{-1} &=& {\cal H}_{(1,1,1)}(k_{2},k_{1}), \ {\cal M}=\tfrac{1}{\sqrt 2}\mathbbm{1}_N\!\otimes\!(\sigma_x-\sigma_y)\!\otimes\!\left[(L_x-L_y)^2-\mathbbm{1}_3\right]\!\otimes\!\mathbbm{1}_2
\end{eqnarray}
and the time-reversal symmetry
\begin{eqnarray}
{\cal T}{\cal H}_{(1,1,1)}(k_{1},k_{2}){\cal T}^{-1} &=& {\cal H}_{(1,1,1)}(-k_{1},-k_{2}), \ {\cal T}=i{\cal K} \mathbbm{1}_N\!\otimes\!\sigma_y\!\otimes\!\mathbbm{1}_3\!\otimes\!\mathbbm{1}_2.
\end{eqnarray}

Finally, the orthogonal surface quasimomenta $k_{\tilde{x}}$ and $k_{\tilde{y}}$, used in Figs.~1 (main text), \ref{s_fig:loops+berry} and \ref{s_fig:loops+berry2}, are defined as

\begin{equation}
k_{\tilde{x}}=k_2,\quad k_{\tilde{y}}=\frac{1}{\sqrt 3}(2k_1-k_2).
\end{equation}

\section{Mirror and time-reversal symmetry protected quantization of the Berry phases for non-degenerate bands}\label{Berry}

In this section, we assume that all bands are non-degenerate due to the absence of the inversion symmetry. We show that in the presence mirror symmetry the Berry phases for all mirror-symmetric Fermi loops (the Fermi loop maps back to itself in the mirror symmetry operation) are quantized to $0$ or $\pi$. Then we show that in the presence of time-reversal symmetry the Berry phases for all time-reversal-symmetric Fermi loops (the Fermi loop maps back to itself in the time-reversal symmetry operation) are quantized to $\pi$.

Consider the eigenstates $|\psi_{\theta}^{(n)}\rangle$ belonging to the $n$th energy band forming a Fermi loop around $\overline{\Gamma}$ (or $\overline{M}_1$) point parametrized by angle $\theta\in [-\pi,\pi]$. Since $\theta$ is periodic variable $|\psi_{\theta}^{(n)}\rangle$ can acquire a Berry phase by a parallel shift along the loop. The gauge-invariant form of the Berry phase is given by,

\begin{equation}
\varphi_{n}=\arg\left[\left\langle \psi_{-\pi}^{(n)}\right|\!\left.\psi_{-\pi+\delta\theta}^{(n)}\right\rangle \!\left\langle \psi_{-\pi+\delta\theta}^{(n)}\right|\!\left.\psi_{-\pi+2\delta\theta}^{(n)}\right\rangle \!\dots\!\left\langle \psi_{\pi-2\delta\theta}^{(n)}\right|\!\left.\psi_{\pi-\delta\theta}^{(n)}\right\rangle \negmedspace\left\langle \psi_{\pi-\delta\theta}^{(n)}\right|\!\left.\psi_{-\pi}^{(n)}\right\rangle \right].
\end{equation}

where $\delta\theta$ is an infinitesimal step in angle $\theta$. Now we will consider the impact of time-reversal and mirror symmetries on possible values of $\varphi_{n}$.

First, we consider the mirror symmetry. We assume that for $0\leq \theta \leq \pi$ we obtain all $|\psi_{\theta}^{(n)}\rangle$ states by diagonalizing the Hamiltonian and we define $-\pi \leq \theta < 0$ states as

\begin{equation}
\left|\psi_{-\theta}^{(n)}\right\rangle ={\cal M}\left|\psi_{\theta}^{(n)}\right\rangle.
\end{equation}

We can decompose the Berry phase as $\varphi=\arg\left[\chi_{-}\chi_{+}\right]$ with

\begin{eqnarray}
\chi_{-}\!&=&\!\left\langle \psi_{-\pi}^{(n)}\right|\!\left.\psi_{-\pi+\delta\theta}^{(n)}\right\rangle \!\dots\!\left\langle \psi_{-2\delta\theta}^{(n)}\right|\!\left.\psi_{-\delta\theta}^{(n)}\right\rangle \!\left\langle \psi_{-\delta\theta}^{(n)}\right|\!\left.\psi_{0}^{(n)}\right\rangle \!,\nonumber\\
\chi_{+}\!&=&\!\left\langle \psi_{0}^{(n)}\right|\!\left.\psi_{\delta\theta}^{(n)}\right\rangle \!\left\langle \psi_{\delta\theta}^{(n)}\right|\left.\psi_{2\delta\theta}^{(n)}\right\rangle \!\dots\!\left\langle \psi_{\pi-\delta\theta}^{(n)}\right|\!\left.\psi_{-\pi}^{(n)}\right\rangle \!.
\end{eqnarray}

Using mirror symmetry ${\cal M}$ we can relate terms of $\chi_{-}$ and $\chi_{+}$ as,

\begin{equation}
\left\langle \psi_{-\theta-\delta\theta}^{(n)}\right|\!\left.\psi_{-\theta}^{(n)}\right\rangle \!=\!\left\langle \psi_{\theta+\delta\theta}^{(n)}\right|\!{\cal M}^{\dagger}\!{\cal M}\!\left|\psi_{\theta}^{(n)}\right\rangle \!=\!\left\langle \psi_{\theta+\delta\theta}^{(n)}\right|\!\left.\psi_{\theta}^{(n)}\right\rangle \!.
\end{equation}

Therefore most of the phases in $\varphi$ cancel and we get

\begin{equation}
\varphi_{n} = \arg\left[\left\langle \psi_{\pi}^{(n)}\right|\left.\psi_{\pi-\delta\theta}^{(n)}\right\rangle \!\left\langle \psi_{\pi-\delta\theta}^{(n)}\right|\!{\cal M}\!\left|\psi_{\pi}^{(n)}\right\rangle \left\langle \psi_{\delta\theta}^{(n)}\right|\!{\cal M}^{\dagger}\!\left|\psi_{0}^{(n)}\right\rangle \!\left\langle \psi_{0}^{(n)}\right|\!\left.\psi_{\delta\theta}^{(n)}\right\rangle \right]\!.
\end{equation}

On the other hand, we know that $\left|\psi_{0}^{(n)}\right\rangle $ and $\left|\psi_{\pi}^{(n)}\right\rangle $ are eigenstates of ${\cal M}$ with eigenvalues $\pm1$. Hence we have

\begin{eqnarray}
\left\langle \psi_{\pi-\delta\theta}^{(n)}\right|\!{\cal M}\!\left|\psi_{\pi}^{(n)}\right\rangle & \!=\! & \left\langle \psi_{\pi-\delta\theta}^{(n)}\right|\!\left.\psi_{\pi}^{(n)}\right\rangle \!\left\langle \psi_{\pi}^{(n)}\right|\!{\cal M}\!\left|\psi_{\pi}^{(n)}\right\rangle \!,\nonumber\\
\left\langle \psi_{\delta\theta}^{(n)}\right|\!{\cal M}^{\dagger}\!\left|\psi_{0}^{(n)}\right\rangle & \!=\! & \left\langle \psi_{\delta\theta}^{(n)}\right|\!\left.\psi_{0}^{(n)}\right\rangle \!\left\langle \psi_{0}^{(n)}\right|\!{\cal M}\!\left|\psi_{0}^{(n)}\right\rangle \!,
\end{eqnarray}
and consequently

\begin{equation}
\varphi_{n}\!=\!\arg\left[\left\langle \psi_{\pi}^{(n)}\right|\!{\cal M}\!\left|\psi_{\pi}^{(n)}\right\rangle \!\left\langle \psi_{0}^{(n)}\right|\!{\cal M}\!\left|\psi_{0}^{(n)}\right\rangle \right]\!.
\end{equation}

This proves that in the presence of the mirror symmetry $\cal M$ the Berry phase is quantized as $\varphi_{n}=0,\pi$.

Now consider a time-reversal symmetry ${\cal T}={\cal U_{{\cal T}}} {\cal K}$, where ${\cal U_{{\cal T}}}$ is unitary operator and ${\cal K}$ is complex conjugation operator. We assume that the time-reversal symmetry satisfies ${\cal T}^{2}=-1$ so that ${\cal U_{{\cal T}}}{\cal \overline{U_{{\cal T}}}=}-1$, where bar means complex conjugate. We assume that for $-\pi \leq \theta < 0$ we obtain all $|\psi_{\theta}^{(n)}\rangle$ states by diagonalizing the Hamiltonian and we define $0 \leq \theta<\pi$ states as

\begin{equation}
\left|\psi_{\theta+\pi}^{(n)}\right\rangle ={\cal T}\left|\psi_{\theta}^{(n)}\right\rangle ={\cal U_{{\cal T}}}\left|\overline{\psi_{\theta}^{(n)}}\right\rangle \!,
\end{equation}

Now we want to express terms in $\chi_{+}$ by those in $\chi_{-}$. We have that

\begin{equation}
\left\langle \psi_{\theta+\pi}^{(n)}\right|\!\left.\psi_{\theta+\pi+\delta\theta}^{(n)}\right\rangle \!=\!\left\langle \overline{\psi_{\theta}^{(n)}}\right|\!{\cal U_{{\cal T}}}^{\!\dagger}{\cal U_{{\cal T}}}\!\left|\overline{\psi_{\theta+\delta\theta}^{(n)}}\right\rangle \!=\!\left\langle \psi_{\theta+\delta\theta}^{(n)}\right|\!\left.\psi_{\theta}^{(n)}\right\rangle \!.
\end{equation}

Therefore most of the phases in $\varphi$ cancel and we get

\begin{equation}
\varphi_{n} = \arg\left[\left\langle \psi_{-\delta\theta}^{(n)}\right|\!\left.\psi_{0}^{(n)}\right\rangle \!\left\langle \psi_{\pi-\delta\theta}^{(n)}\right|\!\left.\psi_{-\pi}^{(n)}\right\rangle \right]= \arg\left[\left\langle \psi_{-\delta\theta}^{(n)}\right|\!\left.\psi_{0}^{(n)}\right\rangle \!\left\langle \overline{\psi_{-\delta\theta}^{(n)}}\right|\!{\cal U_{{\cal T}}}^{\!\dagger}\overline{{\cal U_{{\cal T}}}^{\!\dagger}\!\left|\psi_{0}^{(n)}\right\rangle }\right] = \arg\left[-\left|\left\langle \psi_{-\delta\theta}^{(n)}\right|\!\left.\psi_{0}^{(n)}\right\rangle \right|^{2}\right]=\pi.
\end{equation}

Hence, we have proved that in the presence of time-reversal symmetry satisfying ${\cal T}^{2}=-1$ the Berry phases are quantized to $\pi$. We point out that if the time-reversal symmetry satisfies ${\cal T}^{2}=1$ the Berry phases are quantized to $0$.

\ \\

\ \\
\section{Berry phases in the symmetry-broken cases} \label{method}

In this section, we discuss how various perturbations breaking the symmetries of the model are included in the theory, and we calculate the effect of these perturbations on the Berry phases. The mirror symmetry breaking can appear due to intentional structural distortions or unintentional inhomogeneities and time-reversal symmetry breaking perturbations can be present due to various mechanisms \citeS{SM_Maestro2013,SM_Pikulin2014,SM_Wang2017,SM_Xue2018,SM_Vaeyrynen2014}. Here, our aim is not to realistically model the breaking of these symmetries in real materials but rather to demonstrate the important role of the time-reversal and mirror symmetries in the quantization of the Berry phase and the WAL effect.

\begin{figure}[tb]
    \includegraphics[width= 0.8\columnwidth]{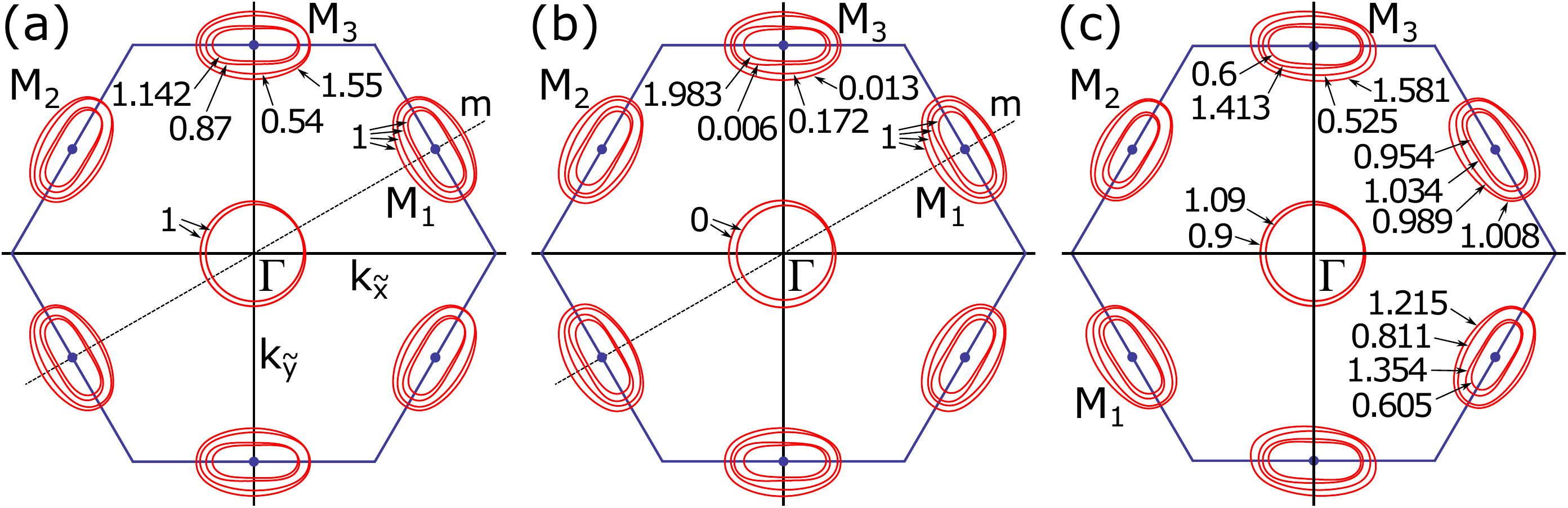}
    \caption{{\bf Fermi loops and their Berry phases (in the units of $\pi$) in presence of the symmetry-breaking terms}. (a) Mirror symmetry is preserved and time-reversal symmetry is weakly broken $\vec{h}=(-0.01,0.01,0)$. (b) Mirror symmetry is preserved and time-reversal symmetry is strongly broken $\vec{h}=(-0.02,0.02,0)$. (c) Mirror and time-reversal symmetries are strongly broken $\gamma=0.4$ and $\vec{h}=(-0.02,0.02,0)$. In the presence of mirror symmetry the Berry phases of mirror-symmetric Fermi loops are quantized to $0$ or $\pi$, depending on the strength of the Zeeman field. If both symmetries are broken all Berry phases are arbitrary. The other parameters in all cases are: $N_{\text{L}}=10$, $m=1.91$, $t_{12}=0.9$, $t_{11}=0.5$ $\lambda=0.3$ and $\mu=-0.25$ (all energies in eV).}
    \label{s_fig:loops+berry}
\end{figure}

\begin{figure}[tb]
    \includegraphics[width=0.7 \columnwidth]{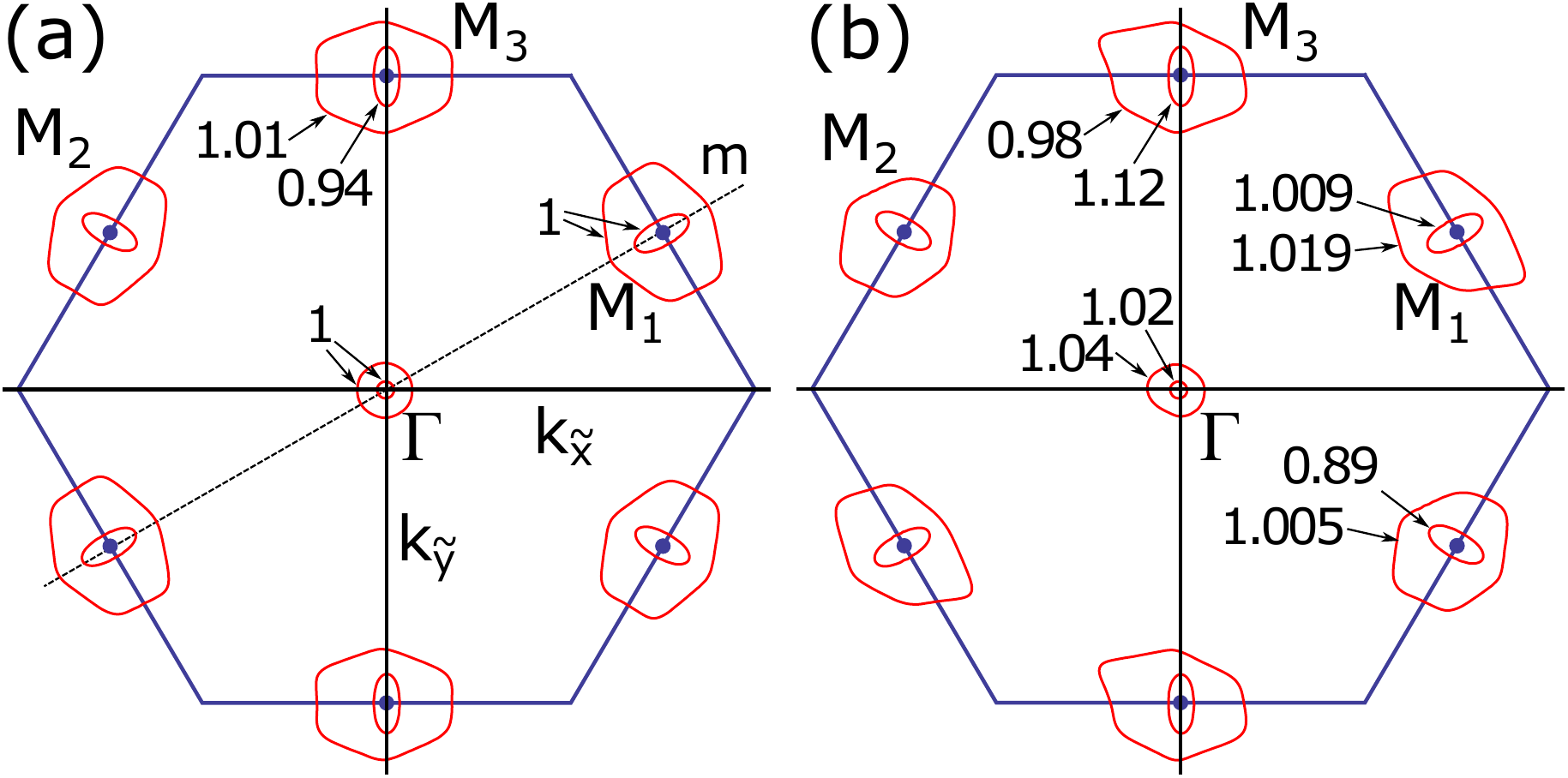}
    \caption{{\bf Fermi loops and their Berry phases (in the units of $\pi$) in presence of the symmetry-breaking terms.} (a) Mirror symmetry is preserved and time-reversal symmetry is weakly broken $\vec{h}=(-0.01,0.01,0)$. (b) Mirror and time-reversal symmetries are strongly broken $\gamma=0.4$ and $\vec{h}=(-0.02,0.02,0)$. In the presence of mirror symmetry the Berry phases of mirror-symmetric Fermi loops are quantized to $0$ or $\pi$, depending on the strength of the Zeeman field. If both symmetries are broken all Berry phases are arbitrary. The other parameters in all cases are: $N_{\text{L}}=10$, $m=0.8$, $t_{12}=0.9$, $t_{11}=0.1$, $\lambda=0.5$ and $\mu=-0.2$ (all energies in eV).}
    \label{s_fig:loops+berry2}
\end{figure}

To break time-reversal symmetry we consider a Zeeman field $\vec{h}$ coupling to spins $\vec{\sigma}$,

\begin{equation}
{\cal H}_{mag}=\mathbbm{1}_N\!\otimes\!\vec{h}\cdot\vec{\sigma}\!\otimes\!\mathbbm{1}_3\!\otimes\!\mathbbm{1}_2.
\label{eq: Zeeman}
\end{equation}

\begin{figure}[tb]
    \includegraphics[width=13cm]{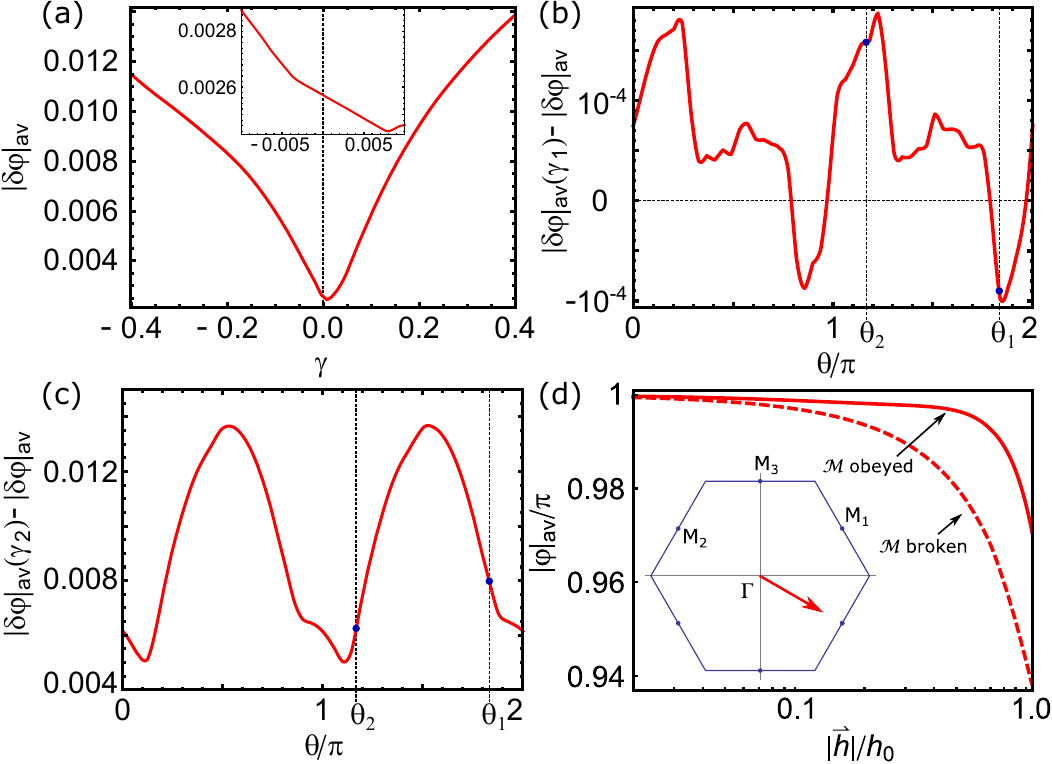}
    \caption{{\bf Dependencies of deviation from the quantized value of the Berry phase (averaged over Fermi loops)} on: (a) strength of mirror symmetry breaking $\gamma$ in presence of a weak Zeeman field $\vec{h}=(0,0.001,0.001)$ and (b,c) in-plane Zeeman field angle $\theta$ for $|\vec{h}|=1$ in presence of weak (b) $\gamma = 0.008$ and strong (c) $\gamma = 0.4$ mirror symmetry breaking. In (b,c) we plot the difference with respect to the $\gamma = 0$ case and angles $\theta_1$ and $\theta_2$ correspond to $\vec{h}\propto -\vec{a}_1$ and  $\vec{h}\propto \vec{a}_2$, respectively. (d) Deviation of the Berry phase from the quantized value $\varphi = \pi $ as a function of the Zeeman field oriented in the same direction as in experiment for the case of unbroken (solid lines) and broken (dashed lines) crystalline mirror symmetry. The other parameters in all cases are: $N_{\text{L}}=10$, $m=1.91$, $t_{12}=0.9$, $t_{11}=0.5$, $\lambda=0.3$, $\mu=-0.25$  and $h_0=0.01$ (all energies in eV).}
    \label{s_fig:loops_g_thet}
\end{figure}

To break mirror symmetry $\cal M$ on one surface of the system we modify the $N_{\text{L}}$-th diagonal block ${\cal H}_{in}$ in ${\cal H}_{(1,1,1)}(k_{1},k_{2})$ of Eq.~(\ref{Hmat}) by setting hopping amplitude in $x$ direction as different than in $y$ direction, i.e.,

\begin{eqnarray}
h_{x,in}^{(1)}\to(1+\gamma)h_{x,in}^{(1)},&\quad&
h_{xz,in}^{(2)}\to(1+\gamma)h_{xz,in}^{(2)},\quad
h_{zx,in}^{(2)}\to(1+\gamma)h_{zx,in}^{(2)}\nonumber\\
h_{y,in}^{(1)}\to(1-\gamma)h_{y,in}^{(1)},&\quad&
h_{yz,in}^{(2)}\to(1-\gamma)h_{yz,in}^{(2)},\quad
h_{zy,in}^{(2)}\to(1-\gamma)h_{zy,in}^{(2)},\nonumber
\end{eqnarray}

where $\gamma$ describes the strength of mirror symmetry breaking.

For Fig.~1 (main text) we have diagonalized ${\cal H}_{(1,1,1)}(k_{1},k_{2})$ with $N_{\text{L}}=10$ layers taking the following parameters (all in eV, $\mu$ is the Fermi level)

\begin{itemize}
    \item Fully symmetric
    \begin{itemize}
        \item Non-trivial: $m=1.91$, $t_{12}=0.9$, $t_{11}=0.5$ $\lambda=0.3$ and chemical potential $\mu=-0.25$.
        \item Trivial: $m=2.208$, $t_{12}=0.9$, $t_{11}=0.5$ $\lambda=0.3$, $V_s=0.5$ and $\mu=-0.35$.
    \end{itemize}
     \item Broken time-reversal
    \begin{itemize}
        \item Non-trivial: $m=1.91$, $t_{12}=0.9$, $t_{11}=0.5$ $\lambda=0.3$, $\vec{h}=(-0.01,0.01,0)$ and $\mu=-0.25$.
        \item Trivial: $m=2.208$, $t_{12}=0.9$, $t_{11}=0.5$ $\lambda=0.3$, $V_s=0.25$, $\vec{h}=(-0.01,0.01,0)$, and $\mu=-0.35$.
    \end{itemize}
    \item Broken mirror and time-reversal
    \begin{itemize}
        \item Non-trivial: $m=1.91$, $t_{12}=0.9$, $t_{11}=0.5$ $\lambda=0.3$, $\gamma=0.4$, $\vec{h}=(-0.01,0.01,0)$ and $\mu=-0.25$.
        \item Trivial: $m=2.208$, $t_{12}=0.9$, $t_{11}=0.5$ $\lambda=0.3$, $V_s=0.25$, $\gamma=0.4$, $\vec{h}=(-0.01,0.01,0)$ and $\mu=-0.35$.
    \end{itemize}
\end{itemize}

The exact values of the parameters are unimportant for our qualitative considerations. They just determine the exact shape and the size of the Fermi loops. In the trivial side we have introduced non-zero surface potential $V_s$. Due to this potential some of the low-energy states are localized close to the surface forming topologically trivial surface states in qualitative agreement with the experimental observations.

In Fig.~\ref{s_fig:loops+berry} we show we show Fermi loops and Berry phases in the case of broken time-reversal symmetry. When the mirror symmetry is preserved the Berry phase for mirror-symmetric Fermi loops around $\overline{\Gamma}$ and $\overline{M}_1$ points are quantized as either $0$ or $\pi$. In the case of weakly broken time-reversal symmetry, the Berry phases remain quantized to $\pi$ (Fig.~\ref{s_fig:loops+berry}a). By increasing the strength of the time-reversal symmetry breaking the Berry phases around the $\overline{\Gamma}$ point become $0$ (Fig.~\ref{s_fig:loops+berry}b). The changes of Berry phases occur at topological transitions where energies of two bands become degenerate at a particular momentum within the Fermi loops. The Fermi loops around $\overline{M}_1$ remain non-trivial up to the larger value of the field and the Berry phases for Fermi loops around other high-symmetry points take non-quantized values. In the case when all symmetries are broken (Fig.~\ref{s_fig:loops+berry}c), all the Berry phases are non-quantized. Fig.~\ref{s_fig:loops_g_thet}d shows the dependencies of deviation of the Berry phase from the quantized value $\varphi = \pi $ as a function of the inplane Zeeman field for the cases of preserved and broken mirror symmetry. The curves follow similar trend as the one observed in the experiment, shown in Fig.~6 (main text) and calculated from the Cooperon propagator, see Fig.~3f (main text).

The Fermi loops around $\overline{M}$ points in Figs.~1 (main text) and \ref{s_fig:loops+berry} take form of the ellipses elongated in the direction perpendicular to $\overline{\Gamma}-\overline{M}$ line. It is however possible in the present model to obtain the elongation parallel to the $\overline{\Gamma}-\overline{M}$ line.  In Fig.~\ref{s_fig:loops+berry2} we show that effect of the time-reversal and mirror symmetry breaking perturbations on the Berry phases is the same as before.

It is important to determine how robust is the tendency of the Berry phase, in the presence of an in-plane magnetic field, to deviate stronger from the quantized value when the mirror symmetry is simultaneously broken. In Fig.~\ref{s_fig:loops_g_thet}a we show a representative dependence of the average deviation of the Berry phase $|\delta \varphi|_{\rm av}$ on $\gamma$ in the presence of a small Zeeman field in the in-plane direction $\vec{a}_2$. We notice that apart from the small interval between $\gamma = 0$ and $\gamma =0.018$ the deviation is always greater compared to the mirror symmetric case $\gamma = 0$. We also note that the curve shown in Fig.~\ref{s_fig:loops_g_thet}a gets transformed as $|\delta \varphi|_{\rm av}(\gamma)\to|\delta \varphi|_{\rm av}(-\gamma)$ when the Zeeman field is transformed by a mirror symmetry ${\cal M}: \vec{h} \propto \vec{a}_2\to \vec{h} \propto -\vec{a}_1$. Therefore, if the mirror symmetry breaking is small, the deviation can be decreased by choosing the Zeeman field in the $\vec{a}_2$ direction and increased by the field in the  $ -\vec{a}_1$ direction. This is confirmed by determining the dependence of the deviation $|\delta \varphi|_{\rm av}$ on the direction of the in-plane Zeeman field. We parametrize this field as $\vec{h}\propto\cos\theta (\vec{a}_1+\vec{a}_2)/\sqrt 6+\sin\theta (\vec{a}_1+\vec{a}_2)/\sqrt 2$ (note that lattice vectors $\vec{a}_i$ are not orthogonal) and for fixed $\gamma$ we track the difference in deviation of the Berry phase with respect to $\gamma = 0$ case as function of $\theta$. In Figs.~\ref{s_fig:loops_g_thet}b,c we show the results for small $\gamma = 0.008$ and large $\gamma = 0.4$. We have marked the angles $\theta_1$ and $\theta_2$ that correspond to mirror-related field directions $\vec{h}\propto -\vec{a}_1$ and  $\vec{h}\propto \vec{a}_2$ (in general case mirror symmetry relates $\vec{h}(\theta)$ with $\vec{h}(\pi-\theta)$). We see that for small $\gamma$ the deviation with respect to mirror-symmetric case is larger for $\theta=\theta_2$ and smaller for  $\theta=\theta_1$ but for large $\gamma$ in both cases the deviation is larger.

\section{Quantum correction to the conductivity \label{sec:coop}}

In the vicinity of the band crossings appearing at the high-symmetry points we derive a low-energy 2D Hamiltonian for a single pair of subbands in a form of
\begin{equation}
H_{\vec{k},\vec{\sigma}}=\frac{\hbar^{2}}{2m_{e}}\vec{k}^{2}+\alpha_{\text{so}}\left(\sigma_{x}k_{y}-\sigma_{y}k_{x}\right)+g\sigma_{z}.
\end{equation}
Here $m_{e}$ is the effective mass of the electron, $\alpha_{\text{so}}$ is an effective spin-orbit-like coupling that arises from breaking of the inversion symmetry (always present due to the surface in these samples), $g$ is the mass term induced by the breaking of the mirror symmetry and weak breaking of the time-reversal symmetry, and $\sigma$ is an effective pseudospin variable which describes entangled spin and orbital degrees of freedom.

The quantum correction to the conductivity can be written as \citeS{SM_Wenk2010}
\begin{equation}
\Delta\sigma_{xx}=-\frac{e^{2}}{\pi\hbar}\frac{D_{e}}{L^{2}}\sum_{\vec{Q}}\sum_{\alpha,\beta=\pm}C_{\alpha\beta\beta\alpha}(\vec{Q})\label{eq:ds}
\end{equation}
where $D_{e}=v_{F}l/2$ is the 2D diffusion constant, $l=v_{F}\tau$ is the elastic mean-free path, $v_{F}$ is the Fermi velocity, $\tau$ is the elastic scattering time, $L^{2}$ is the area of the sample, and $\alpha,\beta=\pm$ are the pseudospin indices of the Cooperon propagator $C$. For weak disorder ($\tau E_F/\hbar\gg1$ where $ E_F$ is the Fermi energy) the Cooperon propagator can be approximated as
\begin{equation}
C(\vec{Q})=\tau\left(1-\int\frac{d\Omega}{2\pi}\frac{1}{1-i\tau\Sigma/\hbar}\right)^{-1},\label{eq:coop}
\end{equation}
where
\begin{equation}
\Sigma(\vec{Q})=H_{\vec{Q}-\vec{k},\vec{\sigma}'}-H_{\vec{k},\vec{\sigma}}
\end{equation}
is a $4\times4$ matrix describing two interfering electrons with pseudospins $\vec{\sigma}'$ and $\vec{\sigma}$. The integral in Eq.\,(\ref{eq:coop}) is over all angles of velocity
\begin{equation}
\vec{v}=\frac{\hbar\vec{k}}{m_{e}}
\end{equation}
on the Fermi surface. To the lowest order in $\vec{Q}$ and $\alpha_{\text{so}}$, we get
\begin{equation}
\Sigma=-\hbar\vec{v}\cdot\vec{Q}-\hbar Q_{\text{so}}\left(S_{x}v_{y}-S_{y}v_{x}\right)+g\left(\sigma_{z}'-\sigma_{z}\right)=-\hbar\vec{v}\cdot(\vec{Q}+Q_{\text{so}}\hat{a}\vec{S})+g\left(\sigma_{z}'-\sigma_{z}\right),
\end{equation}
where $Q_{\text{so}}=2m_{e}\alpha_{\text{so}}/\hbar^{2}$,
\begin{eqnarray}
\vec{S} & = & \frac{1}{2}\left(\vec{\sigma}+\vec{\sigma}'\right),
\end{eqnarray}
and
\begin{equation}
\hat{a}=\begin{pmatrix}0 & -1\\
1 & 0
\end{pmatrix}.
\end{equation}

Thus, we can write the Cooperon as
\begin{equation}
C(\vec{Q})^{-1}=\frac{1}{\tau}\left(1-\int\frac{d\Omega}{2\pi}\frac{1}{1+i\frac{\tau}{\hbar}\left[\hbar\vec{v}\cdot\left(\vec{Q}+Q_{\text{so}}\hat{a}\vec{S}\right)-g\left(\sigma_{z}'-\sigma_{z}\right)\right]}\right).
\end{equation}
Expanding the Cooperon to the second oder in $\left(\vec{Q}+Q_{\text{so}}\hat{a}\vec{S}\right)$ and performing angular integral over $\vec{v}$ (having $|\vec{v}|=v_{F}$) we get for $g/\hbar\ll1/\tau$
\begin{equation}
C(\vec{Q})=\frac{1}{D_{e}\left(\vec{Q}+Q_{\text{so}}\hat{a}\vec{S}\right)^{2}-i\frac{g}{\hbar}\left(\sigma_{z}'-\sigma_{z}\right)}.
\end{equation}
Thus the problem comes down to inverting a non-Hermitian Cooperon Hamiltonian in a form of:
\begin{equation}
H_{c}\equiv\frac{C^{-1}}{D_{e}}=\vec{Q}^{2}+2Q_{\text{so}}\vec{Q}\cdot\hat{a}\vec{S}+Q_{\text{so}}^{2}\left(S_{x}^{2}+S_{y}^{2}\right)-i\frac{g}{\hbar D_{e}}\left(\sigma_{z}'-\sigma_{z}\right).\label{eq:Hc}
\end{equation}
In the basis of triplet $|S=1,m=1\rangle=|\uparrow\uparrow\rangle$,
$|S=1,m=0\rangle=(|\uparrow\downarrow\rangle+|\downarrow\uparrow\rangle)/\sqrt{2}$,
$|S=1,m=-1\rangle=|\downarrow\downarrow\rangle$ and singlet $|S=0,m=0\rangle=(|\uparrow\downarrow\rangle-|\downarrow\uparrow\rangle)/\sqrt{2}$
states, the Cooperon Hamiltonian can be written as
\begin{equation}
H_{c}=\begin{pmatrix}Q_{\text{so}}^{2}+\vec{Q}^{2} & \sqrt{2}Q_{\text{so}}Q_{+} & 0 & 0\\
\sqrt{2}Q_{\text{so}}Q_{-} & 2Q_{\text{so}}^{2}+\vec{Q}^{2} & \sqrt{2}Q_{\text{so}}Q_{+} & -i\eta\\
0 & \sqrt{2}Q_{\text{so}}Q_{-} & Q_{\text{so}}^{2}+\vec{Q}^{2} & 0\\
0 & -i\eta & 0 & \vec{Q}^{2}
\end{pmatrix},
\end{equation}
with
\begin{equation}
Q_{\pm}=Q_{y}\pm iQ_{x},\qquad\eta=\frac{2g}{\hbar D_{e}}.
\end{equation}
Equation\,(\ref{eq:ds}) can now be written as:
\begin{equation}
\Delta\sigma_{xx}=\frac{e^{2}}{\pi\hbar}\frac{1}{L^{2}}\sum_{\vec{Q}}{\rm Tr}\left[\Gamma\left(\frac{1}{D_{e}\tau_{\phi}}+H_{c}\right)^{-1}\right],
\end{equation}
where we have also included the dephasing time $\tau_{\phi}$ and
$\Gamma$ is given by
\begin{equation}
\Gamma=\begin{pmatrix}-1 & 0 & 0 & 0\\
0 & -1 & 0 & 0\\
0 & 0 & -1 & 0\\
0 & 0 & 0 & 1
\end{pmatrix}.
\end{equation}

Following the standard approach (e.g., ref.~\citeS{SM_Hikami1980,SM_Iordanskii1994,SM_Knap1996}), we introduce the perpendicular magnetic field in the Cooperon Hamiltonian by minimal substitution $\vec{Q}\to\vec{Q}+2e\vec{A}/\hbar$, where the vector potential can be chosen as $\vec{A}=(0,xB,0)$. This way we obtain
\begin{eqnarray}
Q_{+} & \to & Q_{y}+iQ_{x}+\frac{2eB}{\hbar}x=2\sqrt{\frac{eB}{\hbar}}\frac{1}{\sqrt{2}}\bigg(\sqrt{\frac{\hbar}{2eB}}\partial_{x}+\sqrt{\frac{2eB}{\hbar}}x+Q_{y}\sqrt{\frac{\hbar}{2eB}}\bigg)=2\sqrt{\frac{eB}{\hbar}}a,\nonumber \\
Q_{-} & \to & Q_{y}-iQ_{x}+\frac{2eB}{\hbar}x=2\sqrt{\frac{eB}{\hbar}}\frac{1}{\sqrt{2}}\bigg(-\sqrt{\frac{\hbar}{2eB}}\partial_{x}+\sqrt{\frac{2eB}{\hbar}}x+Q_{y}\sqrt{\frac{\hbar}{2eB}}\bigg)=2\sqrt{\frac{eB}{\hbar}}a^{\dagger},\nonumber \\
\vec{Q}^{2} & \to & 2\frac{eB}{\hbar}(-\partial_{\xi}^{2}+\xi^{2})=4\frac{eB}{\hbar}\left(a^{\dagger}a+\frac{1}{2}\right),\nonumber \\
a & = & \frac{1}{\sqrt{2}}(\partial_{\xi}+\xi),\ a^{\dag}=\frac{1}{\sqrt{2}}(-\partial_{\xi}+\xi),\ \xi=\sqrt{\frac{2eB}{\hbar}}x+Q_{y}\sqrt{\frac{\hbar}{2eB}}.
\end{eqnarray}
Here $a^{(\dagger)}$ are the harmonic oscillator ladder operators which obey bosonic commutation relations.

The Cooperon Hamiltonian becomes:
\begin{equation}
H_{c}=2Q_{\text{so}}\sqrt{\frac{eB}{\hbar}}\left(aS_{+}+H.c.\right)+4\frac{eB}{\hbar}\left(a^{\dagger}a+\frac{1}{2}\right)+Q_{\text{so}}^{2}\left(2P-S_{z}^{2}\right)-i\eta R,
\end{equation}
with matrices
\begin{equation}
\frac{S_{+}}{\sqrt{2}}=\begin{pmatrix}0 & 1 & 0 & 0\\
0 & 0 & 1 & 0\\
0 & 0 & 0 & 0\\
0 & 0 & 0 & 0
\end{pmatrix},\quad S_{z}=\begin{pmatrix}1 & 0 & 0 & 0\\
0 & 0 & 0 & 0\\
0 & 0 & -1 & 0\\
0 & 0 & 0 & 0
\end{pmatrix},\quad P=\begin{pmatrix}1 & 0 & 0 & 0\\
0 & 1 & 0 & 0\\
0 & 0 & 1 & 0\\
0 & 0 & 0 & 0
\end{pmatrix},\quad R=\begin{pmatrix}0 & 0 & 0 & 0\\
0 & 0 & 0 & 1\\
0 & 0 & 0 & 0\\
0 & 1 & 0 & 0
\end{pmatrix}.
\end{equation}
We can now utilize the fact that the Cooperon Hamiltonian satisfies a symmetry $[H_{c},M_{z}]=0$, where
\begin{equation}
M_{z}=a^{\dagger}a+S_{z}.
\end{equation}
Namely, the spectrum of $M_{z}$ is composed of eigenvalues $\left\{ -1,0,0,0,1,1,1,1,2,2,2,2,\dots\right\} $, and the Hamiltonian has a block diagonal form in the eigenbasis of the $M_{z}$. The first two blocks can be combined into a $4\times4$ Hamiltonian
\begin{equation}
\frac{H_{c}^{(0)}}{Q_{\text{so}}^{2}}=\begin{pmatrix}1+2b & 0 & 0 & 0\\
0 & 2+2b & \sqrt{8b} & -i\eta Q_{\text{so}}^{-2}\\
0 & \sqrt{8b} & 1+6b & 0\\
0 & -i\eta Q_{\text{so}}^{-2} & 0 & 2b
\end{pmatrix},
\end{equation}
and the other blocks are given by
\begin{equation}
\frac{H_{c}^{(n)}}{Q_{\text{so}}^{2}}=\begin{pmatrix}1+\left(4n-2\right)b & \sqrt{8nb} & 0 & 0\\
\sqrt{8nb} & 2+\left(4n+2\right)b & \sqrt{8\left(n+1\right)b} & -i\eta Q_{\text{so}}^{-2}\\
0 & \sqrt{8\left(n+1\right)b} & 1+\left(6+4n\right)b & 0\\
0 & -i\eta Q_{\text{so}}^{-2} & 0 & \left(2+4n\right)b
\end{pmatrix},\ (n=1,2,\dots n_{max}-1)
\end{equation}
where $b=\frac{eB}{\hbar Q_{\text{so}}^{2}}$. The maximum number of Landau levels $n_{max}$ is restricted by the condition that the Cooperon cyclotron radius should be larger than the elastic mean-free path. This can be written as
\begin{equation}
n_{max}=\frac{1}{4\tau B}\frac{\hbar}{eD_{e}}.
\end{equation}
Taking also into account that the degeneracy of the Cooperon Landau levels is $eBL^{2}/(\pi\hbar)$, we obtain
\begin{equation}
\Delta\sigma_{xx}=\frac{e^{2}}{4\pi^{2}\hbar}\left(\frac{4Be}{\hbar}\right)\sum_{n=0}^{n_{max}}{\rm Tr}\left[\Gamma\left(\frac{1}{D_{e}\tau_{\phi}}+H_{c}^{(n)}\right)^{-1}\right].\label{eq:ds_fin}
\end{equation}
Assuming that $\tau\ll\tau_{\phi}$, we can take the limit $n_{max}\to\infty$. In this case, we recover the Hikami-Larkin-Nagaoka (HLN) formula in the limit $\eta=Q_{\text{so}}=0$, i.e.,
\begin{equation}
\Delta\sigma_{xx}(B)-\Delta\sigma_{xx}(0)=\frac{e^{2}}{2\pi^{2}\hbar}\left[\psi\left(\frac{1}{2}+\frac{1}{4\tau_{\phi}B}\frac{\hbar}{eD_{e}}\right)-\log\left(\frac{1}{4\tau_{\phi}B}\frac{\hbar}{eD_{e}}\right)\right].
\end{equation}
To obtain $Q_{\text{so}}$ we use the low energy expansion around $\overline{\Gamma}$ point of the 10-layer model in the topological phase (i.e. $m=1.91$, $t_{12}=0.9$, $t_{11}=0.5$, $\lambda=-0.3$ eV) with lattice constant $a=6.12\times10^{-10}$ m to get $Q_{\text{so}}=1.022\times10^{8}$ m$^{-1}$. This means that the pseudospin precession length is $l_{\text{so}}=2\pi/Q_{\text{so}}\approx60$\,nm.

\begin{figure}
\includegraphics[width=0.6\textwidth]{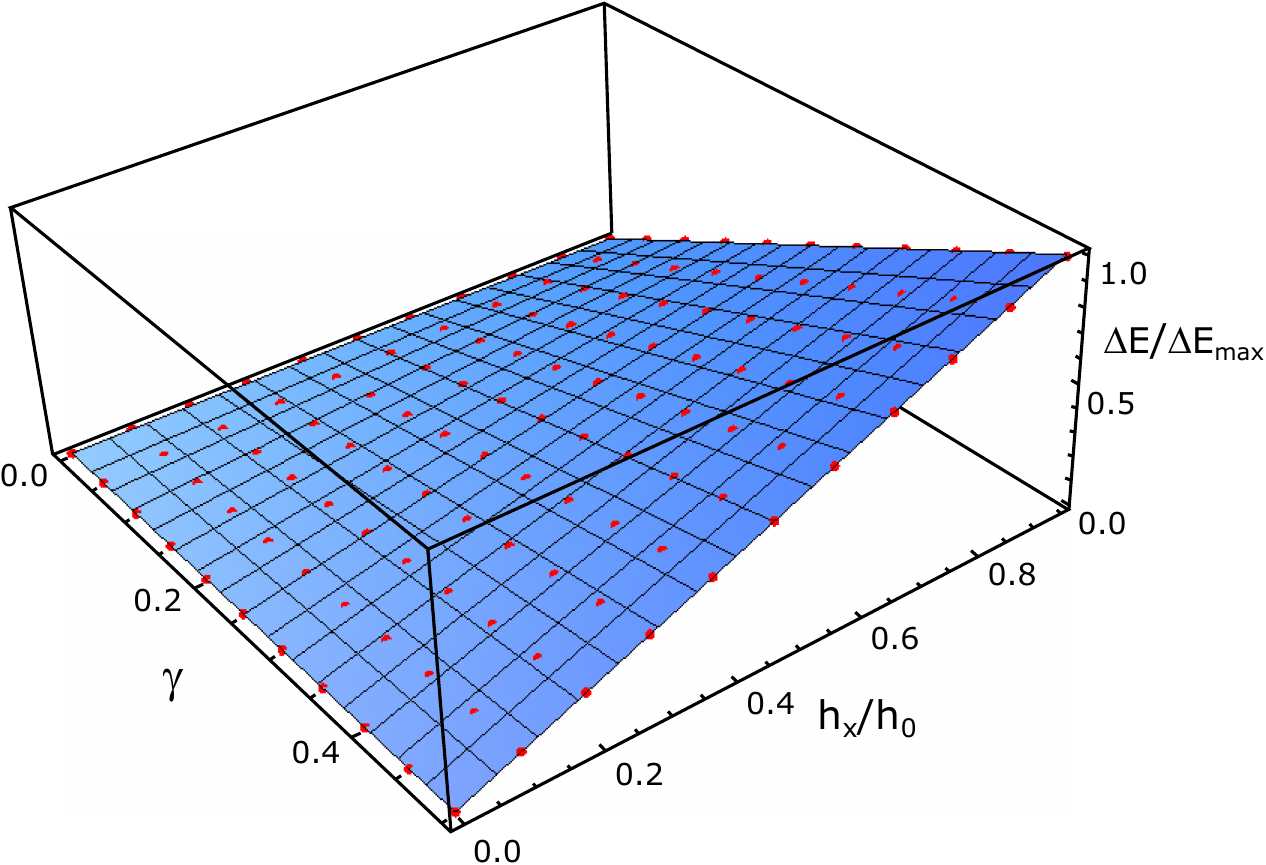}
	\caption{Gap $\Delta E$ between two subbands that form Fermi surface around $\overline{\Gamma}$ points as function of mirror breaking term $\gamma$ and time reversal Zeeman field $h_x = -h_y$ that preserves mirror symmetry. The maximal gap $\Delta E_{\text{max}}$ is defined as $\Delta E$ at $\gamma = 0.5$ and $h_x = h_0$ with $h_0=0.01$\,eV. The other parameters are: $N_{\text{L}}=10$, $m=1.91$, $t_{12}=0.9$, $t_{11}=0.5$, $\lambda=0.3$ and $\mu=-0.25$  (all energies in eV).}
\label{fig:gap}
\end{figure}

The parameter $g=0$ unless both time-reversal and mirror symmetry are simultaneously broken, see Fig.~\ref{fig:gap}. We can assume that the time-reversal symmetry is always weakly broken by the same amount due to the intrinsic mechanisms, impurities and environment, but the breaking of the mirror symmetry is tunable and depends on how the sample is covered.

\section{Structural characterization}
\begin{figure}[tb]
    \centering
    \includegraphics[width=0.55\columnwidth]{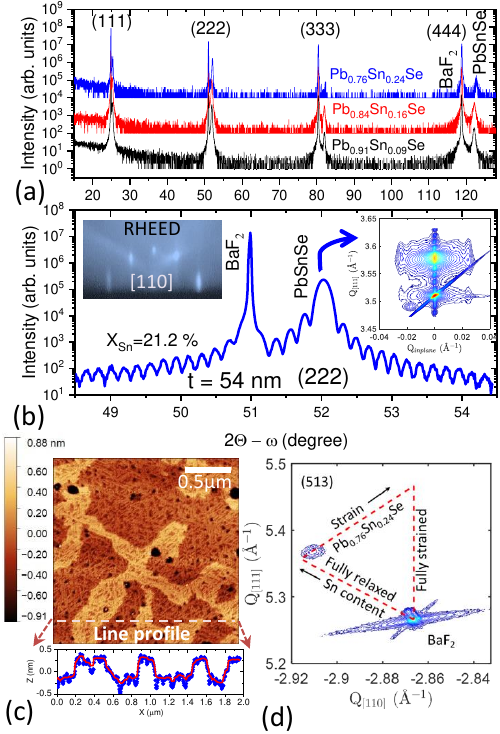}
    \caption{{\bf Structural characterization of obtained samples}. (a) XRD spectra of three samples with different Sn content indicating of (111) oriented single phase from films and substrate. (b) High resolution XRD spectrum of Pb$_{0.78}$Sn$_{0.22}$Se film around (222) reflection showing high order of thickness fringes; the insets represent streaky RHEED pattern obtained along [110] azimuth for the same sample (left) and XRD RSM in the vicinity of symmetric (222) reflection (right). (c) Typical AFM of the film with extracted line profile demonstrating monolayer thick roughness. (d) XRD (513) asymmetric RSM with relaxation triangle evidencing the presence of small strains in the films.}
    \label{fig:struc}
\end{figure}

Our structural and surface morphology investigations revealed the high structural quality of obtained samples. The XRD studies show the presence of a single phase from the (111) oriented film and substrate (Fig.~\ref{fig:struc}a). The presence of atomically sharp surface of the films, which is indicated by streaky RHEED reflections situated on Laue semicircle (inset to Fig.~\ref{fig:struc}b), is further confirmed by high-resolution XRD measurements (Fig.~\ref{fig:struc}b). A large number of well-developed Kissing fringes, the result of interference between surface and interface reflected beams, guarantees a presence of smooth surface and interface of the films. From the relative position of the fringes, the film thickness is precisely determined. AFM investigations revealed a monolayer flat surface as one can deduce from the AFM line profile (Fig.~\ref{fig:struc}c). The contours of monolayer steps are clearly defined in the presented AFM image. The average RMS roughness of 0.345~nm is also consistent with an elevation level difference of 1 monolayer (1 atomic layer of metal + 1 atomic layer of Se). Due to relatively large lattice constant misfit between the film and substrate ($\Delta a/a=1.2-2$\%), one can expect that films are not completely relaxed. Indeed, small in-plane tensile strains of order 0.2\% were determined from XRD reciprocal space maps (RSM) of asymmetric (513) reflection at room temperature (Fig.~\ref{fig:struc}d). Such small changes of the lattice constant can not result in significant alternations of band structure \citeS{SM_zhao2015tuning}. Because of negligible difference in thermal expansion coefficient (TEC) of film and substrate (TEC$_{\text{BaF$_{2}$}}$=18.1 10$^{-6}$/K, TEC$_{\text{PbSe}}$=19.4 10$^{-6}$/K at room temperature \citeS{SM_Villars2016,SM_LandoltBornstein1998}, we do not expect big changes in the value of strain upon film cooling. In addition to residual strain determination, asymmetric RSM is used to calculate a fully relaxed lattice constant and verify the composition of the films according to Vegard's law \citeS{SM_mccann1987phase,SM_assaf2017magnetooptical}.

\section{List of samples and magnetotransport characterization}

\begin{figure}[t]
    \includegraphics[width=0.85\columnwidth]{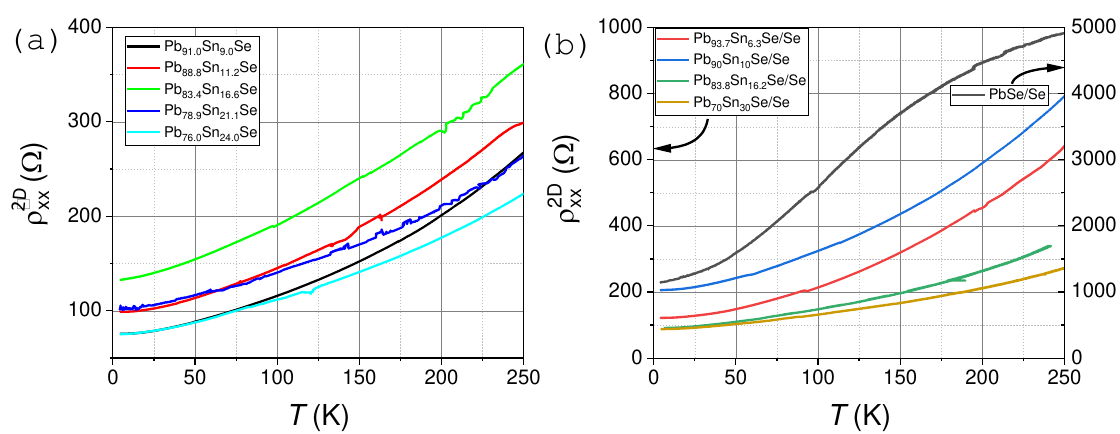}
    \caption{{\bf Temperature dependence of the resistivity.} (a) Uncovered samples; (b) samples covered by Se.}
    \label{s_fig:RvsT}
\end{figure}

\begin{table*}[b]
\caption{Parameters of the studied epilayers: tin content $x$; film thickness $d$; apparent hole concentration $p$, carrier mobility $\mu_p$ and mean free path $l$ determined from the Hall and resistivity data; the phase coherence length extracted from the HLN fit, Eq.~(\ref{s_eq:HLN}) at 1.5~K.}
\begin{tabular}{|c|c|c|c|c|c|c|c|c|}
\hline
Epilayer & Se covered & Sn content, \% & d, nm& $p$, $10^{19}$cm$^{-3}$ & $\mu_p$, cm$^2$/Vs & $l$, nm & $l_{\phi}$ at 1.5~K, nm \\
\hline
A & no & 9.0 & 52 & 1.20 & 1200 & 34.9 & 1080 \\
B & no & 11.2 & 50 & 1.37 & 950 & 21.7 & 1890 \\
C & no & 16.6 & 50 & 1.57 & 490 & 7.9 & 760 \\
D & no & 19.5 & 52 & 2.32 & 540 & 17.7 & 910 \\
E & no & 24.0 & 50 & 1.71 & 930 & 25.1 & 1830 \\
\hline
F & yes & 0.0 & 44 & 0.03 & 3850 & 125 & 400 \\
G & yes & 6.5 & 47 & 0.49 & 2080 & 63 & 310 \\
H & yes & 13.6 & 53 & 1.26 & 480 & 5 & 270 \\
I & yes & 15.9 & 54 & 1.26 & 1120 & 18 & 370 \\
J & yes & 30.0 & 50 & 2.1 & 690 & 47 & 150 \\
\hline
\end{tabular}
\label{tab:info}
\end{table*}

\begin{figure}[tb]
    \includegraphics[width=\columnwidth]{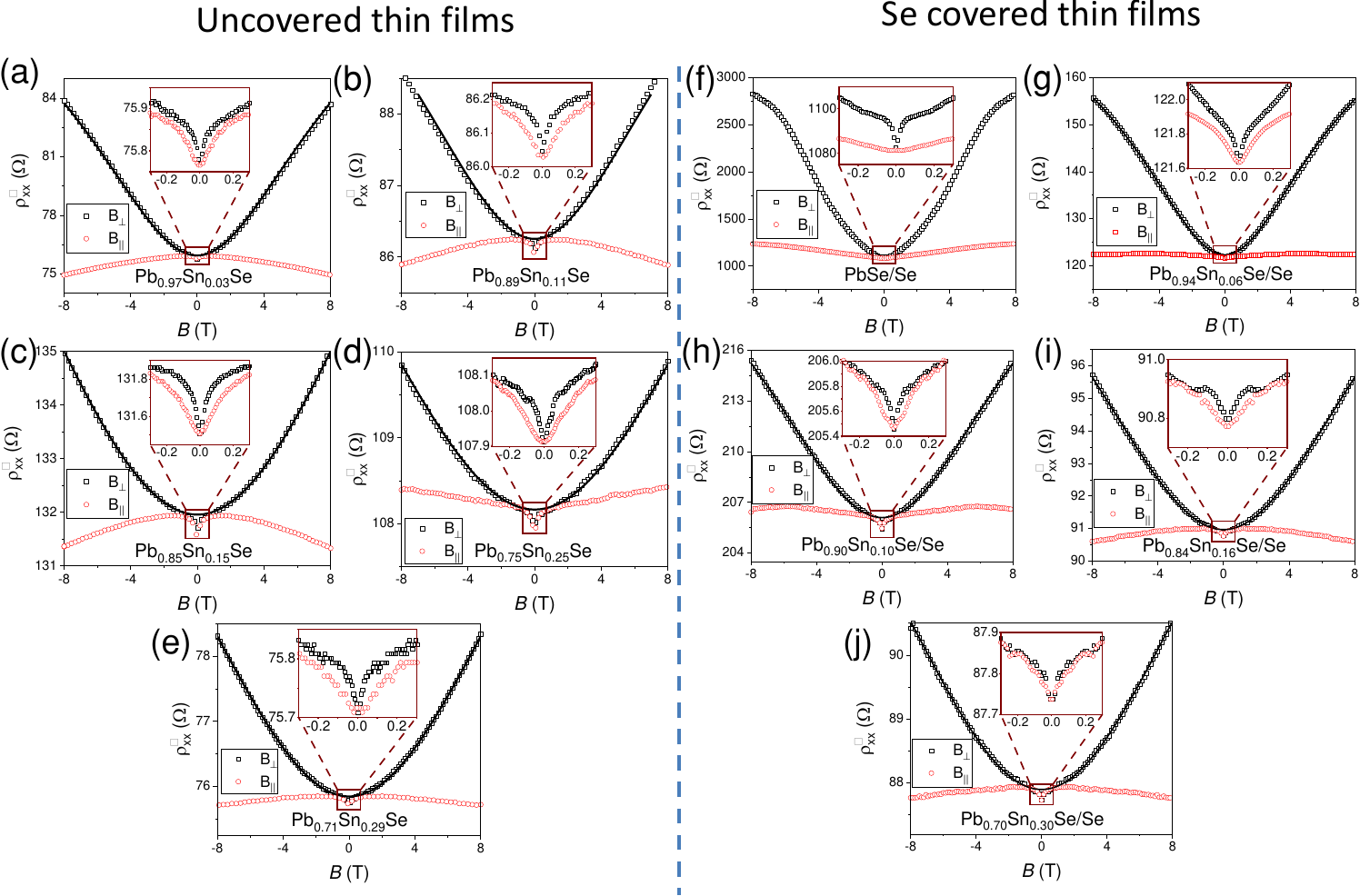}
    \caption{{\bf Magnetoresistivity in a high field range}. (a-e) Uncovered samples; (f-j) samples covered by Se. The data for the magnetic field perpendicular and parallel to the film plane are shown by black and red symbols, respectively. Black lines present the best fit to Eq.~(\ref{eq:MR}).}
    \label{s_fig:RvsB}
\end{figure}

Upon cooling, all grown epilayers exhibit metallic behaviour (i.e. $dR_{xx}/dT>0$). The dependencies $R(T)$ follow power law for most samples, except for the PbSe epilayer (Fig.~\ref{s_fig:RvsT}). Carrier density and low-temperature mobility are extracted from the slope of the Hall resistivity measured up to 2~T and zero-field resistivity. According to such measurements, the PbSe epilayer has relatively low hole density, $3\times10^{17}$~cm$^{-3}$, and high mobility, $3.8\times10^3$~cm$^2$/Vs. Addition of Sn rapidly increases the Hall carrier density, thus all studied Pb$_{1-x}$Sn$_x$Se films are $p$-type and their carrier densities and mobilities are in the range of (0.5 -- $2.3)\times10^{19}$~cm$^{-3}$ and (0.5 -- $2)\times10^3$~cm$^2$/Vs, respectively. High carrier densities and relatively low mobilities are associated with the presence of highly disordered $p$-type region at the interface with BaF$_2$ \citeS{SM_Tranta1988,SM_Kolwas:2013_pss}. Table \ref{tab:info} contains detailed information on the studied epilayers.

According to data collected in Fig.~\ref{s_fig:RvsB}, magnetoresistance (MR) in all studied epilayers is similar. In particular in the case of the perpendicular field, there is a pronounced dip of resistance around $B = 0$, which is associated with the Berry phase positive WAL MR. In the higher field range, MR is parabolic, and then smoothly changes to a linear dependence in the strongest fields. Such high-field linear MR is usually explained by semiclassical models \citeS{SM_Parish2005,SM_Ping2014,SM_Ramakrishnan2017}, assigning it to spatial variations of carrier density. As shown in  Fig.~\ref{s_fig:RvsB}, such a behavior is well described by the semi-empirical model \citeS{SM_Cho2008,SM_Ping2014,SM_Rosen:2019_PRB},
\begin{equation}
    \rho_{xx}(B)=\frac{\rho_{xx}(0)}{1-2A+\frac{2A}{\sqrt{1+(\mu B)^2}}},
    \label{eq:MR}
\end{equation}
where $\mu$ is carrier mobility treated together with $A$ as fitting parameters.

In high in-plane magnetic fields ($B\parallel I$), most of the studied epilayers exhibit negative  MR, as shown in Fig.~\ref{s_fig:RvsB} by red points. However, in a few cases  resistance increases with the in-plane field (Fig.~\ref{s_fig:RvsB}). We have not found any correlation between the MR sign and composition of the films. Such behaviour may be associated with non-uniformities; model calculations \citeS{SM_Hu2007} showed that inhomogeneity along the growth direction can induce negative MR in the case of the in-plane field.

\section{Fitting experimental data with simplified and full HLN expression}
\label{s_sec:HLN}

\begin{figure}[tb]
    \includegraphics[width=\columnwidth]{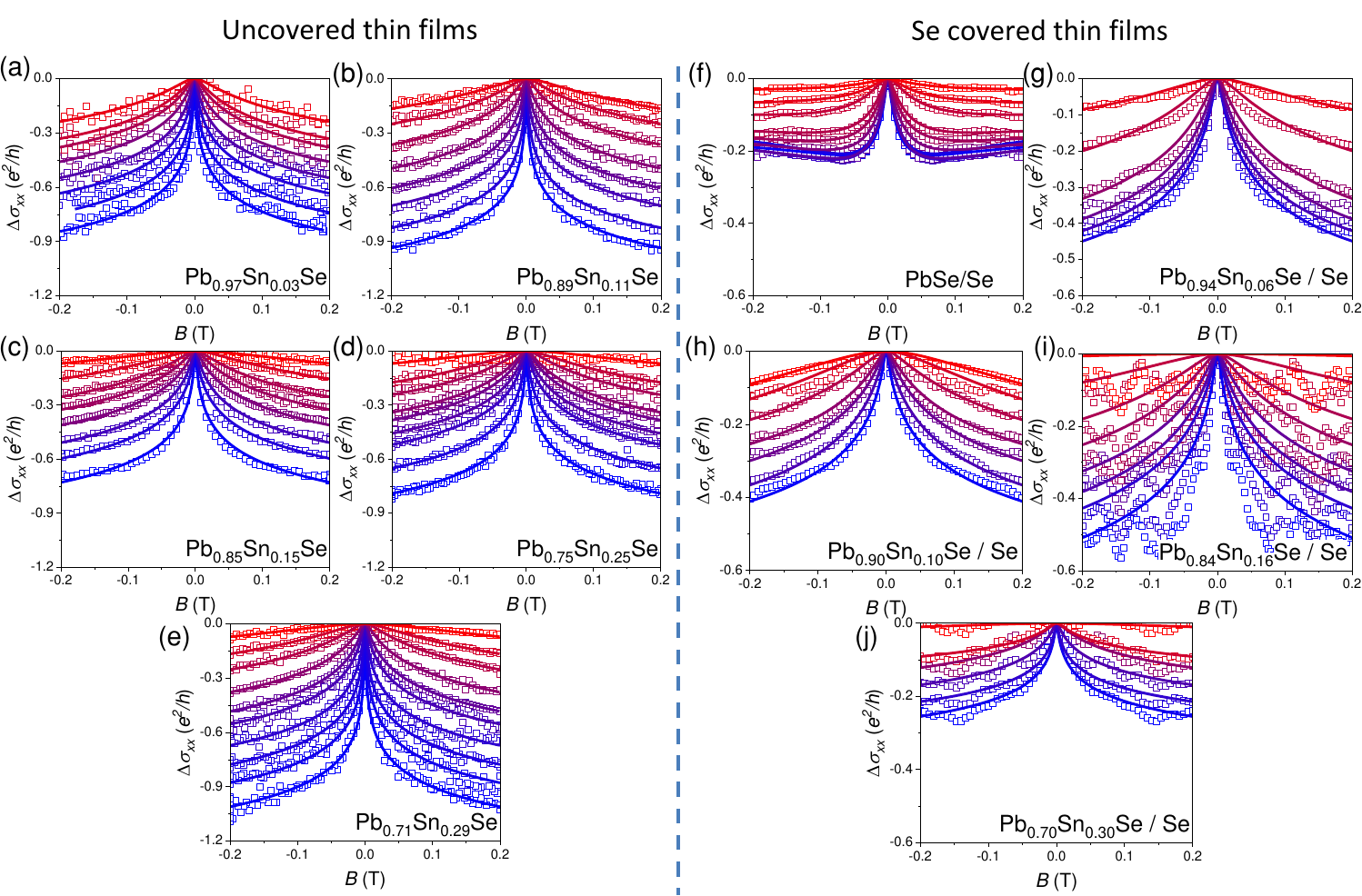}
    \caption{Temperature evolution of WAL magnetoconductivity   $\sigma_{xx}(B)$ in (a-e) uncovered and (f-j) in Se-covered epilayers. Lines present the best fit of the experimental data points to the HLN equation in the strong spin-orbit approximation, Eq.\,(\ref{s_eq:HLN} for $n_v = 1$ (except PbSe/Se epilayer for which full HLN was employed), treating the phase coherence length $l_{\phi}(T)$ as the only fitting parameter and assuming  $\alpha = -1/2$. The determined values of $l_{\phi}(T)$ are shown in Fig.~5(e) of the main text. Note that the range of  $\sigma_{xx}$ in the (a-e) panels  is twice larger than in the (d-j) panels. Temperature varies from 1.6\,K (blue points) to 25-30\,K (red points).}
    \label{s_fig:WAL_sHLN}
\end{figure}

\begin{figure}[tb]
    \includegraphics[width=\columnwidth]{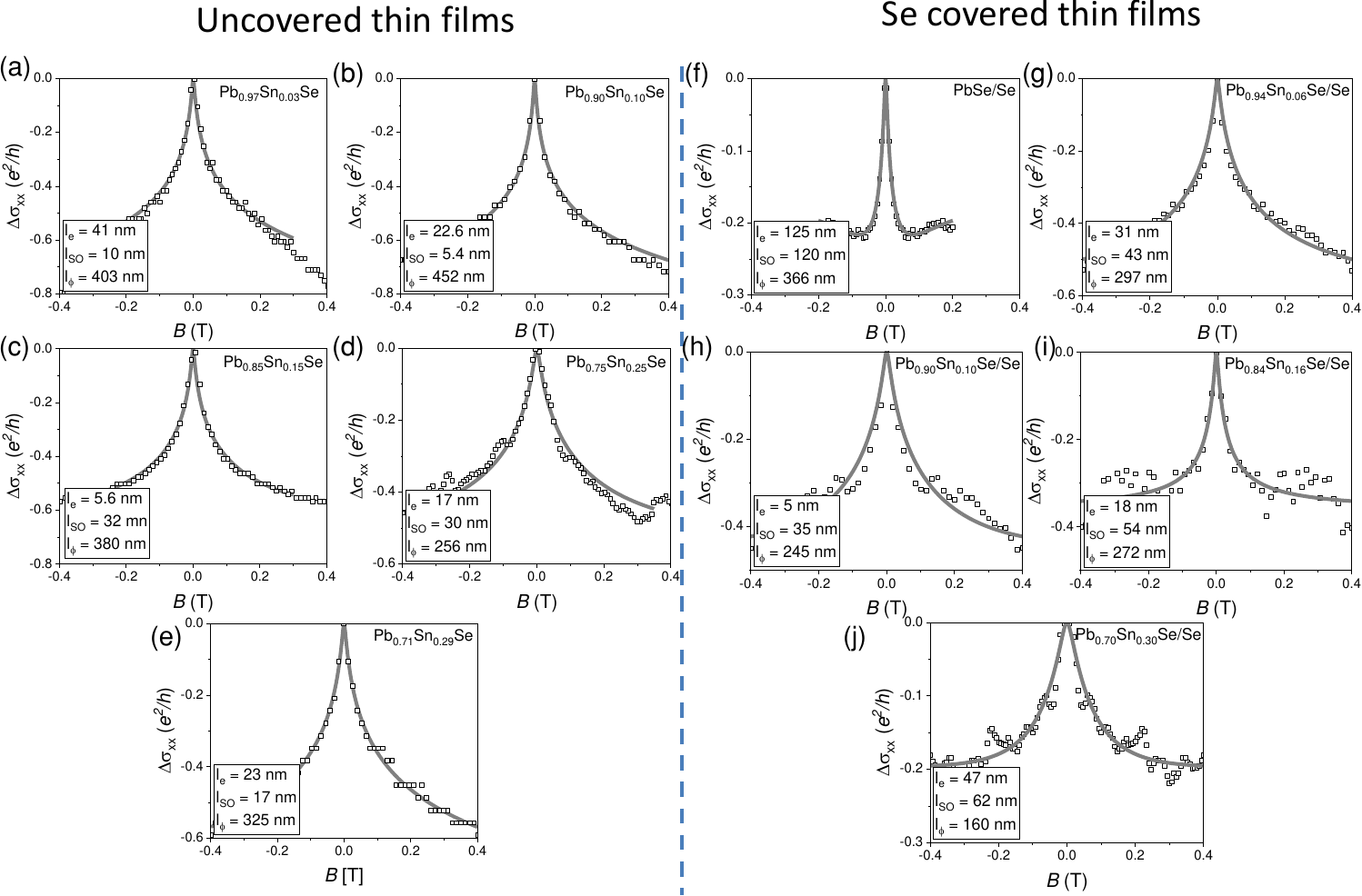}
    \caption{Low-field magnetoconductivity at 4.2~K fitted with the full HLN expression [Eq.~(\ref{s_eq:fHLN})].}
    \label{s_fig:WAL_fHLN}
\end{figure}

According to the theory developed here, the low-field MR is assigned to the Berry phase quantization brought about by the mirror and time reversal symmetries rather than by a non-trivial character of the topological phase. Within this scenario and by noting that the phase coherence length $l_{\phi}$ is larger than the film thickness $d = 50$\,nm, the MR has been described by the HLN  theory in the limit $l_{\phi} \gg l{\text{so}}$, where $l{\text{so}}$ is the mean free path for spin-orbit scattering \citeS{SM_Hikami1980},
\begin{equation}
    \Delta\sigma_{xx}(B)=\alpha\sum_{i}^{n_v}\frac{e^2}{2\pi^2\hbar}f\left(\frac{\hbar c}{4el_{\phi,i}^2B}\right),
    \label{s_eq:HLN}
\end{equation}
where the prefactor $\alpha = -1/2$; the magnetic field $B$ is perpendicular to the film plane; the summation is over independent conduction channels, $f(x)\equiv \psi(1/2+x)-\ln(x)$, $\psi(x)$ is the digamma function, $l_{\phi,i}$ is the phase coherence length in the $i$'th conduction channel. In our case, as shown in Fig.~3d in the main text and in Fig.~\ref{s_fig:WAL_sHLN}, MR for all epilayers can be fitted by the one-channel formula, $n_v = 1$ treating $l_{\phi}(T)$, as the only fitting parameter [whose values are shown in Fig.~5(e) in the main text]. This means that $l_{\phi,i}$ are longer than length scales characterizing scattering between subbands and valleys (including surface ones in the topological case) as well as between $n$ and $p$-type layers. Alternatively, and more probably, because of short length scales characterizing the $p$-type region, the corresponding WAL or WL MR is shifted to a high field region, so the low field features are solely due to electrons residing closer to the outer surface. It is important recalling that if only one layer shows MR, Eq.~(\ref{s_eq:HLN}) with $n_v = 1$ remains valid even in the multichannel case.

The simplified HLN formula [Eq.~(\ref{s_eq:HLN})] is assumed to be valid in the case of the topological surface state for which the Berry phase is quantized. Resulted presented in Fig.~5 of the main text show that this is also in the case of bulk-dominated transport (Fermi level in the conduction band) and even in the topologically trivial materials. To check the validity of the simplified HLN expression, we have fitted experimental data employing the full HLN expression,
\begin{equation}
    \Delta\sigma_{xx}(B)=\frac{e^2}{2\pi^2\hbar}\left[f\left(\frac{B_1}{B}\right)+f\left(\frac{B_2}{B}\right)+f\left(\frac{B_3}{B}\right)\right],
    \label{s_eq:fHLN}
\end{equation}
where
\begin{equation}
    \begin{split}
    & B_1=B_e+B_{SO}+B_s, \\
    & B_2=\frac43B_{\text{so}}+\frac23B_{s}+B_{\phi}, \\
    & B_3=2B_s+B_{\phi}.
    \end{split}
    \label{s_eq:fHLN_Bs}
\end{equation}
The fields $B_e$, $B_{\text{so}}$, $B_s$ and $B_{\phi}$ are related to the mean free path $l$, the spin diffusion length limited by spin disorder scattering $l_s$ and spin-orbit coupling $l_{\text{so}}$, and the phase coherence length $l_{\phi}$ according to $B_i=\hbar/4el^2_i$. Results of the best fit of experimental data at 4.2~K are presented in Fig.~\ref{s_fig:WAL_fHLN}. Since there is no magnetic doping, we assume that $B_s=0$; values of the mean free path extracted from zero-field mobility are used for $B_e$, thus there are only two adjustable parameters, $l_{\phi}$ and $l_{\text{so}}$. Fitting experimental data at 4.2~K results in $l_{\text{so}}$ values below 60~nm, i.e., comparable to $l$ and several times shorter than $l_{\phi}$. Moreover, fitting results do not change appreciably with varying $l_{\text{so}}$ towards even lower values, so that using the simplified HLN expression in a strong SO coupling limit is justified. The only exception here is PbSe/Se epilayer, for which it was impossible to fit with the simplified HLN expression, as a large magnitude of the mean free path $l= 125$\,nm makes that the condition $B <B_e$ is violated for this sample \citeS{SM_Golub:2005_PRB,SM_Nestoklon:2011_SSC}. It worth noting that fitting of WAL MR data at different temperatures with the full HLN expression (\ref{s_eq:fHLN}) results in $l_{\text{so}}$ values that are temperature independent within the accuracy of the fitting.
\begin{figure}[tb]
    \includegraphics[width=0.9\columnwidth]{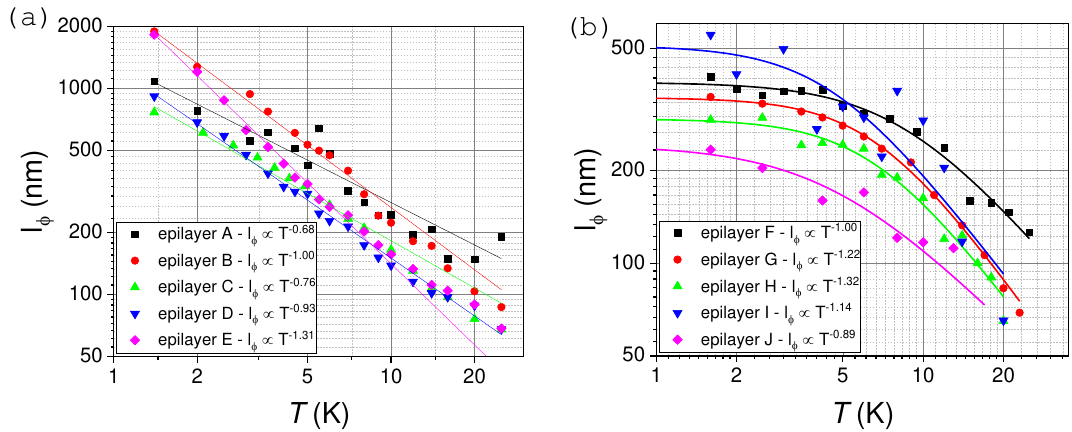}
    \caption{{\bf Temperature dependence of the phase coherence length obtained from fit to the full HLN formula}. (a) Se-uncovered epilayers. (b) Se-covered epilayers. No saturation is observed for uncovered epilayers down to 1.5~K.}
    \label{s_fig:L_phi_vs_T}
\end{figure}

\section{Comparison of coherence length values to published data}

\begin{table*}
\caption{Published values for $l_{\phi}$ in topological insulators}
\begin{tabular}{|c|c|c|c|c|c|}
\hline
Material & Thickness & Temperature & $l_{\phi}$ & Temperature dependence \\
\hline
\multicolumn{5}{|c|}{TI materials} \\
\hline
Bi$_2$Te$_3$ \citeS{SM_Kim2011} & 5-100~QL & 1.5~K & 100-1000~nm & --- \\
Bi$_2$Te$_3$ \citeS{SM_Kim2011} & 5-100~QL & 1.5~K & 100-1000~nm & --- \\
Bi$_2$Se$_2$Te \citeS{SM_Bao2012} & 50~nm & 1.9~K & 320~nm & $\approx T^{-0.44}$ \\
Bi$_2$Te$_3$ \citeS{SM_Bansal2012} & 5-128~QL & 1.5~K & 800~nm & --- \\
Bi$_2$SeTe$_2$ \citeS{SM_Assaf2013} & 15~nm & 7~K & 60~nm & $\approx T^{-0.75}$ \\
Bi$_{1.5}$Sb$_{0.5}$Te$_2$ \citeS{SM_Lee2012a} & 90~nm & 4.2~K & 160~nm & $\approx T^{-0.5}$ \\
Bi$_2$Te$_3$ \citeS{SM_Chiu2013} & 65~nm & 1~K & 300~nm & $\approx T^{-0.24}$ \\
Bi$_2$Te$_3$ \citeS{SM_Taskin2012} & 40~QL & 2~K & 1000~nm & $\approx T^{-0.75}$ \\
Bi$_2$SeTe$_2$ \citeS{SM_Shekhar2014} & bulk ($\sim$0.4~mm) & 2~K & 140~nm & $\approx T^{-0.5}$ \\
Bi$_{1.46}$Sb$_{0.54}$Te$_{1.7}$Se$_{1.3}$ \citeS{SM_Pan2014} & 130~nm & 2~K & 116~nm & $\approx T^{-0.46}$ \\
Bi$_2$Se$_3$ \citeS{SM_Wang2016b} & 6-54~nm & 2~K & 40-120~nm & $\approx T^{-0.5}$ \\
BiSbSe$_2$Te \citeS{SM_Banerjee2014} & 50-90~nm & 2~K & 86~nm & $\approx T^{-0.58}$ \\
Bi$_2$Se$_2$Te \citeS{SM_Gopal2015} & 100-120~nm & 1.8~K & 57~nm & $\approx T^{-0.44}$ \\
Bi$_2$Te$_3$/Te \citeS{SM_Ngabonziza2016} & 15~nm & 0.1~K & 1500~nm & --- \\
Bi$_2$Se$_3$ \citeS{SM_Zhang2016} & 30~nm & 10~K & 318~nm & $\approx T^{-0.51}$ \\
Bi$_2$SeTe$_2$ \citeS{SM_Assaf2012} & 15~nm & 2~K & 120~nm & $\approx T^{-0.44}$ \\
Sb$_2$Te$_3$ \citeS{SM_Rosen:2019_PRB} & 8~QL & 1.8~K & 150~nm & $\approx T^{-0.37}$ \\
\hline
\multicolumn{5}{|c|}{TCI materials} \\
\hline
SnTe \citeS{SM_Assaf2014} & 30-60~nm & 2~K & 200-400~nm & --- \\
SnTe \citeS{SM_Akiyama2015} & 46~nm & 4~K & 200~nm & $\approx T^{-0.51}$ \\
SnTe \citeS{SM_Akiyama2015} & 74~nm & 4~K & 600~nm & $\approx T^{-1.4}$ \\
Pb$_{1-x}$Sn$_x$Se \citeS{SM_Zhang2015a} & 10-16~nm & 2~K & 250-350~nm & from $T^{-0.65}$ to $T^{-1.65}$ \\
(Pb$_{0.65}$Sn$_{0.35}$)$_{0.98}$In$_{0.02}$ \citeS{SM_Zhong2015} & bulk (0.2-0.6~mm) & 5~K & 123~nm & --- \\
Pb$_{0.6}$Sn$_{0.4}$Te \citeS{SM_Zhang2019} & 20~nm & 2~K & 147~nm & --- \\
SnTe \citeS{SM_Zou2019} & 10~u.c. & 1.8~K & 200~nm & --- \\
SnTe \citeS{SM_Dybko:2018_arXiv} & 10-100~nm & 1.7~K & 350-1020~nm & from $T^{-0.87}$ to $T^{-1.03}$ \\
Pb$_{0.7}$Sn$_{0.3}$Se \citeS{SM_Wang:2020_PRB}&55~nm&3~K&900-1700~nm&from $T^{-0.74}$ to $T^{-0.83}$\\
\hline
\end{tabular}
\label{s_tab:lphi}
\end{table*}

Reported values for $l_{\phi}$ for topological materials differ significantly, and depend, most likely, on particular growth and processing conditions. However, one can notice (see Table \ref{s_tab:lphi}) that despite variation in particular values of $l_{\phi}$, data for topological insulators (TIs) show a similar temperature dependence $\approx T^{-p/2}$ (above low temperature saturation) with $p\approx1$, which points to the dominance of e-e interactions in 2D \cite{SM_Altshuler1982}. Published data for TCI materials are more scarce, but there are several works \citeS{SM_Akiyama2015,SM_Zhang2015a,SM_Dybko:2018_arXiv}, which report $l_{\phi}$ temperature dependence with $p$ are in the range from 1 to 3. This suggests that the role of the e-e dephasing mechanism is reduced. Though at low temperatures the dominant mechanism of dephasing is e-e scattering \citeS{SM_Datta1995}, this source of dephasing is suppressed due to a large magnitude of the dielectric constant in the studied materials \citeS{SM_Prinz1999}. Thus, in our case electron-phonon (e-ph) interactions dominate in the dephasing process. Indeed, e-ph interactions results in $p$ = 2, 3, or 4; with particular value depending on the details of the studied system \citeS{SM_Prinz1999,SM_Lin:2002_JPC}). Our results agree with the published data for TCI materials. Also, values of $l_{\phi}$ rarely exceed 1~$\mu$m at 1-2~K. Thus values of $l_{\phi}$ in the current study, obtained in samples without Se cap at 1.5~K (1-2~$\mu$m, see Fig.~\ref{s_fig:L_phi_vs_T} and Table~\ref{tab:info}) are one of the largest among topological materials \citeS{SM_Akiyama2015,SM_Dybko:2018_arXiv}. With the support of the published data, we can argue that the dephasing mechanism in IV-VI TCI compounds differs from the one in V-VI TI materials and, most likely, the role of e-e interactions is diminished.

\section{Fitting MR in parallel fields}

\begin{figure}[tb]
    \includegraphics[width=0.75\columnwidth]{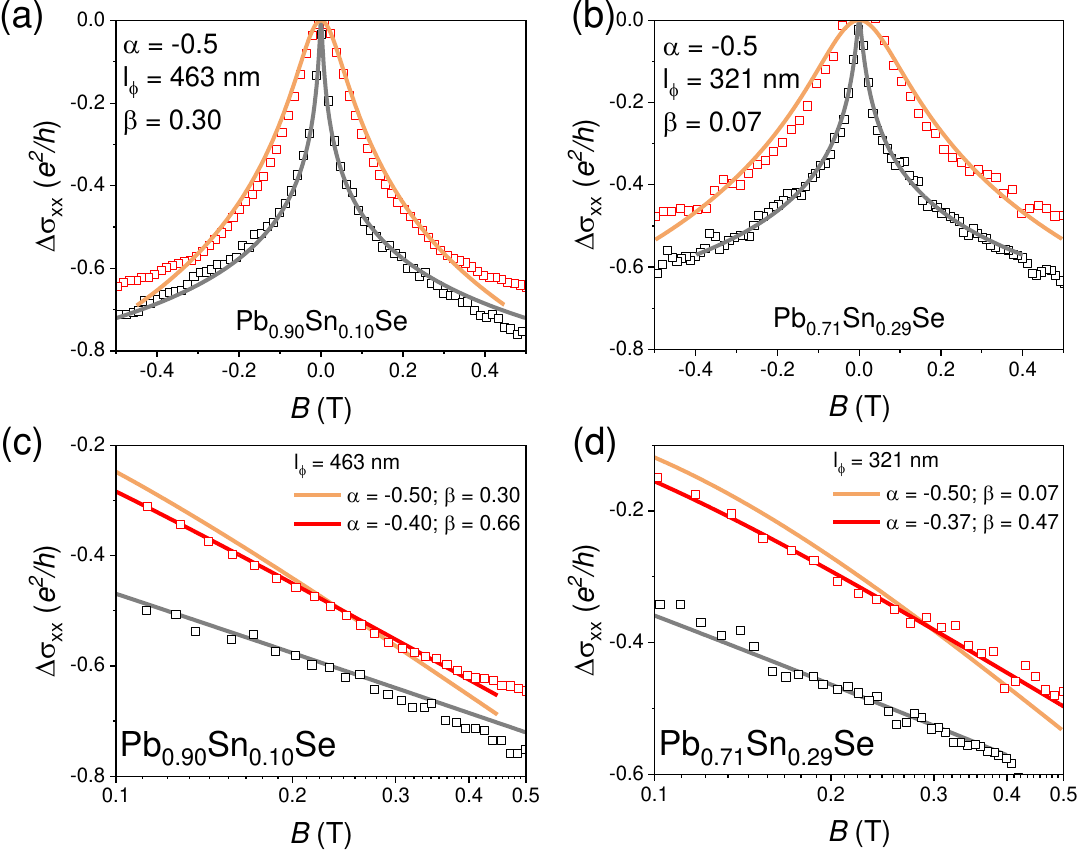}
    \caption{{\bf Comparison of low-field magnetoconductance in perpendicular and parallel configurations}. The data (black symbols -- perpendicular field; red symbols -- parallel field) have been taken at 4.2\,K for samples with Sn content corresponding to the topologically trivial (a,c) and non-trivial (b,d) phases. Grey lines are best fits to black symbols employing the HLN formula [Eq.~(\ref{s_eq:HLN})]; orange and red lines: fitted WAL theory in the parallel field [Eq.\,((\ref{s_eq:parallel})] with $\beta$ as a single adjustable parameter (assuming $\alpha = -1/2$) and with $\alpha$ and $\beta$ as adjustable parameters,  respectively, employing $l_{\phi}$ values determined in the perpendicular configuration (grey curves). Two-parameter fitting results in lower values of $|\alpha|<0.5$ and much higher values of $\beta>1/3$. A comparison of the data for samples uncovered and covered by amorphous Se is shown in Fig.~6 of the main text.}
    \label{s_fig:WAL_par}
\end{figure}

As discussed in the main text and shown there in Fig.~6, we have also studied low-field MR for the magnetic field applied in-plane, along the current direction. In the parallel field, the WAL effect does not disappear and exhibits a similar magnitude as in the perpendicular field, as presented in Fig.~\ref{s_fig:WAL_par}. In the past, conventional WL MR in a parallel field was theoretically considered for diffusive films, $l\ll d$, where $d$ is the film thickness \cite{Altshuler1981}, later for the clean films, i.e, quantum wells, $l\gg d$ \cite{Dugaev1984}, and finally for the intermediate regime \cite{Beenakker1988} ($l \simeq d$). All these theories predict a similar logarithmic shape of $\Delta\sigma_{xx}(B_{\parallel})$,
\begin{equation}
    \Delta\sigma_{xx}(B_{\parallel})=\alpha\frac{e^2}{2\pi^2\hbar}\ln\left(1+\beta\frac{e^2d^2l_{\phi}^2}{\hbar^2}B_{\parallel}^2\right),
    \label{s_eq:parallel}
\end{equation}
where the prefactor $\alpha= 1$ for the WL and one channel case; the dimensionless parameter $\beta = 1/3$ in the diffusive limit, $d \ll l$ \cite{Altshuler1981} $\beta=(1/16)d/l$ in the clean limit \cite{Dugaev1984}, and in the intermediate regime $d/16l<\beta<1/3$ \cite{Beenakker1988}. These theories neglect the presence of spin-orbit scattering and Zeeman splitting. Such additional breaking of spin-rotation and time-reversal symmetries turn $\alpha$ to $-1/2$ and generates an additional non-universal term proportional to the square of the Zeeman energy $(g^*\mu_{\text{B}}B)^2$ in the argument of the logarithm function in Eq.~(\ref{s_eq:parallel}) \cite{Maekawa:1981_JPSJ,Malshukov:1997_PRB,Glazov:2009_SST}, which may dominate in the  small $d$ or large $g^*$ limit ($g^*= 35\pm5$ in PbSe \cite{Bauer1992}). Recently WAL in a parallel field was considered for topological insulators for which $\alpha = - 1/2$ per channel and $\beta=2\lambda^2/d^2$, where $\lambda$ is the penetration of the surface state into the bulk \cite{Tkachov2011,Tkachov2013}. Furthermore, interfacial roughness enhances the apparent $\beta$ value \cite{Mathur:2001_PRB,Minkov:2004_PRB}.

We have fitted experimental data to the conventional theory [Eq.~(\ref{s_eq:parallel})] with $\alpha = -0.5$ and with $l_{\phi}$ values obtained from the fitting of the data taken in the perpendicular orientation, thus having $\beta$ as the only adjustable parameter. Such a procedure leads to $\beta$: 0.48; 0.30; 0.20; 0.22 and 0.07 for epilayers A ($x_{\text{Sn}}=3.5\%$), B ($x_{\text{Sn}}=10.5\%$), C ($x_{\text{Sn}}=15.5\%$), D ($x_{\text{Sn}}=24.7\%$), and E ($x_{\text{Sn}}=28.7\%$), respectively. Since the actual thickness of the layer giving rise to WAL is presumable thinner than the nominal layer thickness $d$, the determined values have to be regarded as their lower limit. Meanwhile in the DK theory \cite{Dugaev1984} $\beta= d/16l $ is equal to 0.077; 0.138; 0.691; 0.223 and 0.226 for epilayers A, B, C, D, and E. Within the AA theory \cite{Altshuler1981}, the parameter $\beta$ has a fixed value 1/3. The fact that experimental values of $\beta$ are larger than 1/3, was reported for TI materials \cite{Lin2013}. Theory of WAL in TIs in parallel fields \cite{Tkachov2011,Tkachov2013} can explain high magnitudes of $\beta$ in the epilayers with composition corresponding to the inverted band structure by assuming that $\lambda$ extends over almost the whole film thickness \cite{Dybko:2018_arXiv}. However, this approach fails to explain the similar behaviour of MR in epilayers in the topologically trivial phase. We have also fitted MR data in the parallel field to the formula (\ref{s_eq:parallel}) using two adjustable parameters -- $\alpha$ and $\beta$. Again, we have used the value of $l_{\phi}$ determined from fit to the MR data in the perpendicular geometry measured at the same temperature. Using this approach we have obtained $\alpha=-0.40\pm0.03$, and higher values for $\beta$: 0.99; 0.66; 0.76; 0.35 and 0.17 for epilayers A, B, C, D, and E. Moreover, fitting experimental data with two adjustable parameters results, not surprisingly, in a better fit than with a single one (Fig.~\ref{s_fig:WAL_par}).

Importantly, epilayers covered by amorphous Se that breaks the mirror symmetry, behave in a similar way, but show systematically larger $\beta$ values: $\beta=0.52$, 0.32, 0.79, 1.16 are obtained for epilayers G ($x_{\text{Sn}}=6.5\%$), H ($x_{\text{Sn}}=13.6\%$), I ($x_{\text{Sn}}=15.9\%$), and J ($x_{\text{Sn}}=30\%$), respectively, assuming $\alpha=-0.5$. Treating $\alpha$ as an additional fitting parameter even larger values $\beta=1.38$, 0.97, 2.11, 8.82 are found out for the same epilayers together with $\alpha = -0.35\pm0.04$. As emphasized in the main text, these findings, that is a stronger destruction of the Berry phase quantization by breaking time-reversal symmetry in samples in which mirror symmetry is violated by the amorphous overlayer,  provides a strong support to the theory developed here.

\end{widetext}

\bibliographystyleS{naturemag}


\begin{thebibliography}{100}
\expandafter\ifx\csname url\endcsname\relax
  \def\url#1{\texttt{#1}}\fi
\expandafter\ifx\csname urlprefix\endcsname\relax\def\urlprefix{URL }\fi
\providecommand{\bibinfo}[2]{#2}
\providecommand{\eprint}[2][]{\url{#2}}

\bibitem{Beenakker1997}
\bibinfo{author}{Beenakker, C. W.~J.}
\newblock \bibinfo{title}{Random-matrix theory of quantum transport}.
\newblock \emph{\bibinfo{journal}{Rev. Mod. Phys.}}
  \textbf{\bibinfo{volume}{69}}, \bibinfo{pages}{731--808}
  (\bibinfo{year}{1997}).

\bibitem{Altland:1997_PRB}
\bibinfo{author}{Altland, A.} \& \bibinfo{author}{Zirnbauer, M.~R.}
\newblock \bibinfo{title}{Nonstandard symmetry classes in mesoscopic
  normal-superconducting hybrid structures}.
\newblock \emph{\bibinfo{journal}{Phys. Rev. B}} \textbf{\bibinfo{volume}{55}},
  \bibinfo{pages}{1142--1161} (\bibinfo{year}{1997}).

\bibitem{Ryu:2010_NJP}
\bibinfo{author}{Ryu, S.}, \bibinfo{author}{Schnyder, A.~P.},
  \bibinfo{author}{Furusaki, A.} \& \bibinfo{author}{Ludwig, A. W.~W.}
\newblock \bibinfo{title}{Topological insulators and superconductors: tenfold
  way and dimensional hierarchy}.
\newblock \emph{\bibinfo{journal}{New J. Phys.}} \textbf{\bibinfo{volume}{12}},
  \bibinfo{pages}{065010} (\bibinfo{year}{2010}).

\bibitem{Evers:2008_RMP}
\bibinfo{author}{Evers, F.} \& \bibinfo{author}{Mirlin, A.~D.}
\newblock \bibinfo{title}{Anderson transitions}.
\newblock \emph{\bibinfo{journal}{Rev. Mod. Phys.}}
  \textbf{\bibinfo{volume}{80}}, \bibinfo{pages}{1355--1417}
  (\bibinfo{year}{2008}).

\bibitem{Hikami1980}
\bibinfo{author}{Hikami, S.}, \bibinfo{author}{Larkin, A.~I.} \&
  \bibinfo{author}{Nagaoka, Y.}
\newblock \bibinfo{title}{Spin-orbit interaction and magnetoresistance in the
  two dimensional random system}.
\newblock \emph{\bibinfo{journal}{Prog. Theor. Phys.}}
  \textbf{\bibinfo{volume}{63}}, \bibinfo{pages}{707--710}
  (\bibinfo{year}{1980}).

\bibitem{Wojtowicz:1986_PRL}
\bibinfo{author}{Wojtowicz, T.}, \bibinfo{author}{Dietl, T.},
  \bibinfo{author}{Sawicki, M.}, \bibinfo{author}{Plesiewicz, W.} \&
  \bibinfo{author}{Jaroszy\'{n}ski, J.}
\newblock \bibinfo{title}{Metal-insulator transition in semimagnetic
  semiconductors}.
\newblock \emph{\bibinfo{journal}{Phys. Rev. Lett.}}
  \textbf{\bibinfo{volume}{56}}, \bibinfo{pages}{2419--2422}
  (\bibinfo{year}{1986}).

\bibitem{Finkelstein:1990_SSR}
\bibinfo{author}{Finkelstein, A.~M.}
\newblock \bibinfo{title}{Electron liquid in disordered conductors}.
\newblock \emph{\bibinfo{journal}{Soviet Sci. Rev.}}
  \textbf{\bibinfo{volume}{14}}, \bibinfo{pages}{1--101}
  (\bibinfo{year}{1990}).

\bibitem{Lu:2011_PRB}
\bibinfo{author}{Lu, H.-Z.} \& \bibinfo{author}{Shen, S.-Q.}
\newblock \bibinfo{title}{Weak localization of bulk channels in topological
  insulator thin films}.
\newblock \emph{\bibinfo{journal}{Phys. Rev. B}} \textbf{\bibinfo{volume}{84}},
  \bibinfo{pages}{125138} (\bibinfo{year}{2011}).

\bibitem{Garate:2012_PRB}
\bibinfo{author}{Garate, I.} \& \bibinfo{author}{Glazman, L.}
\newblock \bibinfo{title}{Weak localization and antilocalization in topological
  insulator thin films with coherent bulk-surface coupling}.
\newblock \emph{\bibinfo{journal}{Phys. Rev. B}} \textbf{\bibinfo{volume}{86}},
  \bibinfo{pages}{035422} (\bibinfo{year}{2012}).

\bibitem{Adroguer:2015_PRB}
\bibinfo{author}{Adroguer, P.}, \bibinfo{author}{Liu, W.~E.},
  \bibinfo{author}{Culcer, D.} \& \bibinfo{author}{Hankiewicz, E.~M.}
\newblock \bibinfo{title}{Conductivity corrections for topological insulators
  with spin-orbit impurities: {Hikami-Larkin-Nagaoka} formula revisited}.
\newblock \emph{\bibinfo{journal}{Phys. Rev. B}} \textbf{\bibinfo{volume}{92}},
  \bibinfo{pages}{241402(R)} (\bibinfo{year}{2015}).

\bibitem{Wang:2020_PRL}
\bibinfo{author}{Wang, H.-W.}, \bibinfo{author}{Fu, B.} \&
  \bibinfo{author}{Shen, S.-Q.}
\newblock \bibinfo{title}{Anomalous temperature dependence of quantum
  correction to the conductivity of magnetic topological insulators}.
\newblock \emph{\bibinfo{journal}{Phys. Rev. Lett.}}
  \textbf{\bibinfo{volume}{124}}, \bibinfo{pages}{206603}
  (\bibinfo{year}{2020}).

\bibitem{Golub:2016_PRB}
\bibinfo{author}{Golub, L.~E.}, \bibinfo{author}{Gornyi, I.~V.} \&
  \bibinfo{author}{Kachorovskii, V.~Y.}
\newblock \bibinfo{title}{Weak antilocalization in two-dimensional systems with
  large {Rashba} splitting}.
\newblock \emph{\bibinfo{journal}{Phys. Rev. B}} \textbf{\bibinfo{volume}{93}},
  \bibinfo{pages}{245306} (\bibinfo{year}{2016}).

\bibitem{Chatterjee:2019_PRB}
\bibinfo{author}{Chatterjee, S.} \emph{et~al.}
\newblock \bibinfo{title}{Weak antilocalization in quasi-two-dimensional
  electronic states of epitaxial {LuSb} thin films}.
\newblock \emph{\bibinfo{journal}{Phys. Rev. B}} \textbf{\bibinfo{volume}{99}},
  \bibinfo{pages}{125134} (\bibinfo{year}{2019}).

\bibitem{Wu:2007_PRL}
\bibinfo{author}{Wu, X.}, \bibinfo{author}{Li, X.}, \bibinfo{author}{Song, Z.},
  \bibinfo{author}{Berger, C.} \& \bibinfo{author}{de~Heer, W.~A.}
\newblock \bibinfo{title}{Weak antilocalization in epitaxial graphene: Evidence
  for chiral electrons}.
\newblock \emph{\bibinfo{journal}{Phys. Rev. Lett.}}
  \textbf{\bibinfo{volume}{98}}, \bibinfo{pages}{136801}
  (\bibinfo{year}{2007}).

\bibitem{McCann:2006_PRL}
\bibinfo{author}{McCann, E.} \emph{et~al.}
\newblock \bibinfo{title}{Weak-localization magnetoresistance and valley
  symmetry in graphene}.
\newblock \emph{\bibinfo{journal}{Phys. Rev. Lett.}}
  \textbf{\bibinfo{volume}{97}}, \bibinfo{pages}{146805}
  (\bibinfo{year}{2006}).

\bibitem{Fu:2011_PRL}
\bibinfo{author}{Fu, L.}
\newblock \bibinfo{title}{Topological crystalline insulators}.
\newblock \emph{\bibinfo{journal}{Phys. Rev. Lett.}}
  \textbf{\bibinfo{volume}{106}}, \bibinfo{pages}{106802}
  (\bibinfo{year}{2011}).

\bibitem{Chiu:2016_RMP}
\bibinfo{author}{Chiu, C.-K.}, \bibinfo{author}{Teo, J. C.~Y.},
  \bibinfo{author}{Schnyder, A.~P.} \& \bibinfo{author}{Ryu, S.}
\newblock \bibinfo{title}{Classification of topological quantum matter with
  symmetries}.
\newblock \emph{\bibinfo{journal}{Rev. Mod. Phys.}}
  \textbf{\bibinfo{volume}{88}}, \bibinfo{pages}{035005}
  (\bibinfo{year}{2016}).

\bibitem{Hsieh2012}
\bibinfo{author}{Hsieh, T.~H.} \emph{et~al.}
\newblock \bibinfo{title}{Topological crystalline insulators in the {SnTe}
  material class}.
\newblock \emph{\bibinfo{journal}{Nat. Commun.}} \textbf{\bibinfo{volume}{3}},
  \bibinfo{pages}{982} (\bibinfo{year}{2012}).

\bibitem{Safaei:2013_PRB}
\bibinfo{author}{Safaei, S.}, \bibinfo{author}{Kacman, P.} \&
  \bibinfo{author}{Buczko, R.}
\newblock \bibinfo{title}{Topological crystalline insulator {(Pb,Sn)Te}:
  Surface states and their spin polarization}.
\newblock \emph{\bibinfo{journal}{Phys. Rev. B}} \textbf{\bibinfo{volume}{88}},
  \bibinfo{pages}{045305} (\bibinfo{year}{2013}).

\bibitem{Dziawa2012}
\bibinfo{author}{Dziawa, P.} \emph{et~al.}
\newblock \bibinfo{title}{Topological crystalline insulator states in
  {Pb$_{1-x}$Sn$_{x}$Se}}.
\newblock \emph{\bibinfo{journal}{Nat. Mater.}} \textbf{\bibinfo{volume}{11}},
  \bibinfo{pages}{1023--1027} (\bibinfo{year}{2012}).

\bibitem{Tanaka:2012_NP}
\bibinfo{author}{Tanaka, Y.} \emph{et~al.}
\newblock \bibinfo{title}{Experimental realization of a topological crystalline
  insulator in {SnTe}}.
\newblock \emph{\bibinfo{journal}{Nat. Phys.}} \textbf{\bibinfo{volume}{8}},
  \bibinfo{pages}{800--803} (\bibinfo{year}{2012}).

\bibitem{Xu:2012_NC}
\bibinfo{author}{Xu, S.-Y.} \emph{et~al.}
\newblock \bibinfo{title}{Observation of a topological crystalline insulator
  phase and topological phase transition in {Pb$_{1-x}$Sn$_x$Te}}.
\newblock \emph{\bibinfo{journal}{Nat. Commun.}} \textbf{\bibinfo{volume}{3}},
  \bibinfo{pages}{1192} (\bibinfo{year}{2012}).

\bibitem{Tanaka2013}
\bibinfo{author}{Tanaka, Y.} \emph{et~al.}
\newblock \bibinfo{title}{Two types of {D}irac-cone surface states on the (111)
  surface of the topological crystalline insulator {SnTe}}.
\newblock \emph{\bibinfo{journal}{Phys. Rev. B}} \textbf{\bibinfo{volume}{88}},
  \bibinfo{pages}{235126} (\bibinfo{year}{2013}).

\bibitem{Polley2014}
\bibinfo{author}{Polley, C.~M.} \emph{et~al.}
\newblock \bibinfo{title}{Observation of topological crystalline insulator
  surface states on (111)-oriented {Pb$_{1-x}$Sn$_{x}$Se} films}.
\newblock \emph{\bibinfo{journal}{Phys. Rev. B}} \textbf{\bibinfo{volume}{89}},
  \bibinfo{pages}{075317} (\bibinfo{year}{2014}).

\bibitem{Brzezicki2019}
\bibinfo{author}{Brzezicki, W.}, \bibinfo{author}{Wysoki{\'{n}}ski, M.~M.} \&
  \bibinfo{author}{Hyart, T.}
\newblock \bibinfo{title}{Topological properties of multilayers and surface
  steps in the {SnTe} material class}.
\newblock \emph{\bibinfo{journal}{Phys. Rev. B}}
  \textbf{\bibinfo{volume}{100}}, \bibinfo{pages}{121107(R)}
  (\bibinfo{year}{2019}).

\bibitem{assaf2017magnetooptical}
\bibinfo{author}{Assaf, B.~A.} \emph{et~al.}
\newblock \bibinfo{title}{Magnetooptical determination of a topological index}.
\newblock \emph{\bibinfo{journal}{npj Quant. Mater.}}
  \textbf{\bibinfo{volume}{2}}, \bibinfo{pages}{26} (\bibinfo{year}{2017}).

\bibitem{Krizman2018}
\bibinfo{author}{Krizman, G.} \emph{et~al.}
\newblock \bibinfo{title}{Dirac parameters and topological phase diagram of
  {Pb$_{1-x}$Sn$_{x}$Se} from magnetospectroscopy}.
\newblock \emph{\bibinfo{journal}{Phys. Rev. B}} \textbf{\bibinfo{volume}{98}},
  \bibinfo{pages}{245202} (\bibinfo{year}{2018}).

\bibitem{Liu2012d}
\bibinfo{author}{Liu, M.} \emph{et~al.}
\newblock \bibinfo{title}{Crossover between weak antilocalization and weak
  localization in a magnetically doped topological insulator}.
\newblock \emph{\bibinfo{journal}{Phys. Rev. Lett.}}
  \textbf{\bibinfo{volume}{108}}, \bibinfo{pages}{036805}
  (\bibinfo{year}{2012}).

\bibitem{Tkac:2019_PRL}
\bibinfo{author}{Tk\'a\v{c}, V.} \emph{et~al.}
\newblock \bibinfo{title}{Influence of an anomalous temperature dependence of
  the phase coherence length on the conductivity of magnetic topological
  insulators}.
\newblock \emph{\bibinfo{journal}{Phys. Rev. Lett.}}
  \textbf{\bibinfo{volume}{123}}, \bibinfo{pages}{036406}
  (\bibinfo{year}{2019}).

\bibitem{Sawicki:1986_PRL}
\bibinfo{author}{Sawicki, M.} \emph{et~al.}
\newblock \bibinfo{title}{Influence of s-d exchange interaction on the
  conductivity of {Cd$_{1-x}$Mn$_x$Se:In} in the weakly localized regime}.
\newblock \emph{\bibinfo{journal}{Phys. Rev. Lett.}}
  \textbf{\bibinfo{volume}{56}}, \bibinfo{pages}{508--511}
  (\bibinfo{year}{1986}).

\bibitem{Adhikari:2019_PRB}
\bibinfo{author}{Adhikari, R.}, \bibinfo{author}{Volobuev, V.~V.},
  \bibinfo{author}{Faina, B.}, \bibinfo{author}{Springholz, G.} \&
  \bibinfo{author}{Bonanni, A.}
\newblock \bibinfo{title}{Ferromagnetic phase transition in topological
  crystalline insulator thin films: Interplay of anomalous {Hall} angle and
  magnetic anisotropy}.
\newblock \emph{\bibinfo{journal}{Phys. Rev. B}}
  \textbf{\bibinfo{volume}{100}}, \bibinfo{pages}{134422}
  (\bibinfo{year}{2019}).

\bibitem{Assaf2014}
\bibinfo{author}{Assaf, B.~A.} \emph{et~al.}
\newblock \bibinfo{title}{Quantum coherent transport in {SnTe} topological
  crystalline insulator thin films}.
\newblock \emph{\bibinfo{journal}{Appl. Phys. Lett.}}
  \textbf{\bibinfo{volume}{105}}, \bibinfo{pages}{102108}
  (\bibinfo{year}{2014}).

\bibitem{Akiyama2015}
\bibinfo{author}{Akiyama, R.}, \bibinfo{author}{Fujisawa, K.},
  \bibinfo{author}{Yamaguchi, T.}, \bibinfo{author}{Ishikawa, R.} \&
  \bibinfo{author}{Kuroda, S.}
\newblock \bibinfo{title}{Two-dimensional quantum transport of multivalley
  (111) surface state in topological crystalline insulator {SnTe} thin films}.
\newblock \emph{\bibinfo{journal}{Nano Research}} \textbf{\bibinfo{volume}{9}},
  \bibinfo{pages}{490--498} (\bibinfo{year}{2015}).

\bibitem{Yan:2020_JMST}
\bibinfo{author}{Yan, C.} \emph{et~al.}
\newblock \bibinfo{title}{Structure and topological transport in {Pb}-doping
  topological crystalline insulator {SnTe} (001) film}.
\newblock \emph{\bibinfo{journal}{J. Mater. Sci. Techn.}}
  \textbf{\bibinfo{volume}{44}}, \bibinfo{pages}{223--228}
  (\bibinfo{year}{2020}).

\bibitem{Wang:2020_PRB}
\bibinfo{author}{Wang, J.} \emph{et~al.}
\newblock \bibinfo{title}{Weak antilocalization beyond the fully diffusive
  regime in {Pb$_{1-x}$Sn$_{x}$Se} topological quantum wells}.
\newblock \emph{\bibinfo{journal}{Phys. Rev. B}}
  \textbf{\bibinfo{volume}{102}}, \bibinfo{pages}{155307}
  (\bibinfo{year}{2020}).

\bibitem{Peres:2014_JAP}
\bibinfo{author}{Peres, M.~L.} \emph{et~al.}
\newblock \bibinfo{title}{Experimental investigation of spin-orbit coupling in
  n-type {PbTe} quantum wells}.
\newblock \emph{\bibinfo{journal}{J. Appl. Phys.}}
  \textbf{\bibinfo{volume}{115}}, \bibinfo{pages}{093704}
  (\bibinfo{year}{2014}).

\bibitem{Afonin:2002_PRB}
\bibinfo{author}{Afonin, V.~V.}, \bibinfo{author}{Bergli, J.},
  \bibinfo{author}{Galperin, Y.~M.}, \bibinfo{author}{Gurevich, V.~L.} \&
  \bibinfo{author}{Kozub, V.~I.}
\newblock \bibinfo{title}{Possible weak temperature dependence of electron
  dephasing}.
\newblock \emph{\bibinfo{journal}{Phys. Rev. B}} \textbf{\bibinfo{volume}{66}},
  \bibinfo{pages}{165326} (\bibinfo{year}{2002}).

\bibitem{Lin:2002_JPC}
\bibinfo{author}{Lin, J.~J.} \& \bibinfo{author}{Bird, J.~P.}
\newblock \bibinfo{title}{Recent experimental studies of electron dephasing in
  metal and semiconductor mesoscopic structures}.
\newblock \emph{\bibinfo{journal}{J. Phys.: Condensed Matter}}
  \textbf{\bibinfo{volume}{14}}, \bibinfo{pages}{R501--R596}
  (\bibinfo{year}{2002}).

\bibitem{Jing:2016_NS}
\bibinfo{author}{Jing, Y.} \emph{et~al.}
\newblock \bibinfo{title}{Weak antilocalization and electron-electron
  interaction in coupled multiple-channel transport in a {Bi$_2$Se$_3$} thin
  film}.
\newblock \emph{\bibinfo{journal}{Nanoscale}} \textbf{\bibinfo{volume}{8}},
  \bibinfo{pages}{1879--1885} (\bibinfo{year}{2016}).

\bibitem{Islam2019}
\bibinfo{author}{Islam, S.} \emph{et~al.}
\newblock \bibinfo{title}{Low-temperature saturation of phase coherence length
  in topological insulators}.
\newblock \emph{\bibinfo{journal}{Phys. Rev. B}} \textbf{\bibinfo{volume}{99}},
  \bibinfo{pages}{245407} (\bibinfo{year}{2019}).

\bibitem{Rosen:2019_PRB}
\bibinfo{author}{Rosen, I.~T.} \emph{et~al.}
\newblock \bibinfo{title}{Absence of strong localization at low conductivity in
  the topological surface state of low-disorder {Sb}$_{2}${Te}$_{3}$}.
\newblock \emph{\bibinfo{journal}{Phys. Rev. B}} \textbf{\bibinfo{volume}{99}},
  \bibinfo{pages}{201101(R)} (\bibinfo{year}{2019}).

\bibitem{Nakamura:2020_NC}
\bibinfo{author}{Nakamura, H.} \emph{et~al.}
\newblock \bibinfo{title}{Robust weak antilocalization due to spin-orbital
  entanglement in {Dirac} material {Sr$_3$SnO}}.
\newblock \emph{\bibinfo{journal}{Nat. Commun.}} \textbf{\bibinfo{volume}{11}},
  \bibinfo{pages}{1161} (\bibinfo{year}{2020}).

\bibitem{SM}
\bibinfo{note}{See Supplemental Material at [URL will be inserted by publisher]
  for additional information on the Berry phase and conductivity calculations,
  structural characterization and electrical properties of the epilayers,
  fitting of the data with the full HLN formula, phase coherence length in
  various topological systems, and fitting of the magnetoresistance for the
  in-plane magnetic field.}

\bibitem{Wenk2010}
\bibinfo{author}{Wenk, P.} \& \bibinfo{author}{Kettemann, S.}
\newblock \bibinfo{title}{Dimensional dependence of weak localization
  corrections and spin relaxation in quantum wires with {R}ashba spin-orbit
  coupling}.
\newblock \emph{\bibinfo{journal}{Phys. Rev. B}} \textbf{\bibinfo{volume}{81}},
  \bibinfo{pages}{125309} (\bibinfo{year}{2010}).

\bibitem{Wojek2013}
\bibinfo{author}{Wojek, B.~M.} \emph{et~al.}
\newblock \bibinfo{title}{Spin-polarized (001) surface states of the
  topological crystalline insulator {Pb$_{0.73}$Sn$_{0.27}$Se}}.
\newblock \emph{\bibinfo{journal}{Phys. Rev. B}} \textbf{\bibinfo{volume}{87}},
  \bibinfo{pages}{115106} (\bibinfo{year}{2013}).

\bibitem{Xu:2015_NC}
\bibinfo{author}{Xu, S.-Y.} \emph{et~al.}
\newblock \bibinfo{title}{Unconventional transformation of spin {Dirac} phase
  across a topological quantum phase transition}.
\newblock \emph{\bibinfo{journal}{Nat. Commun.}} \textbf{\bibinfo{volume}{6}},
  \bibinfo{pages}{6870} (\bibinfo{year}{2015}).

\bibitem{Preier1979}
\bibinfo{author}{Preier, H.}
\newblock \bibinfo{title}{Recent advances in lead-chalcogenide diode lasers}.
\newblock \emph{\bibinfo{journal}{Appl. Phys.}} \textbf{\bibinfo{volume}{20}},
  \bibinfo{pages}{189--206} (\bibinfo{year}{1979}).

\bibitem{Volobuev2017}
\bibinfo{author}{Volobuev, V.~V.} \emph{et~al.}
\newblock \bibinfo{title}{Giant {R}ashba splitting in {Pb$_{1-x}$Sn$_{x}$Te}
  (111) topological crystalline insulator films controlled by {Bi} doping in
  the bulk}.
\newblock \emph{\bibinfo{journal}{Advanced Materials}}
  \textbf{\bibinfo{volume}{29}}, \bibinfo{pages}{1604185}
  (\bibinfo{year}{2017}).

\bibitem{Mandal2017}
\bibinfo{author}{Mandal, P.~S.} \emph{et~al.}
\newblock \bibinfo{title}{Topologi.al quantum phase transition from mirror to
  time reversal symmetry protected topological insulator}.
\newblock \emph{\bibinfo{journal}{Nat. Commun.}} \textbf{\bibinfo{volume}{8}},
  \bibinfo{pages}{968} (\bibinfo{year}{2017}).

\bibitem{Assaf2017}
\bibinfo{author}{Assaf, B.~A.} \emph{et~al.}
\newblock \bibinfo{title}{Negative longitudinal magnetoresistance from the
  anomalous ${N} = 0$ {L}andau level in topological materials}.
\newblock \emph{\bibinfo{journal}{Phys. Rev. Lett.}}
  \textbf{\bibinfo{volume}{119}}, \bibinfo{pages}{106602}
  (\bibinfo{year}{2017}).

\bibitem{Kolwas:2013_pss}
\bibinfo{author}{Kolwas, K.~A.} \emph{et~al.}
\newblock \bibinfo{title}{Absence of nonlocal resistance in microstructures of
  {PbTe} quantum wells}.
\newblock \emph{\bibinfo{journal}{phys. stat. sol. (b)}}
  \textbf{\bibinfo{volume}{250}}, \bibinfo{pages}{37--47}
  (\bibinfo{year}{2013}).

\bibitem{Ramakrishnan2017}
\bibinfo{author}{Ramakrishnan, N.}, \bibinfo{author}{Lai, Y.~T.},
  \bibinfo{author}{Lara, S.}, \bibinfo{author}{Parish, M.~M.} \&
  \bibinfo{author}{Adam, S.}
\newblock \bibinfo{title}{Equivalence of effective medium and random resistor
  network models for disorder-induced unsaturating linear magnetoresistance}.
\newblock \emph{\bibinfo{journal}{Phys. Rev. B}} \textbf{\bibinfo{volume}{96}},
  \bibinfo{pages}{224203} (\bibinfo{year}{2017}).

\bibitem{Fukuyama:1980_PTPS}
\bibinfo{author}{Fukuyama, H.}
\newblock \bibinfo{title}{{Non-Metallic Behaviors of Two-Dimensional Metals and
  Effect of Intervalley Impurity Scattering}}.
\newblock \emph{\bibinfo{journal}{Prog. Theor. Phys. Suppl.}}
  \textbf{\bibinfo{volume}{69}}, \bibinfo{pages}{220--231}
  (\bibinfo{year}{1980}).

\bibitem{Maekawa:1981_JPSJ}
\bibinfo{author}{Maekawa, S.} \& \bibinfo{author}{Fukuyama, H.}
\newblock \bibinfo{title}{Magnetoresistance in two-dimensional disordered
  systems: Effects of {Zeeman} splitting and spin-orbit scattering}.
\newblock \emph{\bibinfo{journal}{J. Phys. Soc. Jpn.}}
  \textbf{\bibinfo{volume}{50}}, \bibinfo{pages}{2516--2524}
  (\bibinfo{year}{1981}).

\bibitem{Lu:2011_PRL}
\bibinfo{author}{Lu, H.-Z.}, \bibinfo{author}{Shi, J.} \&
  \bibinfo{author}{Shen, S.-Q.}
\newblock \bibinfo{title}{Competition between weak localization and
  antilocalization in topological surface states}.
\newblock \emph{\bibinfo{journal}{Phys. Rev. Lett.}}
  \textbf{\bibinfo{volume}{107}}, \bibinfo{pages}{076801}
  (\bibinfo{year}{2011}).

\bibitem{Altshuler1981}
\bibinfo{author}{Al'tshuler, B.~L.} \& \bibinfo{author}{Aronov, A.~G.}
\newblock \bibinfo{title}{Magnetoresistance of thin films and of wires in a
  longitudinal magnetic field}.
\newblock \emph{\bibinfo{journal}{{JETP} Lett.}} \textbf{\bibinfo{volume}{33}},
  \bibinfo{pages}{499--501} (\bibinfo{year}{1981}).
\newblock \urlprefix\url{http://jetpletters.ac.ru/ps/1510/article_23070.shtml}.

\bibitem{Dugaev1984}
\bibinfo{author}{Dugaev, V.} \& \bibinfo{author}{Khmel'nitskii, D.}
\newblock \bibinfo{title}{Magnetoresistance of metal films with low impurity
  concentrations in a parallel magnetic field}.
\newblock \emph{\bibinfo{journal}{Sov. Phys. JETP}}
  \textbf{\bibinfo{volume}{59}}, \bibinfo{pages}{1038--1041}
  (\bibinfo{year}{1984}).
\newblock \urlprefix\url{http://www.jetp.ac.ru/cgi-bin/dn/e_059_05_1038.pdf}.

\bibitem{Beenakker1988}
\bibinfo{author}{Beenakker, C. W.~J.} \& \bibinfo{author}{van Houten, H.}
\newblock \bibinfo{title}{Boundary scattering and weak localization of
  electrons in a magnetic field}.
\newblock \emph{\bibinfo{journal}{Phys. Rev. B}} \textbf{\bibinfo{volume}{38}},
  \bibinfo{pages}{3232--3240} (\bibinfo{year}{1988}).

\bibitem{Bauer1992}
\bibinfo{author}{Bauer, G.}, \bibinfo{author}{Pascher, H.} \&
  \bibinfo{author}{Zawadzki, W.}
\newblock \bibinfo{title}{Magneto-optical properties of semimagnetic lead
  chalcogenides}.
\newblock \emph{\bibinfo{journal}{Semicond. Sci. Techn.}}
  \textbf{\bibinfo{volume}{7}}, \bibinfo{pages}{703--723}
  (\bibinfo{year}{1992}).

\bibitem{Malshukov:1997_PRB}
\bibinfo{author}{Mal'shukov, A.~G.}, \bibinfo{author}{Chao, K.~A.} \&
  \bibinfo{author}{Willander, M.}
\newblock \bibinfo{title}{Magnetoresistance of a weakly disordered {III-V}
  semiconductor quantum well in a magnetic field parallel to interfaces}.
\newblock \emph{\bibinfo{journal}{Phys. Rev. B}} \textbf{\bibinfo{volume}{56}},
  \bibinfo{pages}{6436--6439} (\bibinfo{year}{1997}).

\bibitem{Glazov:2009_SST}
\bibinfo{author}{Glazov, M.~M.} \& \bibinfo{author}{Golub, L.~E.}
\newblock \bibinfo{title}{Spin{\textendash}orbit interaction and weak
  localization in heterostructures}.
\newblock \emph{\bibinfo{journal}{Semicon. Sci. Techn.}}
  \textbf{\bibinfo{volume}{24}}, \bibinfo{pages}{064007}
  (\bibinfo{year}{2009}).

\bibitem{SM_Altshuler1982}
\bibinfo{author}{Altshuler, B.~L.}, \bibinfo{author}{Aronov, A.~G.} \&
  \bibinfo{author}{Khmelnitsky, D.~E.}
\newblock \bibinfo{title}{Effects of electron-electron collisions with small
  energy transfers on quantum localisation}.
\newblock \emph{\bibinfo{journal}{J. Phys. C: Solid State Phys.}}
  \textbf{\bibinfo{volume}{15}}, \bibinfo{pages}{7367--7386}
  (\bibinfo{year}{1982}).

\bibitem{Tkachov2011}
\bibinfo{author}{Tkachov, G.} \& \bibinfo{author}{Hankiewicz, E.~M.}
\newblock \bibinfo{title}{Weak antilocalization in {HgTe} quantum wells and
  topological surface states: Massive versus massless {D}irac fermions}.
\newblock \emph{\bibinfo{journal}{Phys. Rev. B}} \textbf{\bibinfo{volume}{84}},
  \bibinfo{pages}{035444} (\bibinfo{year}{2011}).

\bibitem{Tkachov2013}
\bibinfo{author}{Tkachov, G.} \& \bibinfo{author}{Hankiewicz, E.~M.}
\newblock \bibinfo{title}{Spin-helical transport in normal and superconducting
  topological insulators}.
\newblock \emph{\bibinfo{journal}{phys. stat. solidi (b)}}
  \textbf{\bibinfo{volume}{250}}, \bibinfo{pages}{215--232}
  (\bibinfo{year}{2013}).

\bibitem{Mathur:2001_PRB}
\bibinfo{author}{Mathur, H.} \& \bibinfo{author}{Baranger, H.~U.}
\newblock \bibinfo{title}{Random {Berry} phase magnetoresistance as a probe of
  interface roughness in {Si} {MOSFET's}}.
\newblock \emph{\bibinfo{journal}{Phys. Rev. B}} \textbf{\bibinfo{volume}{64}},
  \bibinfo{pages}{235325} (\bibinfo{year}{2001}).

\bibitem{Minkov:2004_PRB}
\bibinfo{author}{Minkov, G.~M.} \emph{et~al.}
\newblock \bibinfo{title}{Transverse negative magnetoresistance of
  two-dimensional structures in the presence of a strong in-plane magnetic
  field: Weak localization as a probe of interface roughness}.
\newblock \emph{\bibinfo{journal}{Phys. Rev. B}} \textbf{\bibinfo{volume}{70}},
  \bibinfo{pages}{035304} (\bibinfo{year}{2004}).

\bibitem{Lin2013}
\bibinfo{author}{Lin, C.~J.} \emph{et~al.}
\newblock \bibinfo{title}{Parallel field magnetoresistance in topological
  insulator thin films}.
\newblock \emph{\bibinfo{journal}{Phys. Rev. B}} \textbf{\bibinfo{volume}{88}},
  \bibinfo{pages}{041307(R)} (\bibinfo{year}{2013}).

\bibitem{Dybko:2018_arXiv}
\bibinfo{author}{Dybko, K.} \emph{et~al.}
\newblock \bibinfo{title}{Probing spatial extent of topological surface states
  by weak antilocalization experiments}.
\newblock \emph{\bibinfo{journal}{preprint: arxiv.org/abs/1812.08711}}
  (\bibinfo{year}{2018}).

\end{thebibliography}

\begin{thebibliography}{100}
\expandafter\ifx\csname url\endcsname\relax
  \def\url#1{\texttt{#1}}\fi
\expandafter\ifx\csname urlprefix\endcsname\relax\def\urlprefix{URL }\fi
\providecommand{\bibinfo}[2]{#2}
\providecommand{\eprint}[2][]{\url{#2}}

\bibitem{SM_Hsieh2012}
\bibinfo{author}{Hsieh, T.~H.} \emph{et~al.}
\newblock \bibinfo{title}{Topological crystalline insulators in the {SnTe}
  material class}.
\newblock \emph{\bibinfo{journal}{Nat. Commun.}} \textbf{\bibinfo{volume}{3}},
  \bibinfo{pages}{982} (\bibinfo{year}{2012}).

\bibitem{SM_Brzezicki2019}
\bibinfo{author}{Brzezicki, W.}, \bibinfo{author}{Wysoki{\'{n}}ski, M.~M.} \&
  \bibinfo{author}{Hyart, T.}
\newblock \bibinfo{title}{Topological properties of multilayers and surface
  steps in the {SnTe} material class}.
\newblock \emph{\bibinfo{journal}{Phys. Rev. B}}
  \textbf{\bibinfo{volume}{100}}, \bibinfo{pages}{121107(R)}
  (\bibinfo{year}{2019}).

\bibitem{SM_Maestro2013}
\bibinfo{author}{Maestro, A.~D.}, \bibinfo{author}{Hyart, T.} \&
  \bibinfo{author}{Rosenow, B.}
\newblock \bibinfo{title}{Backscattering between helical edge states via
  dynamic nuclear polarization}.
\newblock \emph{\bibinfo{journal}{Phys. Rev. B}} \textbf{\bibinfo{volume}{87}},
  \bibinfo{pages}{165440} (\bibinfo{year}{2013}).

\bibitem{SM_Pikulin2014}
\bibinfo{author}{Pikulin, D.~I.} \& \bibinfo{author}{Hyart, T.}
\newblock \bibinfo{title}{Interplay of exciton condensation and the quantum
  spin {H}all effect in {InAs}/{GaSb} bilayers}.
\newblock \emph{\bibinfo{journal}{Phys. Rev. Lett.}}
  \textbf{\bibinfo{volume}{112}}, \bibinfo{pages}{176403}
  (\bibinfo{year}{2014}).

\bibitem{SM_Wang2017}
\bibinfo{author}{Wang, J.}, \bibinfo{author}{Meir, Y.} \&
  \bibinfo{author}{Gefen, Y.}
\newblock \bibinfo{title}{Spontaneous breakdown of topological protection in
  two dimensions}.
\newblock \emph{\bibinfo{journal}{Phys. Rev. Lett.}}
  \textbf{\bibinfo{volume}{118}}, \bibinfo{pages}{046801}
  (\bibinfo{year}{2017}).

\bibitem{SM_Xue2018}
\bibinfo{author}{Xue, F.} \& \bibinfo{author}{MacDonald, A.~H.}
\newblock \bibinfo{title}{Time-reversal symmetry-breaking nematic insulators
  near quantum spin {H}all phase transitions}.
\newblock \emph{\bibinfo{journal}{Phys. Rev. Lett.}}
  \textbf{\bibinfo{volume}{120}}, \bibinfo{pages}{186802}
  (\bibinfo{year}{2018}).

\bibitem{SM_Vaeyrynen2014}
\bibinfo{author}{V{\"{a}}yrynen, J.~I.}, \bibinfo{author}{Goldstein, M.},
  \bibinfo{author}{Gefen, Y.} \& \bibinfo{author}{Glazman, L.~I.}
\newblock \bibinfo{title}{Resistance of helical edges formed in a semiconductor
  heterostructure}.
\newblock \emph{\bibinfo{journal}{Phys. Rev. B}} \textbf{\bibinfo{volume}{90}},
  \bibinfo{pages}{115309} (\bibinfo{year}{2014}).

\bibitem{SM_Wenk2010}
\bibinfo{author}{Wenk, P.} \& \bibinfo{author}{Kettemann, S.}
\newblock \bibinfo{title}{Dimensional dependence of weak localization
  corrections and spin relaxation in quantum wires with {R}ashba spin-orbit
  coupling}.
\newblock \emph{\bibinfo{journal}{Phys. Rev. B}} \textbf{\bibinfo{volume}{81}},
  \bibinfo{pages}{125309} (\bibinfo{year}{2010}).

\bibitem{SM_Hikami1980}
\bibinfo{author}{Hikami, S.}, \bibinfo{author}{Larkin, A.~I.} \&
  \bibinfo{author}{Nagaoka, Y.}
\newblock \bibinfo{title}{Spin-orbit interaction and magnetoresistance in the
  two dimensional random system}.
\newblock \emph{\bibinfo{journal}{Prog. Theor. Phys.}}
  \textbf{\bibinfo{volume}{63}}, \bibinfo{pages}{707--710}
  (\bibinfo{year}{1980}).

\bibitem{SM_Iordanskii1994}
\bibinfo{author}{Iordanskii, S.}, \bibinfo{author}{Lyanda-Geller, Y.} \&
  \bibinfo{author}{Pikus, G.}
\newblock \bibinfo{title}{Magnetoresistance of thin films and of wires in a
  longitudinal magnetic field}.
\newblock \emph{\bibinfo{journal}{{JETP} Letters}}
  \textbf{\bibinfo{volume}{33}}, \bibinfo{pages}{206--211}
  (\bibinfo{year}{1994}).
\newblock
  \urlprefix\url{http://www.jetpletters.ac.ru/ps/1323/article_20010.shtml}.

\bibitem{SM_Knap1996}
\bibinfo{author}{Knap, W.} \emph{et~al.}
\newblock \bibinfo{title}{Weak antilocalization and spin precession in quantum
  wells}.
\newblock \emph{\bibinfo{journal}{Phys. Rev. B}} \textbf{\bibinfo{volume}{53}},
  \bibinfo{pages}{3912--3924} (\bibinfo{year}{1996}).

\bibitem{SM_zhao2015tuning}
\bibinfo{author}{Zhao, L.}, \bibinfo{author}{Wang, J.}, \bibinfo{author}{Gu,
  B.-L.} \& \bibinfo{author}{Duan, W.}
\newblock \bibinfo{title}{Tuning surface {Dirac} valleys by strain in
  topological crystalline insulators}.
\newblock \emph{\bibinfo{journal}{Phys. Rev. B}} \textbf{\bibinfo{volume}{91}},
  \bibinfo{pages}{195320} (\bibinfo{year}{2015}).

\bibitem{SM_Villars2016}
\bibinfo{title}{{BaF$_2$ thermal expansion: Datasheet from "PAULING FILE
  Multinaries Edition -- 2012}} (\bibinfo{year}{2012}).
\newblock
  \urlprefix\url{https://materials.springer.com/isp/physical-property/docs/ppp_bf08e5451a8c3231b851a1815ab18e9a}.
\newblock \bibinfo{note}{Part of Springer Materials}.

\bibitem{SM_LandoltBornstein1998}
\bibinfo{title}{{Lead selenide (PbSe) crystal structure, lattice parameters,
  thermal expansion: Datasheet from {L}andolt-{B}{\"o}rnstein - Group {III}
  Condensed Matter {\textperiodcentered} Volume {41C}: "Non-Tetrahedrally
  Bonded Elements and Binary Compounds {I}}} (\bibinfo{year}{1998}).
\newblock \bibinfo{note}{Part of Springer Materials}.

\bibitem{SM_mccann1987phase}
\bibinfo{author}{McCann, P.~J.}, \bibinfo{author}{Fuchs, J.},
  \bibinfo{author}{Feit, Z.} \& \bibinfo{author}{Fonstad, C.~G.}
\newblock \bibinfo{title}{Phase equilibria and liquid-phase-epitaxy growth of
  {PbSnSeTe} lattice-matched to {PbSe}}.
\newblock \emph{\bibinfo{journal}{J. Appl. Phys.}}
  \textbf{\bibinfo{volume}{62}}, \bibinfo{pages}{2994--3000}
  (\bibinfo{year}{1987}).

\bibitem{SM_assaf2017magnetooptical}
\bibinfo{author}{Assaf, B.~A.} \emph{et~al.}
\newblock \bibinfo{title}{Magnetooptical determination of a topological index}.
\newblock \emph{\bibinfo{journal}{npj Quant. Mater.}}
  \textbf{\bibinfo{volume}{2}}, \bibinfo{pages}{26} (\bibinfo{year}{2017}).

\bibitem{SM_Tranta1988}
\bibinfo{author}{Tranta, B.} \& \bibinfo{author}{Clemens, H.}
\newblock \bibinfo{title}{Epitaxial growth of {PbTe} doping superlattices on
  (111) {BaF$_{2}$} and (100) {GaAs}}.
\newblock In \bibinfo{editor}{Ferenczi, G.} \& \bibinfo{editor}{Beleznay, F.}
  (eds.) \emph{\bibinfo{booktitle}{New Developments in Semiconductor Physics}},
  vol. \bibinfo{volume}{301}, \bibinfo{pages}{281--285}
  (\bibinfo{publisher}{Springer-Verlag}, \bibinfo{year}{1988}).

\bibitem{SM_Kolwas:2013_pss}
\bibinfo{author}{Kolwas, K.~A.} \emph{et~al.}
\newblock \bibinfo{title}{Absence of nonlocal resistance in microstructures of
  {PbTe} quantum wells}.
\newblock \emph{\bibinfo{journal}{phys. stat. sol. (b)}}
  \textbf{\bibinfo{volume}{250}}, \bibinfo{pages}{37--47}
  (\bibinfo{year}{2013}).

\bibitem{SM_Parish2005}
\bibinfo{author}{Parish, M.~M.} \& \bibinfo{author}{Littlewood, P.~B.}
\newblock \bibinfo{title}{Classical magnetotransport of inhomogeneous
  conductors}.
\newblock \emph{\bibinfo{journal}{Phys. Rev. B}} \textbf{\bibinfo{volume}{72}},
  \bibinfo{pages}{094417} (\bibinfo{year}{2005}).

\bibitem{SM_Ping2014}
\bibinfo{author}{Ping, J.} \emph{et~al.}
\newblock \bibinfo{title}{Disorder-induced magnetoresistance in a
  two-dimensional electron system}.
\newblock \emph{\bibinfo{journal}{Phys. Rev. Lett.}}
  \textbf{\bibinfo{volume}{113}}, \bibinfo{pages}{047206}
  (\bibinfo{year}{2014}).

\bibitem{SM_Ramakrishnan2017}
\bibinfo{author}{Ramakrishnan, N.}, \bibinfo{author}{Lai, Y.~T.},
  \bibinfo{author}{Lara, S.}, \bibinfo{author}{Parish, M.~M.} \&
  \bibinfo{author}{Adam, S.}
\newblock \bibinfo{title}{Equivalence of effective medium and random resistor
  network models for disorder-induced unsaturating linear magnetoresistance}.
\newblock \emph{\bibinfo{journal}{Phys. Rev. B}} \textbf{\bibinfo{volume}{96}},
  \bibinfo{pages}{224203} (\bibinfo{year}{2017}).

\bibitem{SM_Cho2008}
\bibinfo{author}{Cho, S.} \& \bibinfo{author}{Fuhrer, M.~S.}
\newblock \bibinfo{title}{Charge transport and inhomogeneity near the minimum
  conductivity point in graphene}.
\newblock \emph{\bibinfo{journal}{Phys. Rev. B}} \textbf{\bibinfo{volume}{77}},
  \bibinfo{pages}{081402(R)} (\bibinfo{year}{2008}).

\bibitem{SM_Rosen:2019_PRB}
\bibinfo{author}{Rosen, I.~T.} \emph{et~al.}
\newblock \bibinfo{title}{Absence of strong localization at low conductivity in
  the topological surface state of low-disorder {Sb}$_{2}${Te}$_{3}$}.
\newblock \emph{\bibinfo{journal}{Phys. Rev. B}} \textbf{\bibinfo{volume}{99}},
  \bibinfo{pages}{201101(R)} (\bibinfo{year}{2019}).

\bibitem{SM_Hu2007}
\bibinfo{author}{Hu, J.}, \bibinfo{author}{Parish, M.~M.} \&
  \bibinfo{author}{Rosenbaum, T.~F.}
\newblock \bibinfo{title}{Nonsaturating magnetoresistance of inhomogeneous
  conductors: Comparison of experiment and simulation}.
\newblock \emph{\bibinfo{journal}{Phys. Rev. B}} \textbf{\bibinfo{volume}{75}},
  \bibinfo{pages}{214203} (\bibinfo{year}{2007}).

\bibitem{SM_Golub:2005_PRB}
\bibinfo{author}{Golub, L.~E.}
\newblock \bibinfo{title}{Weak antilocalization in high-mobility
  two-dimensional systems}.
\newblock \emph{\bibinfo{journal}{Phys. Rev. B}} \textbf{\bibinfo{volume}{71}},
  \bibinfo{pages}{235310} (\bibinfo{year}{2005}).

\bibitem{SM_Nestoklon:2011_SSC}
\bibinfo{author}{Nestoklon, M.}, \bibinfo{author}{Averkiev, N.} \&
  \bibinfo{author}{Tarasenko, S.}
\newblock \bibinfo{title}{Weak localization of two-dimensional {Dirac} fermions
  beyond the diffusion regime}.
\newblock \emph{\bibinfo{journal}{Solid State Commun.}}
  \textbf{\bibinfo{volume}{151}}, \bibinfo{pages}{1550--1553}
  (\bibinfo{year}{2011}).

\bibitem{SM_Kim2011}
\bibinfo{author}{Kim, Y.~S.} \emph{et~al.}
\newblock \bibinfo{title}{Thickness-dependent bulk properties and weak
  antilocalization effect in topological insulator {Bi$_{2}$Se$_{3}$}}.
\newblock \emph{\bibinfo{journal}{Phys. Rev. B}} \textbf{\bibinfo{volume}{84}},
  \bibinfo{pages}{073109} (\bibinfo{year}{2011}).

\bibitem{SM_Bao2012}
\bibinfo{author}{Bao, L.} \emph{et~al.}
\newblock \bibinfo{title}{Weak anti-localization and quantum oscillations of
  surface states in topological insulator {Bi$_{2}$Se$_{2}$Te}}.
\newblock \emph{\bibinfo{journal}{Sci. Rep.}} \textbf{\bibinfo{volume}{2}},
  \bibinfo{pages}{726} (\bibinfo{year}{2012}).

\bibitem{SM_Bansal2012}
\bibinfo{author}{Bansal, N.}, \bibinfo{author}{Kim, Y.~S.},
  \bibinfo{author}{Brahlek, M.}, \bibinfo{author}{Edrey, E.} \&
  \bibinfo{author}{Oh, S.}
\newblock \bibinfo{title}{Thickness-independent transport channels in
  topological insulator {Bi$_{2}$Se$_{3}$} thin films}.
\newblock \emph{\bibinfo{journal}{Phys. Rev. Lett.}}
  \textbf{\bibinfo{volume}{109}}, \bibinfo{pages}{116804}
  (\bibinfo{year}{2012}).

\bibitem{SM_Assaf2013}
\bibinfo{author}{Assaf, B.~A.} \emph{et~al.}
\newblock \bibinfo{title}{Linear magnetoresistance in topological insulator
  thin films: Quantum phase coherence effects at high temperatures}.
\newblock \emph{\bibinfo{journal}{Appl. Phys. Lett.}}
  \textbf{\bibinfo{volume}{102}}, \bibinfo{pages}{012102}
  (\bibinfo{year}{2013}).

\bibitem{SM_Lee2012a}
\bibinfo{author}{Lee, J.}, \bibinfo{author}{Park, J.}, \bibinfo{author}{Lee,
  J.-H.}, \bibinfo{author}{Kim, J.~S.} \& \bibinfo{author}{Lee, H.-J.}
\newblock \bibinfo{title}{Gate-tuned differentiation of surface-conducting
  states in {Bi$_{1.5}$Sb$_{0.5}$Te$_{1.7}$Se$_{1.3}$} topological-insulator
  thin crystals}.
\newblock \emph{\bibinfo{journal}{Phys. Rev. B}} \textbf{\bibinfo{volume}{86}},
  \bibinfo{pages}{245321} (\bibinfo{year}{2012}).

\bibitem{SM_Chiu2013}
\bibinfo{author}{Chiu, S.-P.} \& \bibinfo{author}{Lin, J.-J.}
\newblock \bibinfo{title}{Weak antilocalization in topological insulator
  {Bi$_{2}$Te$_{3}$} microflakes}.
\newblock \emph{\bibinfo{journal}{Phys. Rev. B}} \textbf{\bibinfo{volume}{87}},
  \bibinfo{pages}{035122} (\bibinfo{year}{2013}).

\bibitem{SM_Taskin2012}
\bibinfo{author}{Taskin, A.~A.}, \bibinfo{author}{Sasaki, S.},
  \bibinfo{author}{Segawa, K.} \& \bibinfo{author}{Ando, Y.}
\newblock \bibinfo{title}{Achieving surface quantum oscillations in topological
  insulator thin films of {Bi$_{2}$Se$_{3}$}}.
\newblock \emph{\bibinfo{journal}{Adv. Materials}}
  \textbf{\bibinfo{volume}{24}}, \bibinfo{pages}{5581--5585}
  (\bibinfo{year}{2012}).

\bibitem{SM_Shekhar2014}
\bibinfo{author}{Shekhar, C.} \emph{et~al.}
\newblock \bibinfo{title}{Evidence of surface transport and weak
  antilocalization in a single crystal of the {Bi$_{2}$Te$_{2}$Se} topological
  insulator}.
\newblock \emph{\bibinfo{journal}{Phys. Rev. B}} \textbf{\bibinfo{volume}{90}},
  \bibinfo{pages}{165140} (\bibinfo{year}{2014}).

\bibitem{SM_Pan2014}
\bibinfo{author}{Pan, Y.} \emph{et~al.}
\newblock \bibinfo{title}{Low carrier concentration crystals of the topological
  insulator {Bi$_{2-x}$Sb$_{x}$Te$_{3-y}$Se$_{y}$}: a magnetotransport study}.
\newblock \emph{\bibinfo{journal}{New J. Phys.}} \textbf{\bibinfo{volume}{16}},
  \bibinfo{pages}{123035} (\bibinfo{year}{2014}).

\bibitem{SM_Wang2016b}
\bibinfo{author}{Wang, W.~J.}, \bibinfo{author}{Gao, K.~H.} \&
  \bibinfo{author}{Li, Z.~Q.}
\newblock \bibinfo{title}{Thickness-dependent transport channels in topological
  insulator {Bi$_{2}$Se$_{3}$} thin films grown by magnetron sputtering}.
\newblock \emph{\bibinfo{journal}{Sci. Rep.}} \textbf{\bibinfo{volume}{6}},
  \bibinfo{pages}{25291} (\bibinfo{year}{2016}).

\bibitem{SM_Banerjee2014}
\bibinfo{author}{Banerjee, K.} \emph{et~al.}
\newblock \bibinfo{title}{Defect-induced negative magnetoresistance and surface
  state robustness in the topological insulator {BiSbTeSe$_{2}$}}.
\newblock \emph{\bibinfo{journal}{Phys. Rev. B}} \textbf{\bibinfo{volume}{90}},
  \bibinfo{pages}{235427} (\bibinfo{year}{2014}).

\bibitem{SM_Gopal2015}
\bibinfo{author}{Gopal, R.~K.}, \bibinfo{author}{Singh, S.},
  \bibinfo{author}{Chandra, R.} \& \bibinfo{author}{Mitra, C.}
\newblock \bibinfo{title}{Weak-antilocalization and surface dominated transport
  in topological insulator {Bi$_{2}$Se$_{2}$Te}}.
\newblock \emph{\bibinfo{journal}{{AIP} Advances}}
  \textbf{\bibinfo{volume}{5}}, \bibinfo{pages}{047111} (\bibinfo{year}{2015}).

\bibitem{SM_Ngabonziza2016}
\bibinfo{author}{Ngabonziza, P.} \emph{et~al.}
\newblock \bibinfo{title}{Gate-tunable transport properties of in situ capped
  {Bi$_{2}$Te$_{3}$} topological insulator thin films}.
\newblock \emph{\bibinfo{journal}{Adv. Electronic Materials}}
  \textbf{\bibinfo{volume}{2}}, \bibinfo{pages}{1600157}
  (\bibinfo{year}{2016}).

\bibitem{SM_Zhang2016}
\bibinfo{author}{Zhang, M.} \emph{et~al.}
\newblock \bibinfo{title}{Electrical transport properties and morphology of
  topological insulator {Bi$_{2}$Se$_{3}$} thin films with different thickness
  prepared by magnetron sputtering}.
\newblock \emph{\bibinfo{journal}{Thin Solid Films}}
  \textbf{\bibinfo{volume}{603}}, \bibinfo{pages}{289--293}
  (\bibinfo{year}{2016}).

\bibitem{SM_Assaf2012}
\bibinfo{author}{Assaf, B.~A.} \emph{et~al.}
\newblock \bibinfo{title}{Modified electrical transport probe design for
  standard magnetometer}.
\newblock \emph{\bibinfo{journal}{Rev. Sci. Instrum.}}
  \textbf{\bibinfo{volume}{83}}, \bibinfo{pages}{033904}
  (\bibinfo{year}{2012}).

\bibitem{SM_Assaf2014}
\bibinfo{author}{Assaf, B.~A.} \emph{et~al.}
\newblock \bibinfo{title}{Quantum coherent transport in {SnTe} topological
  crystalline insulator thin films}.
\newblock \emph{\bibinfo{journal}{Appl. Phys. Lett.}}
  \textbf{\bibinfo{volume}{105}}, \bibinfo{pages}{102108}
  (\bibinfo{year}{2014}).

\bibitem{SM_Akiyama2015}
\bibinfo{author}{Akiyama, R.}, \bibinfo{author}{Fujisawa, K.},
  \bibinfo{author}{Yamaguchi, T.}, \bibinfo{author}{Ishikawa, R.} \&
  \bibinfo{author}{Kuroda, S.}
\newblock \bibinfo{title}{Two-dimensional quantum transport of multivalley
  (111) surface state in topological crystalline insulator {SnTe} thin films}.
\newblock \emph{\bibinfo{journal}{Nano Research}} \textbf{\bibinfo{volume}{9}},
  \bibinfo{pages}{490--498} (\bibinfo{year}{2015}).

\bibitem{SM_Zhang2015a}
\bibinfo{author}{Zhang, C.} \emph{et~al.}
\newblock \bibinfo{title}{Highly tunable {B}erry phase and ambipolar field
  effect in topological crystalline insulator {Pb$_{1-x}$Sn$_{x}$Se}}.
\newblock \emph{\bibinfo{journal}{Nano Letters}} \textbf{\bibinfo{volume}{15}},
  \bibinfo{pages}{2161--2167} (\bibinfo{year}{2015}).

\bibitem{SM_Zhong2015}
\bibinfo{author}{Zhong, R.} \emph{et~al.}
\newblock \bibinfo{title}{Surface-state-dominated transport in crystals of the
  topological crystalline insulator {In}-doped {Pb$_{1-x}$Sn$_{x}$Te}}.
\newblock \emph{\bibinfo{journal}{Phys. Rev. B}} \textbf{\bibinfo{volume}{91}},
  \bibinfo{pages}{195321} (\bibinfo{year}{2015}).

\bibitem{SM_Zhang2019}
\bibinfo{author}{Zhang, A.} \emph{et~al.}
\newblock \bibinfo{title}{Topological phase transition and highly tunable
  topological transport in topological crystalline insulator
  {Pb$_{1-x}$Sn$_{x}$Te} (111) thin films}.
\newblock \emph{\bibinfo{journal}{Nanotechnology}}
  \textbf{\bibinfo{volume}{30}}, \bibinfo{pages}{275703}
  (\bibinfo{year}{2019}).

\bibitem{SM_Zou2019}
\bibinfo{author}{Zou, K.} \emph{et~al.}
\newblock \bibinfo{title}{Revealing surface-state transport in ultrathin
  topological crystalline insulator {SnTe} films}.
\newblock \emph{\bibinfo{journal}{{APL} Materials}}
  \textbf{\bibinfo{volume}{7}}, \bibinfo{pages}{051106} (\bibinfo{year}{2019}).

\bibitem{SM_Dybko:2018_arXiv}
\bibinfo{author}{Dybko, K.} \emph{et~al.}
\newblock \bibinfo{title}{Probing spatial extent of topological surface states
  by weak antilocalization experiments}.
\newblock \emph{\bibinfo{journal}{preprint: arxiv.org/abs/1812.08711}}
  (\bibinfo{year}{2018}).

\bibitem{SM_Wang:2020_PRB}
\bibinfo{author}{Wang, J.} \emph{et~al.}
\newblock \bibinfo{title}{Weak antilocalization beyond the fully diffusive
  regime in {Pb$_{1-x}$Sn$_{x}$Se} topological quantum wells}.
\newblock \emph{\bibinfo{journal}{Phys. Rev. B}}
  \textbf{\bibinfo{volume}{102}}, \bibinfo{pages}{155307}
  (\bibinfo{year}{2020}).

\bibitem{SM_Datta1995}
\bibinfo{author}{Datta, S.}
\newblock \emph{\bibinfo{title}{Electronic Transport in Mesoscopic Systems}}
  (\bibinfo{publisher}{Cambridge University Press}, \bibinfo{year}{1995}).

\bibitem{SM_Prinz1999}
\bibinfo{author}{Prinz, A.} \emph{et~al.}
\newblock \bibinfo{title}{Electron localization in {$n$-Pb$_{1-x}$Eu$_{x}$Te}}.
\newblock \emph{\bibinfo{journal}{Phys. Rev. B}} \textbf{\bibinfo{volume}{59}},
  \bibinfo{pages}{12983--12990} (\bibinfo{year}{1999}).

\bibitem{SM_Lin:2002_JPC}
\bibinfo{author}{Lin, J.~J.} \& \bibinfo{author}{Bird, J.~P.}
\newblock \bibinfo{title}{Recent experimental studies of electron dephasing in
  metal and semiconductor mesoscopic structures}.
\newblock \emph{\bibinfo{journal}{J. Phys.: Condensed Matter}}
  \textbf{\bibinfo{volume}{14}}, \bibinfo{pages}{R501--R596}
  (\bibinfo{year}{2002}).

\end{thebibliography}
\end{document}